# GenoBERT: A Language Model for Accurate Genotype Imputation

Lei Huang[1], Chuan Qiu[2], Kuan-Jui Su[2], Anqi Liu[2], Yun Gong[2], Weiqiang Lin[2], Lindong Jiang[2], Chen Zhao[3], Meng Song[1,4], Jeffrey Deng[5], Qing Tian[2], Zhe Luo[2], Ping Gong[6], Hui Shen[2], Chaoyang Zhang[1,*], and Hong-Wen Deng[2,*]

[1]School of Computing Sciences and Computer Engineering, University of Southern Mississippi, Hattiesburg, MS, USA, [2]Tulane Center for Biomedical Informatics and Genomics, Deming Department of Medicine, School of Medicine, Tulane University, New Orleans, LA, USA, [3]Kennesaw State University, Atlanta, GA, USA,[4]College of Science, Xi'an Shiyou University, Xi'an, Shaanxi, China, [5]Geisel School of Medicine, Dartmouth College, Hanover, NH, USA, [6]Environmental Laboratory, U.S. Army Engineer Research and Development Center, Vicksburg, MS, USA

*To whom correspondence should be addressed.

## Abstract

Genotype imputation enables dense variant coverage for genome-wide association and risk-prediction studies, yet conventional reference-panel methods remain limited by ancestry bias and reduced rare-variant accuracy. We present Genotype Bidirectional Encoder Representations from Transformers model (GenoBERT), a transformer-based, reference-free framework that tokenizes phased genotypes and uses self-attention mechanism to capture both short- and long-range linkage disequilibrium (LD) dependencies. Benchmarking on two independent datasets—the Louisiana Osteoporosis Study (LOS) and the 1000 Genomes Project (1KGP)—across ancestry groups and multiple genotypes missing levels (5–50%) shows that GenoBERT achieves the highest overall accuracy compared to the other four baselines (Beagle5.4, SCDA, BiU-Net, and STICI). At practical sparsity levels ($\leqslant$ 25% missing), GenoBERT attains high overall imputation accuracies ($r^2 \approx 0.98$) across datasets, and even maintains robust performance ($r^2 > 0.90$) at 50% missing level. Experiment results across different ancestries confirm consistent gains for both datasets, with resilience to small sample sizes and weak LD. The 128-SNP (single-nucleotide polymorphism) context window (≈ 100Kb) was validated through LD-decay analyses as sufficient to span local correlation structures. By eliminating reference-panel dependence while preserving high accuracy, GenoBERT provides a scalable, and robust solution for genotype imputation and a foundation for downstream genomic modeling.

## Introduction

Genotype imputation is a central step in modern genomic analysis, enabling dense variant coverage from genotyping arrays and increasing statistical power for genome-wide association studies (GWAS) and polygenic risk prediction[1,2]. Most conventional imputation methods such as MACH[3], IMPUTE[1,4], Beagle[5,6], and Minimac[7] are derived from the Li & Stephens model[8], which models individual haplotypes as mosaics of reference haplotypes through recombination and mutation processes[9]. Grounded in population genetics, these models have delivered exceptionally high accuracy for common variants, with $r^2 > 0.95$ observed for variants with a minor allele frequency (MAF) above 5% given large, ancestry-matched reference panels[2,10]. The progressive expansion of reference panels—from the HapMap's 270 samples (Phase I & II)[11] to large-scale resources such as the Haplotype Reference Consortium (HRC)[12] and the Trans-Omics for Precision Medicine (TOPMed) program[13]—has enabled accurate imputation down to MAF of 0.1% and improved rare-variant under moderate missing levels[12,14].

Despite these advances, reference-panel-based methods remain constrained by their dependence on the population composition of the reference data and by persistent ancestry-specific biases. Accuracy declines markedly for populations that are under-represented in reference panels and for variants with low MAF[15–17]. Traditional genotype imputation methods rely on the ancestral concordance between study samples and reference panels, and deviations from this alignment substantially degrade imputation accuracy. In African Americans, O'Connell et al.[18] show markedly better performance with ancestry-enriched panels (e.g., TOPMed) than with European-dominated HRC (e.g., at 0.5% MAF, aggregate $r^2$ 0.75 vs 0.35). While modern implementations such as IMPUTE5[4] and Beagle5[6] achieve substantial computational efficiency and can scale to very large reference panels, genotype imputation using classical statistical methods remains fundamentally dependent on external reference panels and first-order Markov assumptions, where information between distant variants propagates sequentially through intermediate sites—a mechanism that falters when linkage disequilibrium is sparse[19].

What's more, most statistical genotype imputation methods, including Beagle, are based on the Li and Stephens haplotype copying framework, which assumes phased haplotypes and optimizes marginal genotype likelihoods. At heterozygous loci, phased configurations are often weakly identifiable[20], as the model is invariant to haplotype exchange in the absence of strong long-range linkage disequilibrium. Consequently, while dosage estimation can be highly accurate, phase orientation—particularly for rare variants—may remain uncertain.

Recent studies have therefore explored deep learning–based approaches that learn LD structure directly from study genotype matrices[19,21–25]. By decoupling imputation from fixed reference panels, deep learning–based methods offer potential advantages in adaptability and scalability across diverse populations. Deep learning–based models, while support both unphased and phased data, could employ asymmetric representations and phase-aware objectives that directly penalize phase errors, enabling more effective discrimination between alternative heterozygous haplotype configurations. Empirical evaluations confirm substantial runtime and memory advantages—up to 4× faster inference compared with traditional tools[23].

However, while existing deep learning models achieve comparable or better results for common variants, they typically perform worse for ultra-rare variants (MAF < 1%) and show reduced generalization across ancestries[24,26]. For example, an RNN-based imputation method evaluated on phased chromosome 22 data from both the 1KGP Phase 3 cohort (2,504 individuals) and a large HRC subset (27,165 individuals; 54,330 haplotypes) achieved overall imputation accuracy (Beagle's $R^2$) comparable to Li and Stephens–based methods such as Impute5 and Minimac3[25]. Notably, for low-frequency variants (MAF < 0.01), This RNN-based model's $R^2$ values were consistently slightly lower than those of Li and Stephens–based approaches and failed to produce estimates for variants with MAF < 0.005. Although this limitation was attributed to computational complexity, it may also reflect the absence of biologically informed priors in the model design.

Prior studies have shown that incorporating biologically meaningful structure—such as haplotype relationships[27] or other domain-specific constraints[20]—can substantially improve performance, particularly for rare variant imputation.

To address these challenges, we developed GenoBERT, a transformer-based, reference-free framework for genotype imputation that models phased genotypes as tokenized sequences and leverages self-attention to capture both short- and long-range LD dependencies. By integrating signals across distant loci within each genomic segment, GenoBERT preserves contextual information even under substantial genotype sparsity. In this study, we benchmark GenoBERT against four representative baselines—Beagle5[6], SCDA[21], BiU-Net[20], and STICI[19]—using the LOS[28] dataset and the 1KGP[29] dataset. We evaluate imputation accuracy across multiple missing-genotype ratios and across diverse population groups to assess both robustness and cross-ancestry generalization.

Our results demonstrate that GenoBERT consistently achieves the highest overall accuracy among all models tested, particularly under realistic conditions up to 25% missingness. Across both LOS and 1KGP, GenoBERT delivers superior $r^2$ (the squared Pearson correlation coefficient computed from observed and imputed genotypes.) and $F_1$ scores, minimal degradation with increasing missingness, and stable performance across MAF ranges and populations. Only under the extreme 50% missing condition does its advantage narrow, primarily for ultra-rare variants (MAF < 1%) in 1KGP/African and LOS/African American cohorts, where LD is highly fragmented and weak. In contrast, for populations with stronger LD such as South Asian and East Asian cohorts in 1KGP, GenoBERT maintained very high imputation accuracy even at 50% missing ($r^2$ = 0.9291 and 0.9561, respectively) for the rarest variants (MAF = 0.1% ~ 0.5%). These findings establish GenoBERT as a robust, and reference-free genotype-imputation framework that consistently delivers state-of-the-art performance across realistic and even moderately high missingness levels, outperforming existing baselines in most scenarios. They further indicate that techniques such as LD-adaptive attention and population-aware modeling could help mitigate the rare under-performance observed in extremely weak-LD settings.

# Materials and methods

## Model architecture

GenoBERT adopts a modular and highly customizable architecture (**Fig. 1)**, designed to flexibly balance parameter efficiency, model capacity, and biologically informed inductive biases. The model consists of a stack of N encoder blocks, each composed of an attention module followed by a one-dimensional CNN bottleneck module. To improve parameter efficiency, GenoBERT supports multiple parameter-sharing strategies across encoder layers. Specifically, users may configure the model to share (i) only the projection matrices for queries, keys, and values, (ii) only the CNN bottleneck module, (iii) both components, or (iv) neither, following a design similar to ALBERT[30]. To stabilize optimization in deep architectures, we incorporate a residual scaling factor $\alpha$ inspired by DeepNet[31], which modulates the contribution of residual connections before each addition. This design improves gradient flow and training stability when scaling to deeper models. For positional encoding, GenoBERT employs Rotary Positional Embeddings (RoPE)[32] by default, while also supporting a Relative Genomic Positional Bias (RGPB) term that can be directly added to attention scores. This bias encodes relative genomic distances and introduces biologically meaningful priors into the attention mechanism. In addition, we introduce a 1D CNN bottleneck module to explicitly capture local genomic patterns that are not easily modeled by attention alone. The attention component can be instantiated as standard multi-head attention (MHA) or extended with RGPB to incorporate genomic structure. Together, these components form a flexible architecture that enables adaptation to diverse genomic modeling tasks by adjusting parameter sharing, positional encoding, and inductive biases.

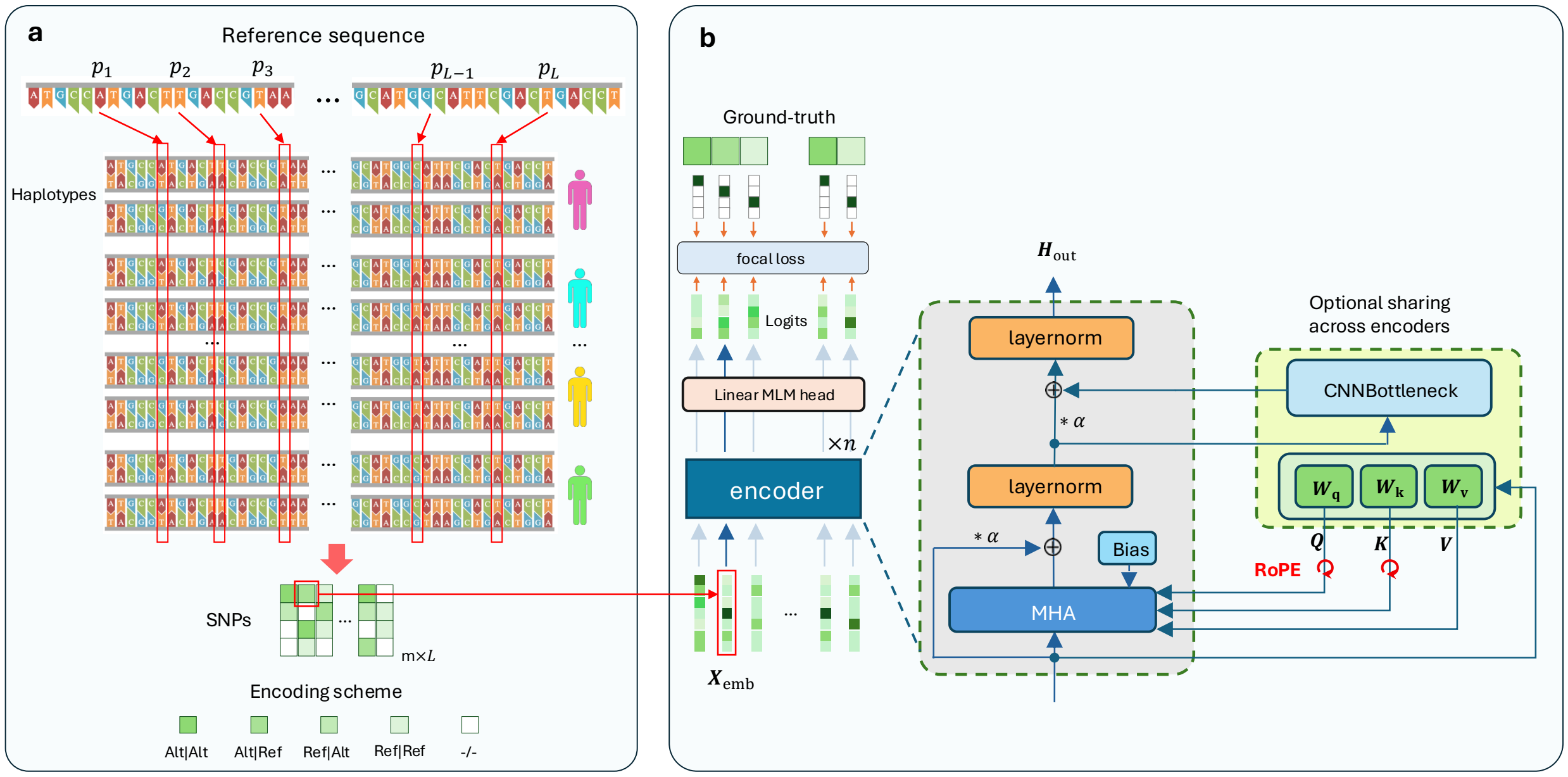

**Figure 1 | The architecture of GenoBERT. (a)** Illustration of the genotype representation used as model input. Genotype sequences are aligned to a reference genome and represented as phased haplotypes, from which single-nucleotide polymorphisms (SNPs) are highlighted and encoded into discrete tokens according to genotype states, e.g., homozygous reference ("Ref|Ref"), heterozygous ("Ref|Alt" or "Alt|Ref"), homozygous alternate ("Alt|Alt"), or missing ("-/-"). These tokens are subsequently mapped to input embeddings ($X_{emb}$). **(b)** Overview of the GenoBERT encoder architecture. The model consists of a stack of $N$ encoder blocks. Within each block, input embeddings are projected into query (Q), key (K), and value (V) representations and processed by a multi-head attention (MHA) module, optionally augmented with Rotary Positional Embeddings (RoPE) and a relative genomic positional bias (RGPB). The attention output is combined with a scaled residual connection (α) and followed by layer normalization. The resulting representation is then passed through a one-dimensional CNN bottleneck module to capture local genomic patterns. A second scaled residual connection and normalization produce the final encoder output. Selected components, including the Q/K/V projections and the CNN bottleneck module, can be shared across layers to improve parameter efficiency.

## Embedding

At the input level, GenoBERT begins with an embedding module that projects discrete SNP genotypes and token type indicators into a continuous latent space of hidden dimension. Although the current pretraining scheme relies solely on the masked language modeling (MLM) objective, we include token type embeddings for future extensibility (e.g., Next Segment Prediction training).

## Positional encoding

Positional information plays a crucial role in the effectiveness of large language models, as the attention mechanism alone does not inherently encode sequence order. GenoBERT leverages RoPE for encoding ordinal information of SNP segments. RoPE modifies the query and key vectors by applying position-dependent rotations in complex space after embedding projection (late combination). These rotations encode relative positions as phase shifts, preserving both local continuity and enabling generalization to unseen sequence lengths. RoPE's functional form is deterministic and scalable, allowing for natural extrapolation beyond the training horizon.

This late combination strategy offers two key benefits for genotype modeling: (i) it maintains the integrity of the original embeddings without magnitude distortion, and (ii) it allows the model to scale to longer genomic segments without architectural changes. Combined with our proposed relative genomic positional bias, this mechanism enables GenoBERT to jointly capture token identity, sequence structure, and positional locality.

## Attention with relative genomic positional bias (RGPB)

To ensure the model can distinguish between locally similar genotype segments from distinct genomic regions, GenoBERT further incorporates genomic position awareness via a relative positional bias mechanism in the attention layer.

Unlike typical NLP corpora, genotype matrices are extremely wide—containing millions of loci—but relatively shallow, with only hundreds or thousands of samples. This poses a challenge for directly applying transformer-based models to genotype data.

To address this, we introduce a segmentation strategy that reinterprets the genotype matrix through an NLP-inspired lens: each individual sample is treated as a "document," and each fixed-length genomic window as a "sentence" or segment. By sliding a window of arbitrary length and overlap across the sequence of SNPs, we transform the data into a corpus of genotype segments. This strategy, originally proposed in our prior work[20], enables efficient and scalable training by dividing the input into manageable, fixed-length segments while still capturing biological signal within each window.

However, segmentation introduces a new challenge: due to the high sparsity and limited variability in local SNP regions, many extracted segments appear identical or nearly so in allele patterns. While it may be tempting to discard such duplicates to reduce redundancy, doing so would eliminate key positional context—since these similar-looking segments may originate from distinct genomic regions. Conversely, retaining them without any positional encoding may confuse the model, as it would receive repeated content with no spatial differentiation.

To resolve this, GenoBERT incorporates genomic position awareness through a relative positional bias mechanism in the attention layer. Instead of using simple contiguous indices as in standard transformer models, we assign positions based on the actual genomic coordinates of SNPs. For each input sequence, the model computes a normalized positional bias vector derived from the actual genomic coordinates of SNPs within the segment. During both training and inference, this normalized bias vector is then incorporated into the transformer's attention computation as an additive positional prior, enabling the model to retain biologically consistent spatial context across segments.

**Fig. 2** illustrates the construction of the GenoBERT model's input. This process includes token conversion, masking, special token handling, and position encoding —establishes the structural framework from which the RGPB term is defined.

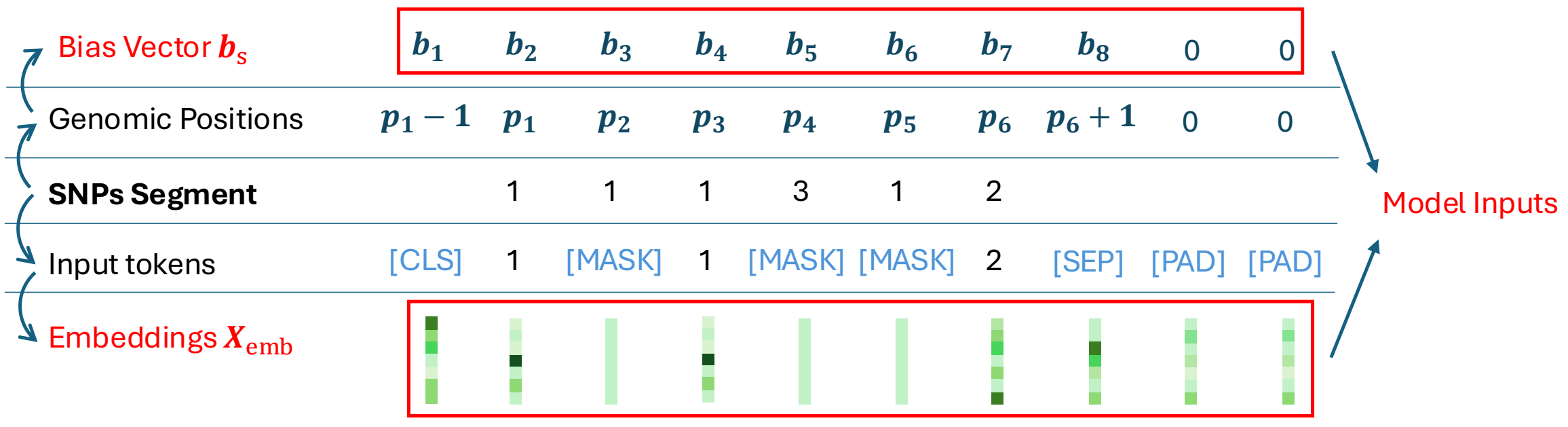

**Figure 2 | GenoBERT needs two parts as the model input: the SNPs segment embedding, and an optional bias vector when relative genomic positional bias is enabled.**

In constructing the model input, Each SNP segment is encoded as a sequence of integers representing phased genotypes: 1 = Ref|Ref , 2 = Ref|Alt, 3 = Alt|Ref, and 4 = Alt|Alt. The value 0 is reserved for the [MASK] token. To support denoising training, random SNPs are masked and replaced with a special [MASK] token. Additional specific tokens [CLS] and [SEP] will be added to the beginning and end of the sequence. To match the model's fixed input length, [PAD] tokens are appended as needed, then each token will be mapped to embedding of given size; genomic coordinates corresponding to each SNP (and interpolated positions for [CLS] and [SEP] tokens) are standardized to be used to compute the bias vector for the construction of the Relative Genomic Positional Bias (RGPB) term, which provides spatial information to the model. The [PAD] tokens are assigned a default position of zero and will be excluded from RGPB calculations later.

The RGPB term enables the model to distinguish structurally identical segments of different regions—preserving spatial context even when local patterns appear identical. This idea draws inspiration from ALiBi[33] which biases the query-key attention scores with a term that is proportional to their distance.

To adapt ALiBi for genotype data, assuming we apply the segmentation on a genotype dataset of $M$ SNPs using $t$ overlap and $L$ segment length, for any segments which start at $i_{th}$ column where $i \in [1, M]$, we extract the corresponding standardized genomic bias vector $\boldsymbol{b}_s \in \mathbb{R}^{\mathrm{L}}$ as in **Fig. 8** and let $\mathbf{1} \in \mathbb{R}^{\mathrm{L}}$ denote a column vector of ones. Then, for segment $s$, the relative positional bias matrix $\boldsymbol{B}_s \in \mathbb{R}^{\mathrm{L \times L}}$ can be constructed as:

$$\boldsymbol{B}_s = \mathbf{1}\boldsymbol{b}_s^{\mathrm{T}} - \boldsymbol{b}_s\mathbf{1}^{\mathrm{T}} \tag{1}$$

This formulation yields a matrix where each entry $B_{s,i,j} = p_j - p_i$, representing the relative genomic distance between SNP positions $i$ and $j$ in the segment. Each unique segment offset (based on its starting SNP index) corresponds to one such bias matrix. The index $s$ for the $s_{th}$ bias matrix is derived as:

$$s = \left\lfloor \frac{i}{L - t} \right\rfloor \tag{2}$$

Depending on runtime constraints, the bias matrices $\boldsymbol{B}_s$ can either be computed on the fly or precomputed and cached. If computational speed is prioritized, they can be generated dynamically at each training step without indexing by $s$. However, if memory reuse is desired, the matrices can be precomputed and stored using their segment index $s$ for retrieval. The computation can be illustrated in **Fig. 3**.

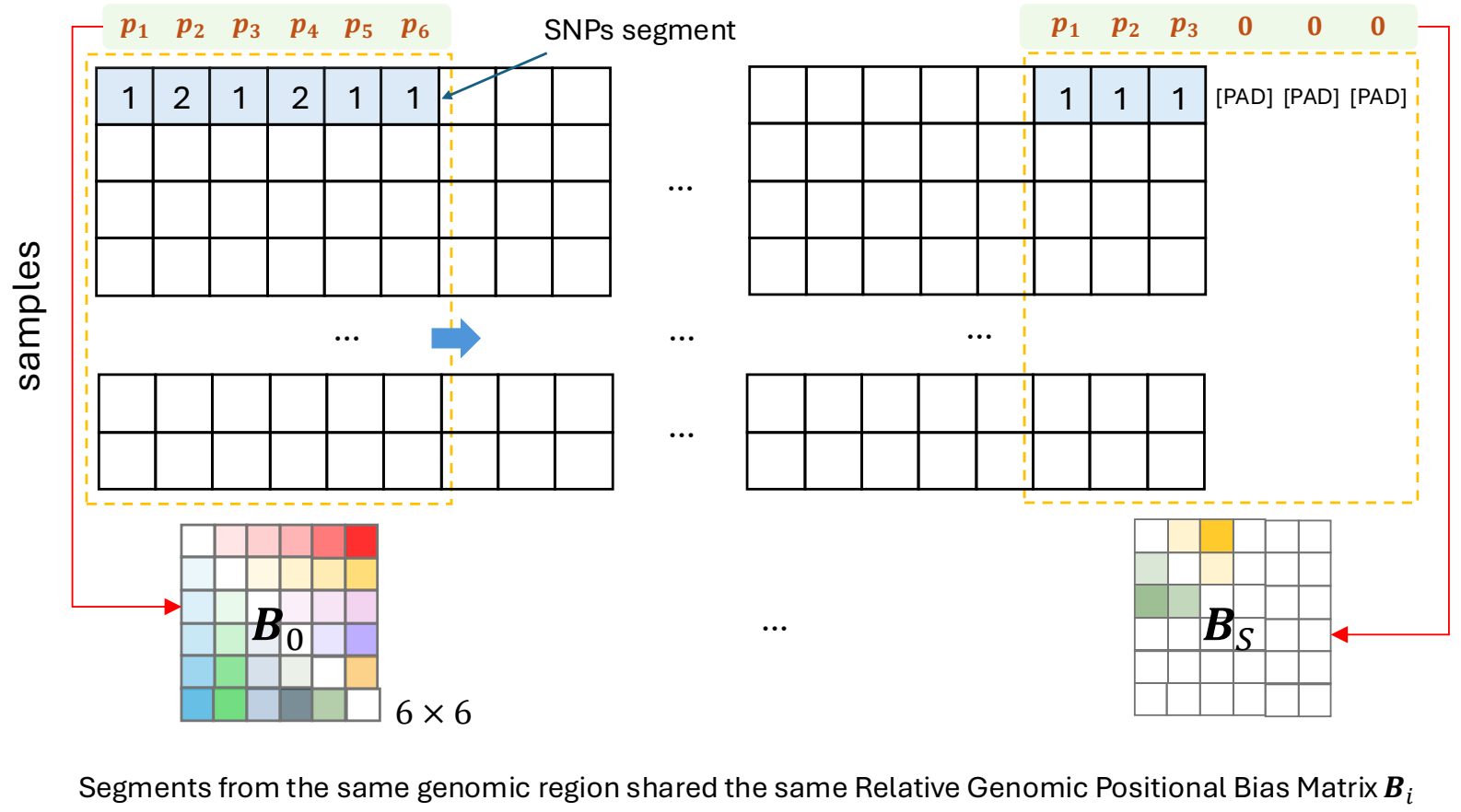

**Figure 3 | Construction of the bias matrices for GenoBERT's relative genomic positional encoding (segment length L=6 in this case).** Each segment's standardized genomic positions are used to compute a relative bias matrix $\boldsymbol{B}_s$ based on the SNP coordinates. The same bias matrix is shared across all segments that start at the same locus. Special tokens [CLS] and [SEP] are excluded from this illustration for simplicity. During the attention calculation, the padding mask will trim down the padding region as shown in the last bias matrix ($\boldsymbol{B}_s$).

An important practical consideration arises from the use of shuffled batches during training: individual segments within a batch may originate from distinct genomic loci. Consequently, the bias matrices—whether dynamically computed or loaded from precomputed storage—must be applied separately for each segment, introducing a substantial computational bottleneck. To address this limitation, we redesigned the batching strategy so that all segments within a batch share the same genomic boundaries. This alignment yields significantly improved training efficiency.

Unlike earlier transformer architectures such as T5[34] or ALiBi-based models where positional biases are globally defined and shared across inputs, GenoBERT employs a batch-level relative genomic positional bias (RGPB) that is computed dynamically for each batch and injected into the attention logits. This bias is further modulated by layer- and head-specific coefficients $\beta_{l,h}$, allowing each attention head in each layer to adaptively control the contribution of genomic distance information during attention computation:

$$\text{Attention}^{(l,h)} = \text{Softmax}\left(\frac{\boldsymbol{Q}^{(h)}\left(\boldsymbol{K}^{(h)}\right)^{\text{T}}}{\sqrt{d/H}} + \beta_{l,h} \cdot \boldsymbol{B}_s\right)\boldsymbol{V}^{(h)} \tag{3}$$

This modification enables the model to more effectively capture genomic structural signals while maintaining parameter efficiency.

Importantly, this mechanism becomes particularly useful during model training. Although the attention module, together with RoPE, can capture local LD structure within each segment, it has no information about the segment's location along the genome. As a result, the model may treat two segments with similar local patterns as equivalent, even if they originate from widely separated genomic regions—an issue that arises frequently in sparse genotype data due to the restricted context window. Introducing a relative genomic positional bias provides each segment with a lightweight "watermark" that encodes its broader genomic context, allowing the model to better distinguish structurally similar patterns that occur in different genomic neighborhoods.

**CNN bottleneck block**

We replace the standard Transformer feed-forward block with a convolutional bottleneck module, which demonstrated consistently better performance in a series of ablation experiments. Let the input embedding dimension be $d$, and define the bottleneck dimension as $d' = d \cdot \gamma$, where $\gamma \in \mathbb{R}^+$ is the bottleneck factor (eg. $\gamma = 0.5$ compression, $\gamma = 2$ implies expansion). The CNN bottleneck applies the following operations:

$$\mathbf{H}_1 = \text{Dropout}(\text{MaxPool1D}(\text{ReLU}(\text{Conv1D}(\boldsymbol{X})))) \tag{4}$$

$$\mathbf{H}_2 = \text{Upsample}(\text{ReLU}(\text{Conv1D}(\mathbf{H}_1))) \tag{5}$$

As illustrated in **Fig. 4**, this architecture simultaneously transforms the input along two orthogonal axes. First, it expands the feature (channel) dimension $\gamma$ times through convolution, increasing the model's representational capacity at each SNP position. After ReLU[35] activation is applied, it then temporarily halves the sequence length using max pooling, enabling local context aggregation, and subsequently restores the original resolution via 1D convolution and up-sampling. Possible dropout is applied at the end of the max-pooling too. This decoupled design allows the model to learn richer, high-capacity representations while also abstracting over local genomic neighborhoods. In contrast to position-wise feed-forward layers, the CNN bottleneck introduces localized receptive fields—analogous to convolutional filters in computer vision—thereby enhancing the model's ability to capture short-range genomic patterns.

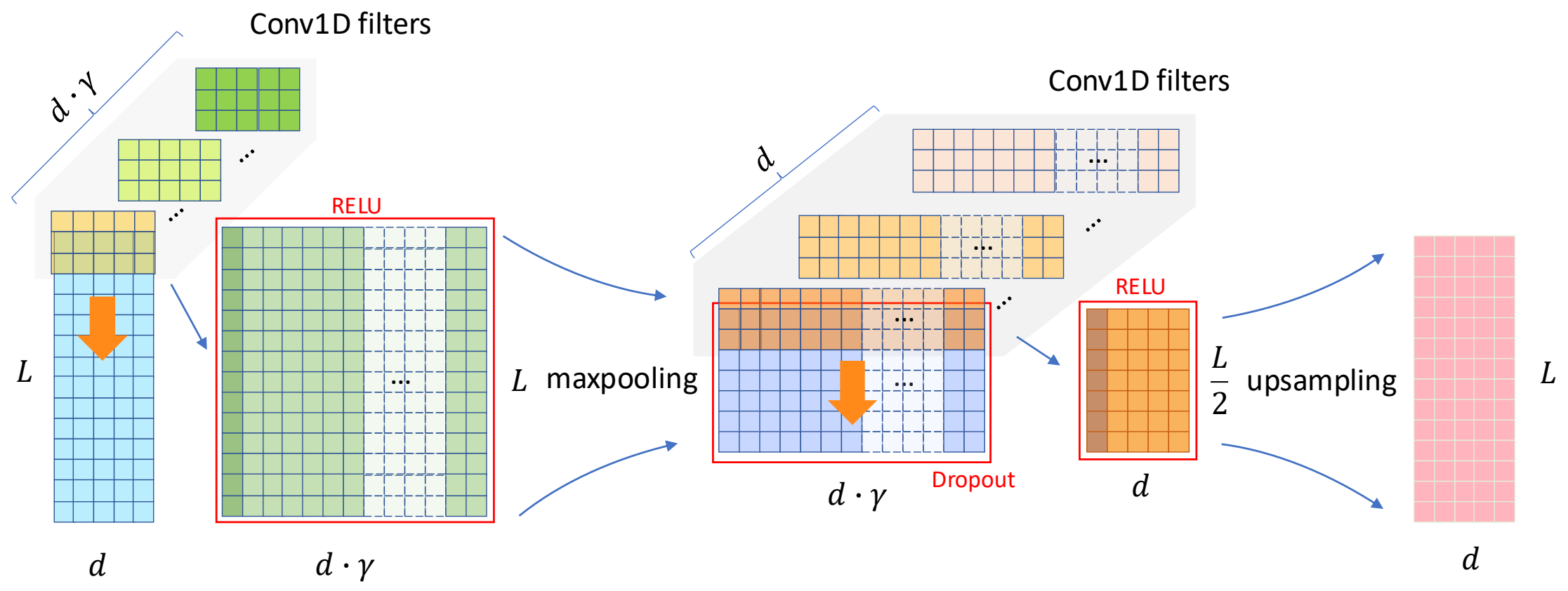

**Figure 4 | Illustration for the CNN bottleneck block.** This CNN bottleneck block expands the channel dimension by a factor of $\gamma$ to create an over-complete representation, applies ReLU activation, and then compresses the sequence length by half through max pooling. A second convolution and up-sampling restore the original shape, with optional dropout applied after pooling.

For a 1D convolution with $C_{in} = d$, $C_{out} = d \cdot \gamma$, and kernel size $k$, the total number of the parameters is $2d^2\gamma k$. By contrast, the number of the weights of a Gated Linear Unit with GELU (Gaussian Error Linear Unit)[36] activation (GeGLU) module[37] used in our ablations employs three position-wise linear layers is $3d^2\gamma$. Although the CNN bottleneck uses $2k/3$ times more parameters than GeGLU for the same expansion factor $\gamma$, it consistently outperformed the GeGLU block in our ablation studies, and the expansion-based design ($\gamma > 1$) yielded further gains.

## Layer normalization

In GenoBERT we adopt DeepNet's strategy to set the scale factor $\alpha$ such that $\alpha = 2N^{\frac{1}{4}}$ ($N$ denotes the depth of GenoBERT), so the layer normalization is like:

$$\text{LayerNorm}(\boldsymbol{X} \cdot \alpha + \phi(\boldsymbol{X})) \tag{6}$$

$\Phi$ can be the attention operation or the CNN bottleneck.

Based on the ablation studies reported in the supplementary materials, we adopt a compact and balanced GenoBERT configuration that provides a favorable trade-off between modeling capacity and computational efficiency. The final model operates on a context window of 130 tokens (128 SNPs plus two special tokens), with 6 Transformer layers and 4 attention heads per layer, and a hidden dimension of 768. To enhance local dependency modeling while controlling parameter growth, the standard feed-forward network is replaced with a CNN-based bottleneck module using a bottleneck factor of 2.0. Relative genomic bias positioning (RGBP) is enabled to encode genomic distance information directly within the attention mechanism, while no inter-layer weight sharing is employed. Under this configuration, the model contains approximately 56.7 million parameters, requires 12.3 GFLOPs per forward pass, and occupies 216 MB of memory, yielding a strong balance between predictive performance, scalability, and computational cost. All experiments in this study are conducted using this final architecture unless otherwise specified.

## Baselines

Beagle is a widely used reference-panel–based genotype imputation method grounded in the Li and Stephens haplotype copying framework. It models phased haplotypes as mosaics of reference haplotypes using a hidden Markov model, enabling efficient inference of missing genotypes and dosages. In this study, we used Beagle v5.4 with default recommended parameters, which represents the current state-of-the-art implementation in terms of computational efficiency and imputation accuracy.

SCDA is a symmetric 1D convolutional autoencoder that processes the full-length SNP sequence without any segmentation, encoding the entire chromosome-level variant array at once using a 5-channel representation of phased genotypes. Its encoder consists of a hierarchy of Conv1d layers with kernel size 3 and doubling channel widths, each followed by ReLU activation and max-pooling, producing a compact latent representation that the decoder reconstructs symmetrically through nearest-neighbor upsampling and convolution, restoring the 5-channel genotype output.

BiU-Net shares the same convolutional backbone but operates on fixed 128-SNP genotype segments with 16-SNP overlap and enhances the autoencoder topology with U-Net–style skip connections that inject encoder features at multiple resolutions directly into the decoder, thereby preserving fine-scale haplotype structure that is often attenuated by repeated pooling. Faithful to its original paper, BiU-Net is optimized using a customized focal loss equipped with predefined penalty coefficients that weight different imputation error types, allowing the model to preferentially penalize misclassification patterns that are biologically or practically more consequential (e.g., confusing heterozygous and homozygous-alternate states).

STICI employs a transformer–convolution hybrid architecture for long-range genomic modeling: each SNP is embedded into a 128-dimensional vector through allele and positional embeddings, and the full sequence is partitioned into 2048-SNP chunks with 128-SNP overlaps. Each chunk is expanded into a local attention window and processed by a composite module integrating multi-head self-attention, multi-kernel convolutions with receptive fields from 3 to 15 bases, residual feed-forward blocks, a convolutional skip pathway, and a cross-attention mechanism linking local and global representations. Following the original STICI study's suggestion, our PyTorch implementation trains the model using random masking ratios between 0.85 and 0.95, exposing it to a distribution of missingness levels similar to those found in real genomic data, and incorporates the MaCH-Rsq (Minimac-style) loss to improve rare-allele imputation.

Except for Beagle, which was used as a reference implementation, all deep learning baselines (SCDA, BiU-Net, and STICI) were re-implemented in PyTorch to match the original architectures and training procedures. Full implementation details are available in the GitHub repository.

## Datasets

This study utilized two complementary datasets: the Louisiana Osteoporosis Study (LOS) and the 1000 Genomes Project (1KGP). The in-house LOS dataset consists of individual-level genotype and phenotype data from collected from African American and Caucasian participants recruited in Southern Louisiana, aimed at investigating genetic and non-genetic determinants of osteoporosis and other musculoskeletal diseases. Detailed inclusion and exclusion criteria have been described in previous publication[28]. In this study, samples from individuals of other ancestries were further excluded due to insufficient sample sizes, resulting in a total of 7,504 quality-controlled samples from an initial set of 7,675 individuals (see **Supplementary Table S36** for specification). The 1KGP dataset comprises 2,548 individuals from five continental populations—European (EUR), African (AFR), Admixed American (AMR), South Asian (SAS), and East Asian (EAS)—capturing broad allele-frequency spectra and global population diversity. Together, these datasets provide complementary contexts: LOS offers larger sample size but comparatively weaker LD, while 1KGP contributes rich population diversity with fewer samples, enabling comprehensive evaluation of model generalization and cross-population transferability.

All experiments were conducted on phased genotypes of chromosome 22 (GRCh38 reference). To simulate different levels of genotype missingness, we applied proportional masking at four ratios (5%, 15%, 25%, and 50%), ensuring that missing sites were distributed proportionally with respect to allele dosage. This approach maintained consistent relative missingness across both rare and common variants, avoiding frequency-dependent bias. Each missing level masking simulation was repeated with three random seeds (0, 42, 1024) to ensure statistical robustness and reproducibility.

Model performance was evaluated using accuracy (concordance rate), $r^2$, precision, recall, and $F_1$-score (macro average), and further stratified by seven MAF ranges (0.1–0.5%, 0.5–1%, 1–10%, 10–20%, 20–30%, 30–40%, and 40–50%), in addition to overall summary metrics. This binning scheme enabled fine-grained assessment of model behavior across the allele frequency spectrum, facilitating comparison of performance between common and rare variants.

## Genotype data quality control (QC) and segmentation

Both datasets were processed using the same cohort construction and data-splitting strategy. Two complementary cohort definitions were considered: (i) population-specific cohorts, in which samples were grouped by ancestry, and (ii) an aggregated ("ALL") cohort, formed by pooling samples across all populations. For the LOS dataset, population-specific cohorts consisted of African American (AA) and Caucasian (CA) groups, while for the 1KGP dataset, cohorts corresponded to the five continental populations (AFR, EUR, AMR, SAS, and EAS). For each cohort definition (population specific and ALL), samples were independently split into training, validation, and test sets using an 8:1:1 ratio, with stratification by age and sex to preserve demographic balance. All subsequent preprocessing, model training, and performance analyses were conducted strictly within these predefined cohort partitions.

All genotype data underwent a unified QC and preprocessing pipeline to ensure compatibility across cohorts and eliminate low-quality variants and samples. Before QC, both the LOS and 1KGP datasets were already phased and aligned to the GRCh38 reference genome, minimizing preprocessing discrepancies.

The main filtering steps were as follows:

1. Keep only biallelic SNPs.
2. Remove individuals with more than 5% missing genotypes.
3. Exclude SNPs with missing genotype calls.

4. Split each dataset into training, validation, and test subsets (8:1:1), stratified by population and sex. For pooled cohorts, splits were performed independently within each population and then aggregated across populations to form the final subsets, preventing cross-population data leakage.
5. Apply Hardy–Weinberg equilibrium (HWE) and MAF filters (HWE $p > 1 \times 10^{-6}$; MAF $\geq$ 0.1%) independently on each subset.
6. Retain only SNPs shared across all subsets within each cohort to ensure consistency in training and evaluation.

The results of dataset preprocessing, including cohort-specific sample size and variant counts, are reported in **Supplementary Table S37**.

For model training, each genotype was numerically encoded to preserve phasing information: missing sites were represented as 0, padding tokens as 5, homozygous reference (Ref|Ref) as 1, heterozygous genotypes (Ref|Alt and Alt|Ref) as 2 and 3, respectively, and homozygous alternate (Alt|Alt) as 4. The encoded samples were first split into chunks to optimize memory usage, then segmented into fixed-length, overlapping windows to create training-ready inputs. Segments were subsequently padded and transposed (with SNPs represented as columns) and stored in HDF5 files, enabling both full and batch-wise data loading during training. A detailed implementation of this segmentation and encoding process is provided in our previous study[20]. Unlike the previous implementation, each genomic segment is adapted for masked language modeling (MLM) task by enclosing it with special tokens—[CLS] at the beginning, [SEP] at the end, and [PAD] tokens appended when the segment length is shorter than the predefined context window.

# Results

We evaluated GenoBERT, an attention-based genotype imputation model, against four representative baselines—Beagle5.4, SCDA, BiU-Net, and STICI—to assess imputation accuracy across diverse population structures and allele-frequency regimes. Experiments were conducted on phased chromosome 22 genotypes aligned to the GRCh38 reference genome from both the LOS and the 1KGP datasets. To simulate realistic missingness, genotypes were randomly masked at four levels (5%, 15%, 25%, and 50%), and performance was evaluated using accuracy (concordance rate), $r^2$, precision, recall, and $F_1$ score, with results further stratified by MAF. Unless otherwise stated, all metrics are reported as the mean over three runs using different random masking seeds applied to the same test set (Comprehensive results are provided in **Supplementary Table S6–S35**).

## GenoBERT outperforms the baselines on the LOS dataset

Focusing on the LOS cohort with African American and Caucasian samples pooled for training and evaluation, GenoBERT consistently achieved the highest imputation accuracy across all MAF ranges up to 25% missingness (**Supplementary Tables S6–S10**). For example, at 5% missingness, GenoBERT reached $r^2 = 0.9966$ and $F_1 = 0.9978$, and maintained high performance at 25% missingness ($r^2 = 0.9778$, $F_1 = 0.9860$), outperforming all baselines (BiU-Net > SCDA > STICI > Beagle). The increase in missingness resulted in only a modest degradation for GenoBERT ($\Delta r^2 \approx$ 2%, $\Delta F_1 \approx$ 1%), indicating strong robustness to moderate sparsity.

Among the baselines, BiU-Net and SCDA exhibited moderate performance declines ($\Delta r^2 \approx 3\%$, $\Delta F_1 \approx 2\%$) while retaining high absolute accuracy, supporting the effectiveness of segmentation-based training strategies (see **Materials and Methods, Genotype data quality control (QC) and segmentation** for details). In contrast, STICI and Beagle showed greater robustness to increasing missingness, with STICI even displaying a slight improvement at 25% missingness despite lower baseline accuracy. Beagle consistently achieved the lowest overall accuracy but remained relatively insensitive to missing-rate variation.

Under severe sparsity (50% missingness), STICI demonstrated the highest robustness ($r^2 = 0.9383$, $F_1 = 0.9554$), followed by GenoBERT ($r^2 = 0.9029$, $F_1 = 0.9452$) and BiU-Net ($r^2 = 0.9101$, $F_1 = 0.9322$), whereas SCDA degraded substantially. This robustness of STICI is consistent with its pretraining strategy using high masking ratios and a composite loss emphasizing tolerance to incomplete genotypes, albeit at the expense of reduced accuracy under lower missingness levels.

To provide a fine-grained evaluation, we include a comprehensive comparison of model performance on the LOS/mixed populations (ALL) across varying missing ratios and MAF ranges (**Fig. 5**). As shown, for variants with MAF $\geqslant$ 10%, all reference-free models consistently outperform Beagle across every missingness level. Furthermore, except for STICI, they also surpass Beagle on the rarest variants, maintaining this advantage at 15% missingness, with GenoBERT extending it to 25%. STICI performs worse than the other reference-free models at low missingness levels, but this margin diminishes as missingness increases, and STICI eventually surpasses the others when missingness reaches 50%. Beagle consistently ranked lowest in overall performance across missing ratios, with the only exception being the $r^2$ of SCDA at the 50% missing level, and its accuracy decreased as MAF increased, indicating a limited capacity to recover common variants reliably. Under severe missing conditions (50%), the reference-based method Beagle achieves higher accuracy on the rarest variants under this setting, likely reflecting its reliance on prior haplotype information embedded in the reference panel. In contrast, all reference-free models exhibit similar declines in accuracy that are likewise concentrated in the rare-variant range (MAF < 1%), primarily due to underfitting in sparsely represented SNPs. Among them, STICI maintained a modest advantage over Beagle down to 1% MAF range.

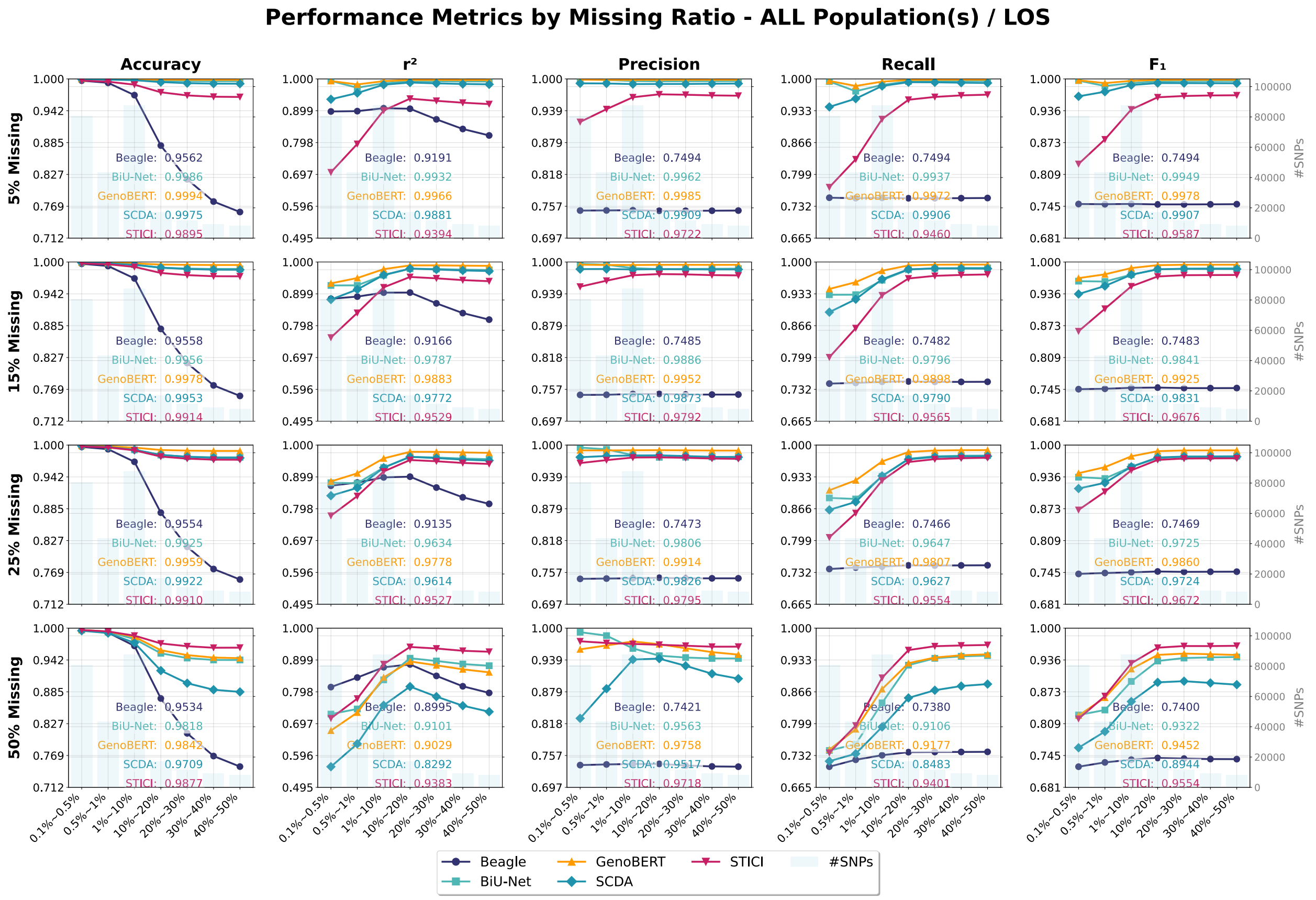

**Figure 5 | Performance comparison of genotype imputation models on the LOS/mixed populations (ALL) cohort across varying missing rates and minor allele frequency (MAF) ranges.** This illustration provides detailed evaluation of GenoBERT, BiU-Net, SCDA, STICI, and Beagle under increasing genotype-missing ratios (rows) and MAF ranges (columns) in the LOS cohort. Accuracy, $r^2$, precision, recall, and $F_1$ are shown for each condition.

These results suggest that applying STICI's high-mask training strategy on GenoBERT could yield complementary benefits. However, training GenoBERT under extreme masking ratios is computationally demanding due to slower convergence. In practice, curriculum learning[38–40] mitigates this limitation by stabilizing optimization and reducing the required epochs. Consistent with this, GenoBERT trained progressively from 15% to 50% missingness achieved the highest performance across all metrics in the 1KGP dataset, including the < 1% MAF bin (**Supplementary Table S25, S27, S29, S31, S33, S35**). Expanding the training dataset by incorporating additional samples from the reference panel would potentially further increase representation of rare variants and help reduce underfitting.

Across the LOS populations, GenoBERT consistently outperformed the baseline models at moderate missing rates (5%–25%) in both the mixed-population (ALL) and ancestry-specific cohorts (**Fig. 6**). At the high missing level of 50%, STICI achieved slightly higher accuracy in the mixed and African American cohorts, where LD structure is more fragmented; On the contrary, in the Caucasian cohort, whose stronger LD patterns enabled GenoBERT to retain the highest $r^2$ across all missingness levels. Overall, segmentation-based deep models (BiU-Net, GenoBERT, and STICI) exhibited clear advantages over non-segmentation methods. This improvement reflects the dual benefit of segmentation in enabling effective dimensionality reduction and, for GenoBERT and BiU-Net, serving as implicit data augmentation by training a unified model across diverse genomic segments. In contrast, STICI applies segmentation for dimensionality reduction only, as its multiple branches are trained separately on individual chunks.

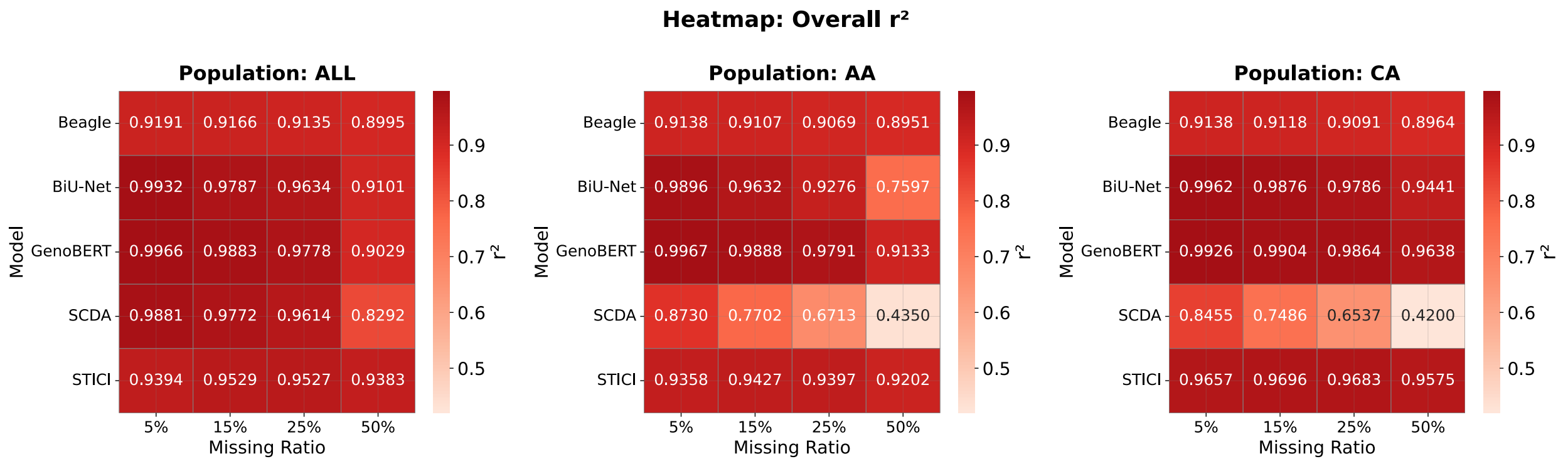

**Figure 6 | Population-specific comparison of imputation accuracy across missing-rate conditions.** Heatmaps showing overall imputation $r^2$ for five models (Beagle, BiU-Net, GenoBERT, SCDA, and STICI) under increasing genotype-missing ratios (5–50%) in the LOS dataset. Each panel corresponds to a population group— mixed-populations (ALL), African American (AA), and Caucasian (CA)—where models were trained and evaluated within the same cohort.

Notably, BiU-Net showed a substantial decline in the AA cohort at 50% missing, reflecting its sensitivity to the combined challenges of weak LD and severe sparsity. In contrast, GenoBERT and STICI remained more stable under extreme missingness. A likely explanation is that self-attention provides a more flexible within-segment dependency structure: unlike CNNs, which rely on fixed local filters, attention enables direct, data-adaptive interactions among all positions within a segment, selectively weighting informative loci and effectively bypassing large missing or uninformative regions. This adaptive aggregation makes attention-based models more robust to irregular sparsity patterns and better at denoising incomplete genotype signals, especially in populations with fragmented LD structure.

## GenoBERT shows consistent imputation accuracy across different cohorts on the 1KGP dataset

**Fig. 7** summarizes overall imputation accuracy ($r^2$) of GenoBERT and STICI across five 1KGP populations: European (EUR), African (AFR), Admixed American (AMR), South Asian (SAS), and East Asian (EAS), under increasing missing-genotype ratios (5%, 15%, 25%, and 50%). As two of the most up-to-date models, GenoBERT consistently achieves higher $r^2$ across all populations and missingness levels in this dataset, demonstrating substantially greater robustness to incomplete genotype input. While both models exhibit the expected monotonic decline in $r^2$ as the missing ratio increases, GenoBERT maintains a markedly higher absolute $r^2$ at every level of missingness, particularly for EUR, SAS, and EAS cohorts. AFR and AMR cohorts present the greatest challenge, with overall lower $r^2$ values for both models. Imputation performance on the AFR cohort declines more sharply with increasing missingness for both models.

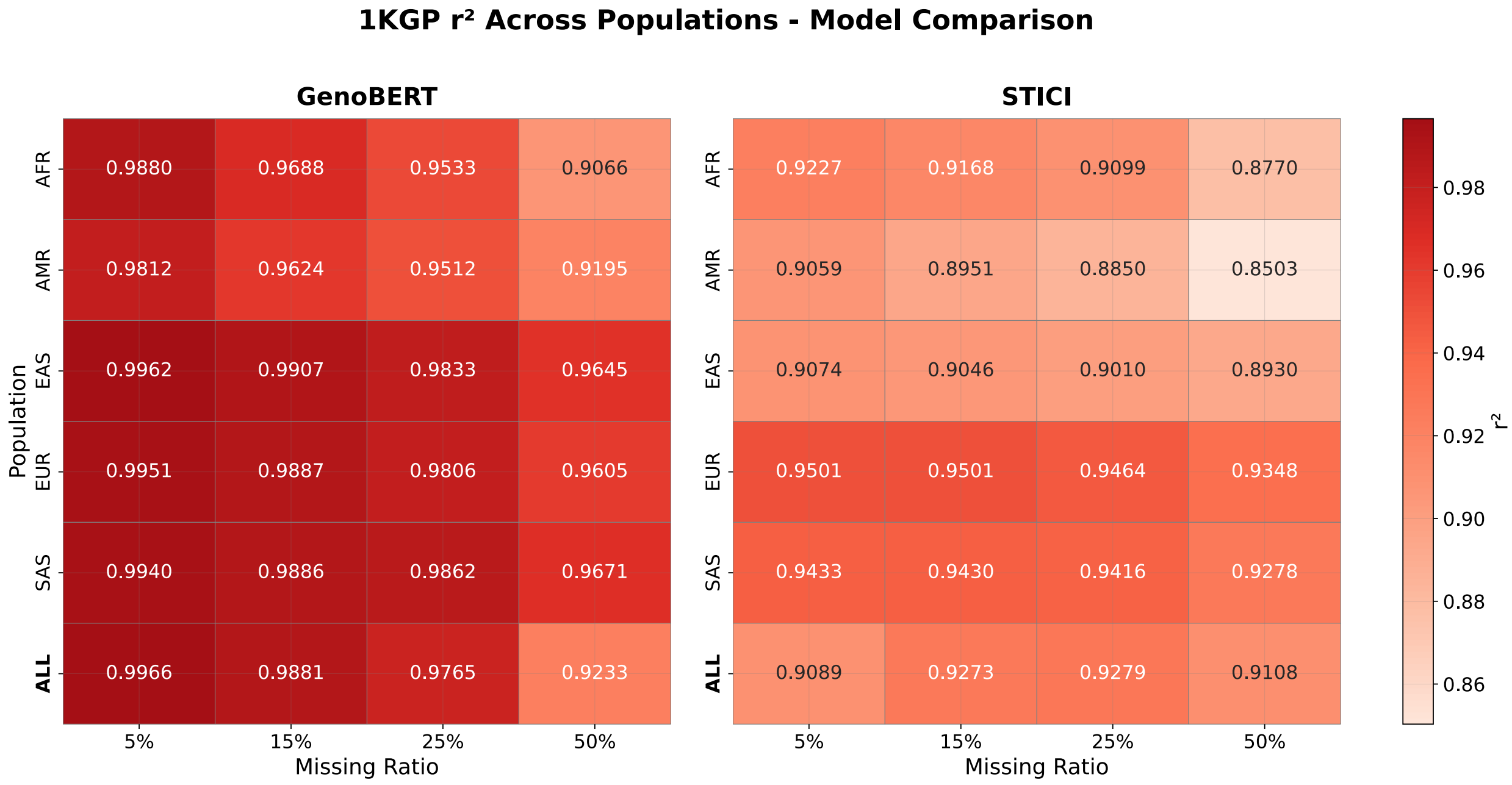

**Figure 7 | Mean imputation accuracy ($r^2$) of GenoBERT and STICI across 1KGP populations under increasing missing genotype ratios.** GenoBERT shows consistently higher accuracy and greater robustness to missingness across all populations and the pooled ALL cohort.

Comparing the performance of the STICI model trained and evaluated on the AFR cohort of the 1KGP dataset (**Fig. 7**) with its performance on the AA cohort of the LOS dataset (**Fig. 6**) reveals a notable divergence. STICI's accuracy dropped sharply in the 1KGP/AFR cohort, whereas GenoBERT remained largely stable across both datasets. This discrepancy can be attributed to the combination of weaker LD structure typical of African ancestry populations and the smaller sample size of 1KGP/AFR, which together limit STICI's ability to learn reliable local patterns, while GenoBERT's architecture remains more robust under such conditions.

**Fig. 8** compares $r^2$ across different MAF ranges for each population and missing ratio. The results reaffirm the population-level trends in **Fig. 7** while revealing finer-grained behavior across allele frequencies. Across every evaluated population, GenoBERT achieves higher $r^2$ values than STICI under all four missingness levels (5%, 15%, 25%, and 50%), on all MAF ranges. This universal advantage indicates that GenoBERT's performance superiority is not confined to specific allele-frequency ranges or genetic backgrounds but extends broadly across ancestries and data sparsity regimes.

a

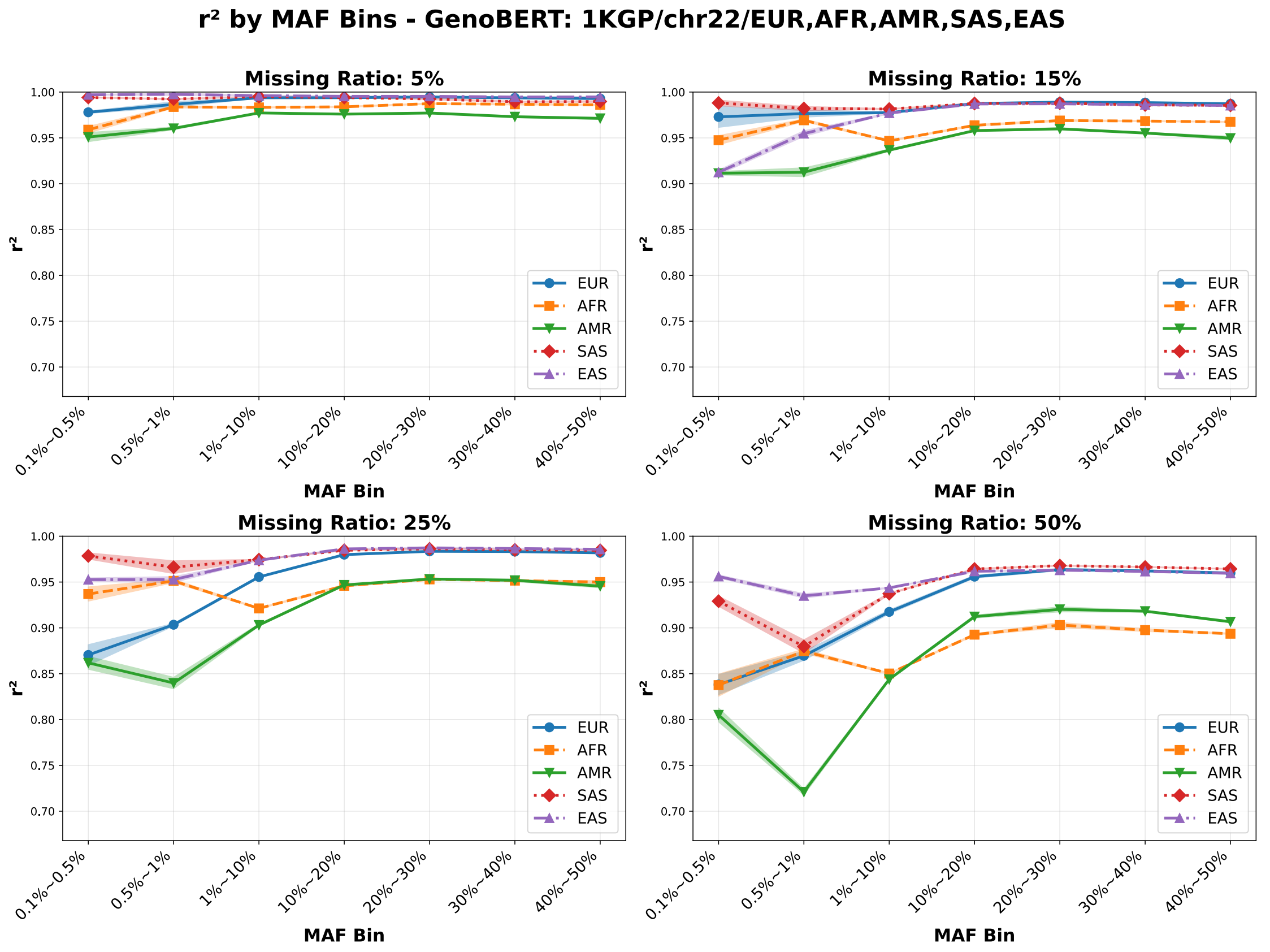
r² by MAF Bins - GenoBERT: 1KGP/chr22/EUR,AFR,AMR,SAS,EAS
Missing Ratio: 5%
Missing Ratio: 15%
Missing Ratio: 25%
Missing Ratio: 50%
r²
MAF Bin
0.1%~0.5%
0.5%~1%
1%~10%
10%~20%
20%~30%
30%~40%
40%~50%
1.00
0.95
0.90
0.85
0.80
0.75
0.70
EUR
AFR
AMR
SAS
EAS

b

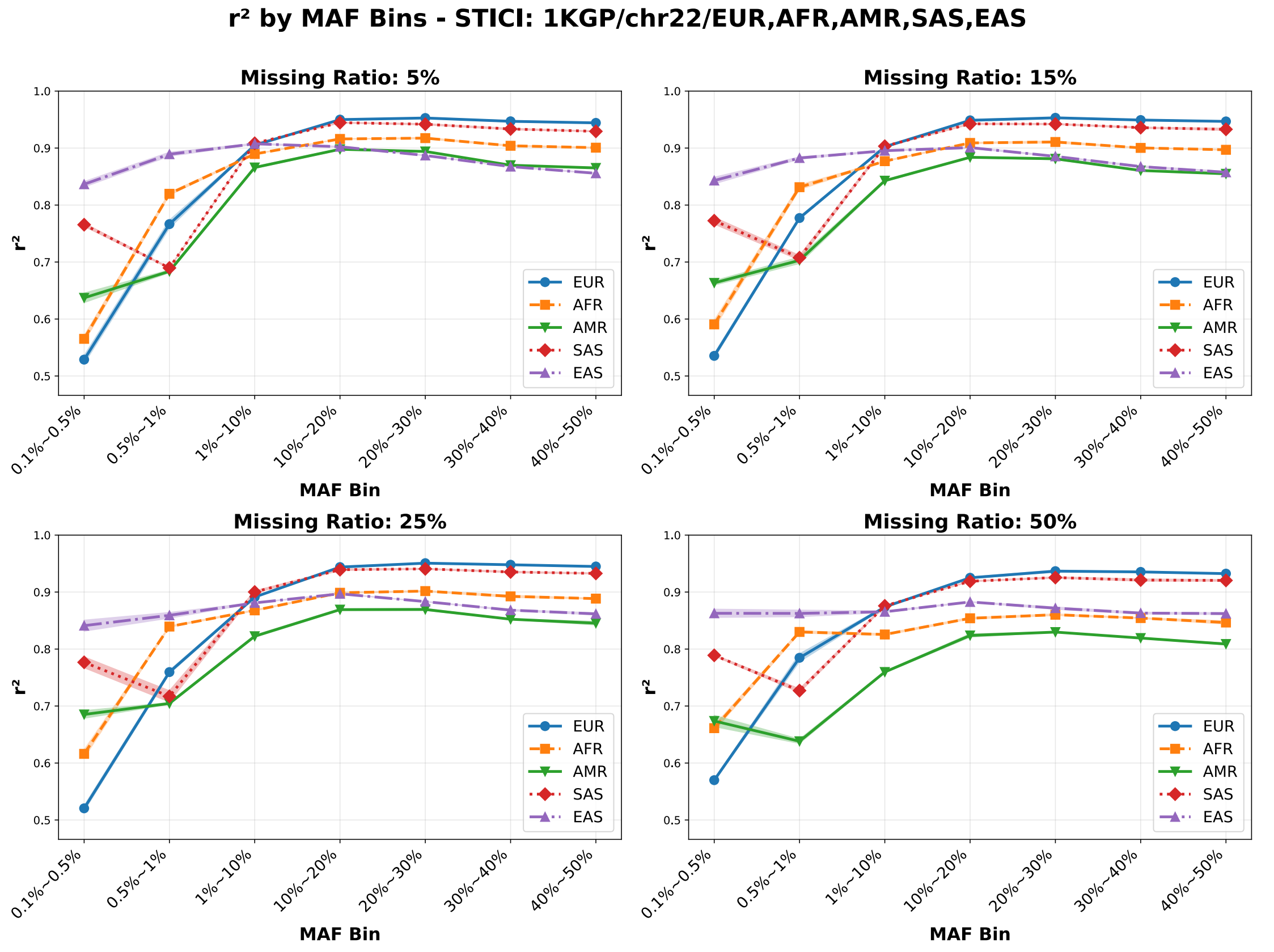
r² by MAF Bins - STICI: 1KGP/chr22/EUR,AFR,AMR,SAS,EAS
Missing Ratio: 5%
Missing Ratio: 15%
Missing Ratio: 25%
Missing Ratio: 50%
r²
MAF Bin
0.1%~0.5%
0.5%~1%
1%~10%
10%~20%
20%~30%
30%~40%
40%~50%
1.0
0.9
0.8
0.7
0.6
0.5
EUR
AFR
AMR
SAS
EAS

**Figure 8 | Population- and frequency-stratified imputation accuracy of GenoBERT and STICI across missing-genotype levels. (a)** GenoBERT and **(b)** STICI performance ($r^2$) across MAF ranges in five 1KGP populations (EUR, AFR, AMR, SAS, EAS) under four simulated missing-genotype ratios (5%, 15%, 25%, 50%). Values represent mean $r^2 \pm$ 95% confidence interval.

As shown in **Fig. 8a**, GenoBERT exhibits a uniform decline in $r^2$ with increasing missingness across all minor allele frequency (MAF) bins and populations, affecting rare, intermediate, and common variants alike; this frequency-independent degradation is consistent across EUR, AFR, EAS, and AMR despite differences in absolute performance levels. By contrast, **Fig. 8b** indicates that STICI's decline is driven primarily by intermediate and common variants (MAF $\geqslant$ 1%), whereas rare variants (MAF < 1%) show stable or increasing $r^2$ as missingness increases.

# Discussion

GenoBERT introduces a transformer-based framework for genotype imputation that performs robustly across diverse populations without relying on reference panels. By modeling phased genotypes through contextual self-attention, it effectively captures both short- and long-range linkage disequilibrium (LD) dependencies. Despite its strong cross-population performance, several areas remain open for refinement and extension.

## The relation between context window size and population-specific linkage structures

To further inspect how model architecture interacts with population-specific linkage structures under varying sparsity, we visualized the overall imputation accuracy ($r^2$) across four missingness levels for five methods on the CA and AA cohorts of the LOS dataset (**Fig. 9**).

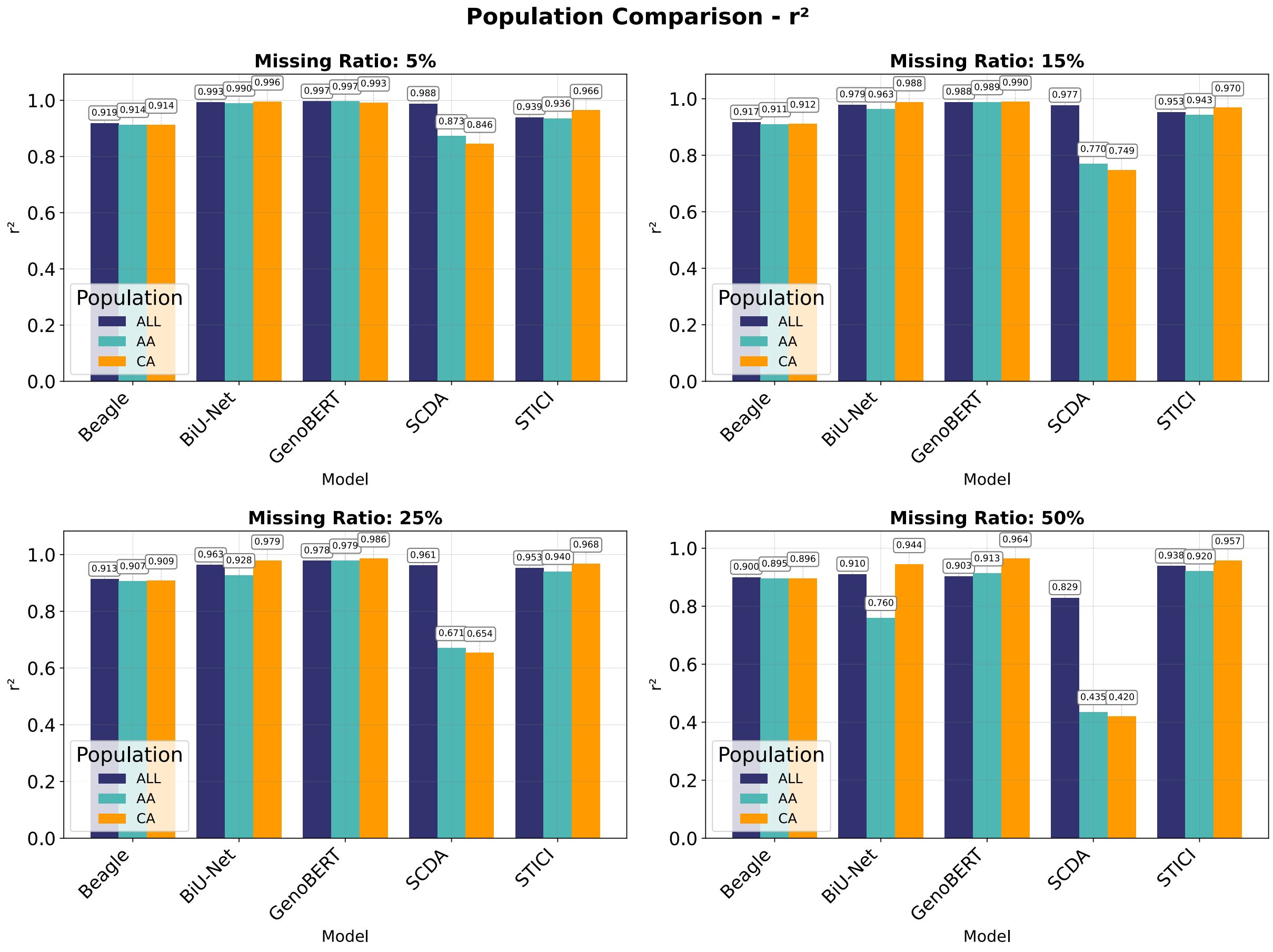

**Figure 9 | Population-specific divergence in imputation accuracy under increasing sparsity.** Overall imputation $r^2$ across three cohorts (mixed populations [ALL], African American [AA], and Caucasian [CA]) and four missing-rate levels (5%, 15%, 25%, 50%), with each model trained and evaluated within the same cohort.

**Fig. 9** compares imputation accuracy across mixed-populations (ALL) and ancestry-specific cohorts (AA and CA) under increasing missingness. At low missingness (5%), population differences are modest for all learned models, whereas Beagle exhibits uniformly lower but stable performance across cohorts. As missingness increases, model-specific divergence patterns emerge. BiU-Net shows a rapidly widening CA–AA gap, increasing from 0.006 at 5% missing to 0.184 at 50%, reflecting strong ancestry sensitivity under sparse observation. In contrast, GenoBERT maintains high accuracy across cohorts with comparatively small population gaps, which remain below 0.01 up to 25% missingness and increase to 0.051 at 50%. STICI exhibits consistently small CA–AA differences across all missingness levels, with gaps remaining below 0.04. SCDA, by contrast, undergoes severe performance degradation in both AA and CA cohorts as missingness increases, with accuracy dropping below 0.45 at 50% missingness, indicating limited robustness to population-specific sparsity. Overall, attention-based models (GenoBERT and STICI) demonstrate substantially improved stability across populations compared with convolutional approaches, while GenoBERT achieves the highest absolute accuracy across all cohorts and missingness regimes.

This pattern aligns with known ancestry differences in LD and with each model's effective receptive field. African-ancestry genomes contain shorter LD blocks and greater haplotypic diversity; when half the markers are masked, a 128-SNP context window retains too few informative observations to reconstruct missing sites accurately. Models that rely mainly on local contiguous context therefore deteriorate faster on AA, explaining BiU-Net's steep decline as missingness grows. GenoBERT employs the same 128-SNP context window but uses self-attention to aggregate information from any surviving informative tokens within the segment. Its higher-dimensional embedding (d = 768) and a proposed bias term for self-attention (See **Materials and Methods, Model architecture** for details) allows richer feature representation and more flexible weighting of context, leading to smoother performance degradation than convolution-based architectures under heavy masking. Alternatively, the broader genomic context provided by STICI's 2048-SNP segments (d = 128) ensures that, even with 50% missingness, sufficient intra-segment signal remains to prevent substantial degradation on the AA cohort, mitigating the impact of its smaller embedding dimension. In contrast, SCDA's divergence between the mixed-population and population-specific cohorts is not driven by linkage disequilibrium heterogeneity but is instead attributable to the absence of genotype segmentation during training. When applied to high-dimensional SNP sequences with limited population-specific sample sizes, SCDA struggles to learn stable representations without explicit SNP-sequence segmentation.

For GenoBERT, increasing context length could further mitigate the AA deficit but at the cost of higher memory, computation, and training time. A promising future direction is LD-adaptive segmentation, in which window boundaries follow local LD structure rather than a fixed SNP count. The concat-and-chunk strategy, widely adopted in large language model pretraining (e.g., RoBERTa[41], LLaMA[42]), could potentially provide an efficient way to handle this variable-length genomic contexts while maintaining fixed-length optimization and training throughput.

To assess whether our chosen input sequence length of 128 SNPs adequately captures local genomic context, we examined the physical spacing of consecutive variants across the five 1KGP populations (EUR/AFR/AMR/SAS/EAS). For each population, we extracted SNP positions on chromosome 22 and calculated the distance between adjacent sites along the genome. The resulting log-scaled histograms characterize variant spacing density within each cohort and provide an estimate of the genomic span covered by a sequence of $L$ consecutive SNPs.

Consistent with variant spacing on chromosome 22 in the 1KGP dataset (**Supplementary Fig. S1**), a 128-SNP window corresponds to approximately 100 kb of genomic distance and is sufficient to capture local LD structure. We thus examined how LD decays with genomic distance to assess whether the 128-SNP sequence length sufficiently spans the local correlation range. For each 1KGP population, we computed pairwise LD ($r^2$) on chromosome 22 using randomly sampled SNPs. For the all-MAF analysis, we sampled $1 \times 10^5$ SNPs per population (roughly 5% to 10% of the SNPs, yielding up to $5 \times 10^7$ candidate SNP pairs), and for each MAF bin we sampled 1000 SNPs (roughly 4% to 25% of the SNPs within that range, yielding up to $5 \times 10^5$ candidate pairs). We then restricted to pairs within 1 Mb distance and with valid $r^2$, resulting in $\approx 2.9 \times 10^6$ pairs for the all-MAF analysis and $\approx 3 \times 10^4$ pairs per MAF bin. SNP pairs were grouped into uniform 10-Kb distance intervals and cut at 100Kb, and mean $r^2$ per interval was plotted to characterize LD decay with physical distance (**Fig. 10**).

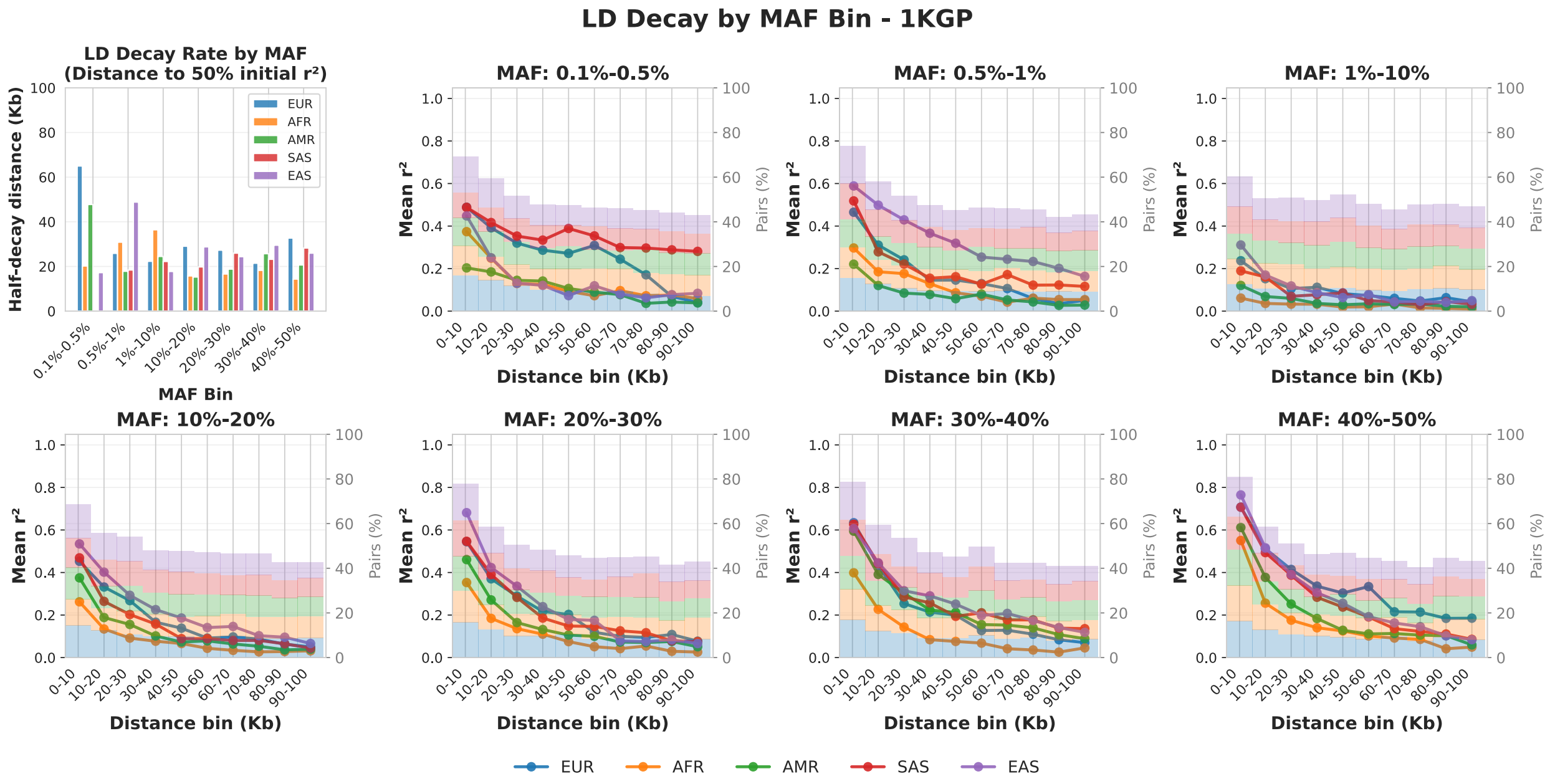

**Figure 10 | LD decay across distance and MAF ranges in 1KGP populations.** Mean pairwise LD ($r^2$) was calculated for sampled SNP pairs on chromosome 22 and averaged within 10-Kb distance bins, separately for seven MAF ranges. The proportion of each distance bin for specific cohort is plot as the stacked histogram. The top-left panel summarizes the half-decay distance—the genomic distance at which each population's mean LD falls to 50% of its initial (0–10 Kb) value—across all MAF bins.

Across all populations and MAF ranges, LD dropped sharply with increasing distance: correlations typically fell below $r^2 = 0.1$ once SNPs were separated by more than 100Kb, and below $r^2 = 0.2$ for most common variants. The strongest LD ($r^2 > 0.5$) was concentrated within the first 10Kb. South Asian (SAS), East Asian (EAS), and European (EUR) cohorts exhibited stronger LD, especially among rare variants, with SAS and EAS maintaining moderate correlations (around 0.2) up to approximately 100Kb. In contrast, the African (AFR) cohort consistently showed the weakest LD across most MAF ranges, reflecting its more heterogeneous ancestral composition. These patterns suggest that different populations retain usable correlation information over different genomic spans, implying that optimal context sizes for LD-aware models may need to be population-specific.

Overall, these results confirm that a 128-SNP sequence—corresponding to roughly 100Kb of genomic distance—captures most of the local correlation structure relevant for genotype modeling, while also indicating that populations with slower LD decay may benefit from proportionally larger context windows.

We further compared LD structure between the 1KGP and LOS cohorts across matched ancestries (**Supplementary Fig. S2**). Consistent with established population-genetic patterns, European ancestry populations showed stronger and more extended LD than African ancestry populations. In both ancestries, LD decayed rapidly with distance, falling below $r^2 = 0.1$ within approximately 30 kb, indicating that a 128-SNP window provides sufficient genomic coverage for both datasets.

Despite the stronger LD in the 1KGP comparing to LOS, which in principle should facilitate imputation, the substantially smaller sample size of the 1KGP/AFR cohort with 410 individuals compared with 2,479 individuals in LOS/AA becomes a limiting factor for both models. This effect is evident for STICI, which achieves an $r^2$ of 0.9202 on LOS/AA at 50% missingness (**Fig. 6, Table S14**) but declines to 0.8770 on 1KGP/AFR under identical conditions (**Fig. 7, Table S28**). In contrast, GenoBERT shows only a minor reduction in accuracy across the same settings, from an $r^2$ of 0.9133 on LOS/AA to 0.9066 on 1KGP/AFR at 50% missingness (**Fig. 6, Fig. 7**).

In STICI, each attention–CNN fusion branch is trained separately on a genotype chunk, a design that reduces input dimensionality but does not perform data augmentation. Under the current configuration, each chunk the model contains roughly 27 million parameters (approximately 24 million chunk-specific and 2.8 million shared) trained on about 400 individuals, resulting in an unfavorable parameter-to-sample ratio that predisposes the model to overfitting and unstable optimization.

In contrast, GenoBERT, a single 56-million-parameter BERT variant (see **Supplementary, Ablation studies/Final architecture** for details) unitarily trained on approximately 152,076 segments for the 1KGP/AFR cohort, achieves much higher data efficiency through global parameter sharing. In addition, GenoBERT incorporates the Relative Genomic Positional Bias (RGPB) term, which encodes each segment's normalized genomic positions as a biologically defined bias within the attention scores, modulated by trainable scalar weights for each head and layer. While this term does not model LD correlations directly, it extends positional encoding to genomic coordinates, effectively providing each variant with a consistent positional context that prevents confusion across segments. As a result, GenoBERT remains stable under extensive data segmentation even with limited sample sizes, whereas STICI's fixed-chunk training suffers from its lack of augmentation and stronger dependence on cohort size.

## Potential Improvements

While GenoBERT achieves strong and scalable performance across diverse populations, several directions remain for enhancing its generalization, biological coverage, and data efficiency. Future advances will likely center on four key areas: (1) improving cross-population adaptability; (2) expanding variant representation beyond single-nucleotide polymorphisms; (3) enhancing data-efficient training; and (4) incorporating reference-based priors to improve rare-variant inference while preserving ancestry neutrality.

### 1. Toward population-invariant modeling.

Achieving population-invariant performance will require architectures capable of adapting to the heterogeneous LD structures and allele-frequency spectra observed across ancestries. Mechanisms such as LD-aware positional encodings could allow the model to dynamically adjust its receptive field according to local haplotype diversity. Similarly, an adaptive sparse attention scheme guided by precomputed MAF or LD matrices could enhance both computational efficiency and rare-variant sensitivity. By reweighting attention toward locally informative SNPs while penalizing weak long-range dependencies, GenoBERT could emulate the locality of convolutional models without sacrificing global contextual capacity.

**2. Expanding variant representation.**

Extending GenoBERT beyond single-nucleotide polymorphisms to include structural variants (SVs), copy-number variants (CNVs), and small indels would substantially broaden its biological scope. SVs collectively account for the majority of base-pair-level differences among human genomes—approximately 18.4 Mbp of sequence per diploid genome, compared with 3.6 Mbp for SNPs[43]—highlighting their major contribution to human genetic diversity. Representing such variants through multi-resolution embeddings could enable the model to capture both discrete allelic events and continuous dosage effects, thereby improving its capacity to integrate diverse forms of variation.

**3. Enhancing data efficiency.**

Improving data efficiency is another key direction. Deep neural models typically require large cohorts for stable convergence, and GenoBERT's behavior in small-sample regimes (e.g., $N < 100$) remains unexplored. Techniques such as contrastive pretraining, parameter-efficient fine-tuning (PEFT)[44,45], or synthetic augmentation could mitigate sample-size constraints and enable applications to disease-specific or underrepresented cohorts.

**4. Integrating reference priors without bias.**

Finally, while GenoBERT's reference-free training paradigm ensures broad applicability across ancestries, selectively incorporating reference-panel priors during pretraining could enhance rare-variant inference. Prior studies have shown that imputation accuracy improves with larger and more diverse reference panels, particularly for low-frequency variants [5,46,47]. Hybrid objectives that combine reference-free embeddings with panel-guided denoising may therefore strengthen cross-cohort transferability while preserving GenoBERT's neutrality to population bias.

In summary, future iterations of GenoBERT should aim to unify adaptability across populations, comprehensive variant representation, and efficient learning under data-limited conditions. By integrating LD-aware modeling, richer variant embeddings, and hybrid pretraining strategies, GenoBERT can evolve toward a population-generalizable and biologically grounded foundation model for human genomics.

## Limitations

While this study demonstrates that GenoBERT can effectively model linkage disequilibrium (LD) structure and achieve strong imputation performance across diverse populations, several limitations should be acknowledged. These limitations reflect both the current experimental scope and the inherent constraints of LD-based sequence modeling, and they point to important directions for future development.

First, current analyses were conducted exclusively on chromosome 22, providing a representative but partial view of genome-wide LD diversity. Pan-genomic validation will be required to evaluate scalability across chromosomes exhibiting distinct recombination rates, variant densities, and structural complexities.

Second, model interpretability remains another limitation. Although attention maps reveal internal dependency structures, they primarily reflect LD correlations—that is, the statistical co-occurrence of alleles due to shared inheritance—rather than direct biological causation. In essence, GenoBERT learns how variants are correlated within local haplotypes, but not why specific variants influence functional or phenotypic outcomes. The non-LD (causal) associations central to genotype–phenotype relationships in GWAS arise from biological mechanisms such as regulatory disruption or coding changes, which cannot be inferred from LD alone. Future extensions could bridge this gap by incorporating functional annotations, causal-inference priors, or LD-aware relevance propagation methods to map model-inferred dependencies onto molecular mechanisms.

Finally, GenoBERT's latent genotype embeddings show promise for downstream functional tasks such as gene-expression prediction, variant-effect prioritization, and regulatory annotation transfer. However, fine-tuning these models requires large, well-aligned multi-omics datasets that jointly profile genotypes, transcriptomes, and epigenomic features (e.g., GTEx[48], ROSMAP[49], or single-cell expression atlases[50,51]). Employing parameter-efficient transfer learning, contrastive cross-modal pretraining, or multi-task fine-tuning could mitigate data requirements and enable broader reuse of the pretrained encoder across diverse biological contexts.

Overall, these limitations highlight that GenoBERT effectively captures LD structure but not causal mechanisms, pointing to clear directions for biologically grounded model extensions and integration with multi-omics data.

## Acknowledgements

Portions of this research were conducted using high-performance computing resources provided by the Louisiana Optical Network Infrastructure (http://www.loni.org).

This research was also supported in part by high-performance computing (HPC) resources and services provided by Information Technology at Tulane University, New Orleans, LA.

We thank Peter Collingridge for providing an online tool for DNA visualization (https://www.petercollingridge.co.uk/tools/draw-dna/), which was used to generate schematic elements in **Fig. 1**.

## Conflicts of interests

None declared.

## Funding

This work was supported in part by the National Institutes of Health (U19AG055373, R01AR069055, R01AG061917, and P20GM109036) and a grant awarded by the U.S. Engineer Research and Development Center (W912HZ20P0023).

# Data availability

Whole-genome sequencing data from the 1000 Genomes Project (1KGP), GRCh38 build, were obtained from the UCSC Genome Browser repository (chr22 example used in this study):

https://hgdownload.soe.ucsc.edu/gbdb/hg38/1000Genomes/ALL.chr22.shapeit2_integrated_snvindels_v2a_27022019.GRCh38.phased.vcf.gz

Sample metadata and demographic information for 1KGP were downloaded from the official project FTP release site:

https://ftp-trace.ncbi.nih.gov/1000genomes/ftp/release/20130502/integrated_call_samples.20130502.ALL.ped

The Louisiana Osteoporosis Study (LOS) datasets supporting the findings of this study are available from the principal investigator (H.W.D.; hdeng2@tulane.edu) upon reasonable request. Access is restricted to academic research use and requires Institutional Review Board (IRB) approval and execution of a data use agreement. LOS whole-genome sequencing data are being deposited in the AgingResearchBiobank and will be accessible to qualified investigators upon application and approval.

# Code availability

The GenoBERT implementation, including model architecture, training, evaluation, and analysis scripts, is available at: https://github.com/learnslowly/genobert The preprocessing pipeline for 1KGP genotype processing is available at: https://github.com/learnslowly/data/ A stable release corresponding to the version used in this study will be archived upon acceptance of the manuscript.

Supplementary material for the paper:

# GenoBERT: A Language Model for Accurate Genotype Imputation

## Table of Contents

## Overview

This supplement provides ablation study, GenoBERT's weight initialization details, the determination of proper segmentation hyperparameter, extended cohort-specific evaluations supporting the findings presented in the main manuscript and dataset specifications. Performance metrics include accuracy (concordance rate), $r^2$, precision, recall, and $F_1$-score (macro), evaluated across varying missingness levels, allele frequency bins, and population cohorts.

The analyses are based on three datasets:

- Louisiana Osteoporosis Study (LOS) – chromosome 22
- 1000 Genomes Project (1KGP) – chromosome 22

Models were evaluated under simulated genotype missingness levels of 5%, 15%, and 25%, with results stratified by minor allele frequency (MAF) bins (e.g., 0.1%≤ MAF ≤ 0.5%, etc.).

## Ablation studies

To determine the optimal configuration, a systematic series of ablation experiments were conducted.

### Alation I: Component-Wise Ablation

Ablation I is performed on two cohorts: the 1KGP/EUR cohort, with training/validation sample sizes of 417/51 and 120,577 SNPs, and the LOS/AA cohort, with 2479/309 samples and 307,599 SNPs.

This ablation study evaluates the contribution of two optional architectural components:

- RGPB: whether the Relative Genomic Positional Bias term is enabled
- CBN: whether the feed-forward module uses the CNN bottleneck (True) or GeGLU (False)

All models share the same core hyperparameters: a hidden dimension of 256, 6 encoder layers, 4 attention heads per layer, 0% dropout, batch size of 128, cosine-annealing learning-rate scheduler, and 40 warm-up epochs with no cool-down phase. Training uses a fixed missing ratio of 25% and learning rate $8\times10^{-4}$ across all settings.

When the CNN bottleneck is used, we set the kernel size to 3 and the expansion factor to 2. Models are trained for 160 epochs on the 1KGP/EUR dataset and 70 epochs on the LOS/AA dataset to account for differences in sample size and convergence speed. Validation accuracy is reported as the primary metric for comparison. **Table S1** summarizes all experiment configurations and results of the validation set in this ablation study.

**Table S1. Alation experiments for model components variation.**

| RGPB | CBN | Params (M) | FLOPs (G) | Size (MB) | Dataset | Accuracy |
|---|---|---|---|---|---|---|
| **T** | **T** | **6.32** | **1.44** | **24.10** | **1KGP** | **98.67%** |
| T | F | 3.96 | 1.14 | 15.10 | 1KGP | 98.37% |
| F | T | 6.32 | 1.44 | 24.10 | 1KGP | 98.10% |
| F | F | 3.96 | 1.13 | 15.10 | 1KGP | 96.52% |
| **T** | **T** | **6.32** | **1.441** | **24.10** | **LOS** | **98.67%** |
| T | F | 3.96 | 1.139 | 15.10 | LOS | 98.12% |
| F | T | 6.32 | 1.441 | 24.10 | LOS | 96.98% |
| F | F | 3.96 | 1.139 | 15.10 | LOS | 96.66% |

"1KGP" stands for 1KGP/EUR cohort, "LOS" stands for LOS/AA, "RGPG" as relative genomic positional bias, "CBN" as CNN bottleneck block, "T" as True, "F" as False. Experiment results achieved the highest imputation performance on the validation set of both datasets have been bolded.

The results reveal some insights:

1. RGPB and CNN Bottleneck each applied alone consistently improved the model performances across datasets.
2. The combination of RGPB + CNN Bottleneck provided the best performance on both datasets.

**Alation II: Depth and hidden dimension scaling**

Building on the findings from Ablation I, Ablation II further examines how model capacity affects performance by varying the hidden dimension and encoder depth, while toggling the RGPB and CNN bottleneck components. All other hyperparameters remain unchanged. Due to computational cost, this ablation is conducted only on the 1KGP/EUR dataset, with each model trained for 160 epochs. Validation accuracy is reported in **Table S2**.

**Table S2. Alation experiments for depth and hidden dimension scaling.**

| Hidden_dim | Depth | Params (M) | FLOPs (G) | Size (MB) | Accuracy |
|---|---|---|---|---|---|
| 128 | 2 | 0.53 | 0.13 | 2.02 | 95.42% |
| 256 | 2 | 2.11 | 0.48 | 8.04 | 96.58% |
| 512 | 2 | 8.41 | 1.85 | 32.09 | 97.77% |
| 768 | 2 | 18.91 | 4.10 | 72.13 | 97.89% |
| 768 | 4 | 37.80 | 8.20 | 144.21 | 98.26% |
| **768** | **6** | **56.70** | **12.29** | **216.29** | **99.01%** |

Experiment results achieved the highest imputation performance on the validation set of the 1KGP dataset have been bolded.

From this ablation study, we observe a clear trend: increasing the hidden dimension and model depth consistently improves performance. However, this comes with a substantial increase in model size. For example, a configuration with a 768-dimensional hidden layer and 6 encoder blocks (4 attention heads by default) contains approximately 57 million parameters, making it the largest model evaluated in our experiments.

### Alation III: Effect of the Bottleneck Expansion Factor

Ablation III investigates how the shape of the convolutional bottleneck affects model performance. We evaluate a range of expansion factors from 0.5 to 2, while keeping all other settings aligned with the Ablation II configuration (768 hidden dimension, 6 encoder layers, 4 attention heads, RGPB enabled, and CBN with kernel size 3). Results for the European cohort of the 1KGP dataset are summarized in **Table S3**.

**Table S3. Shapes of the bottleneck.**

| Bottleneck factor | Params (M) | FLOPs (G) | Size (MB) | Accuracy |
|---|---|---|---|---|
| 0.5 | 24.84 | 6.08 | 94.76 | 98.16% |
| 1 | 35.46 | 8.15 | 135.27 | 98.17% |
| **2** | **56.70** | **12.29** | **216.29** | **99.01%** |

Experiment result achieved the highest imputation performance on the validation set of the 1KGP/EUR dataset have been bolded.

From this ablation, we observe that increasing the bottleneck expansion factor leads to steady improvements in model performance. However, the number of parameters grows almost linearly with the expansion factor, resulting in substantially higher training cost. Considering this trade-off between accuracy and computational efficiency, we chose not to further increase the model size.

### Final architecture

Integrating the findings from all ablation studies, we arrive at the final configuration of the GenoBERT model. The complete architecture is summarized in **Table S4**.

**Table S4. GenoBERT final architecture.**

| Context window | #Depth/#Heads | Hidden dimension | Weight Sharing | RGBP |
|---|---|---|---|---|
| 130 (128+2) | 6/4 | 768 | None | Enabled |
| **Feed-forward** | **Bottleneck factor** | **Params (M)** | **FLOPs (G)** | **Size (MB)** |
| CNN Bottleneck | 2.0 | 56.70 | 12.29 | 216.29 |

## Weight initialization strategies

Effective weight initialization is critical to the stability and convergence of deep networks. Poor initialization can lead to vanishing or exploding gradients, degraded signal propagation, and unstable training dynamics. GenoBERT integrates diverse components—attention layers, the CNN bottlenecks, and layer normalization components—each demanding tailored initialization to ensure stable learning.

At a high level, the goal of initialization is to maintain consistent variance across layers. If weights are initialized too small, signals shrink as they propagate through layers, leading to the vanishing gradient problem. Conversely, large initializations can cause gradients to explode, making optimization unstable.

In GenoBERT, all linear layers—including projection matrices $\boldsymbol{W_Q}$, $\boldsymbol{W_K}$, $\boldsymbol{W_V}$, and output projection head are initialized using Xavier Normal initialization[52]. For attention projection matrices $\boldsymbol{W_Q}, \boldsymbol{W_K}, \boldsymbol{W_V}$, GenoBERT empirically applies a fixed Xavier gain of $\sqrt{2}$, rather than the depth-dependent gain of $(8 * N)^{-1/4}$ proposed in the DeepNet framework for $N$-layer encoder-only Transformers. This adjustment reflects a practical consideration: GenoBERT's shallow architecture would otherwise result in overly large initialization variance when model depth $N$ is small. Without sufficient regularization—such as dropout or warm-up—this could lead to gradient instability and activation explosion in early training.

In initializing the filters of the proposed CNN bottleneck layers, GenoBERT retains PyTorch's default Kaiming Uniform initialization[53,54]:

$$W \sim \mathcal{U}\left(-gain \cdot \sqrt{\frac{3}{\text{fan_in}}}, gain \cdot \sqrt{\frac{3}{\text{fan_in}}}\right) \tag{7}$$

With the default gain as $\sqrt{2}$, and fan_in $= C_{in} \times K$, suitable for ReLU's nonlinearity ($C_{in}$ represents the number of channels, or the hidden dimension $d$ to be more specific, and $K$ represent the kernel size).

For initialize the scale parameters $\gamma$ and the shift parameter $\beta$ for the layer normalization, we use ones and zeros respectively.

The full initialization strategy for all different component of the GenoBERT model can be summarized as in **Table S5**.

**Table S5. The summary of all initializations adopted in GenoBERT model.**

| Components | Initialization | Notes |
|---|---|---|
| ALL linear projections | Xavier Normal (gain varies): $W_{ij} \sim \mathcal{N}\left(0, \frac{gain^2}{n_{in}+n_{out}}\right)$ | gain=$\sqrt{2}$; $n_{in}$ and $n_{out}$ are the input/output dimensions of the projection. |
| CNN Bottleneck filters | Kaiming Uniform: $W \sim \mathcal{U}\left(-gain \cdot \sqrt{\frac{3}{\text{fan_in}}}, gain \cdot \sqrt{\frac{3}{\text{fan_in}}}\right)$ | gain=$\sqrt{2}$; fan_in = $d \times K$, $d$ is the hidden dimension and $K$ is the kernel size. |
| LayerNorm | Scale: $\gamma \leftarrow 1$, shift: $\beta \leftarrow 0$ | |

The detailed experiment settings for model training and performance evaluation can be found in the project's repository.

## On choosing genotype segment length

To guide the selection of genotype segment length for GenoBERT, we examined variant spacing and LD structure in the datasets used in this study.

As shown in **Supplementary Fig. S1**, in the case of chromosome 22 of the 1KGP dataset, most consecutive SNP pairs lie within 1Kb, implying that a sequence of L ≈ 100–128 genotypes correspond to roughly 100Kb of physical distance. This provides an empirical upper bound for the genomic region represented by each model input window and supports the adequacy of the 128-SNP context length for local linkage modeling.

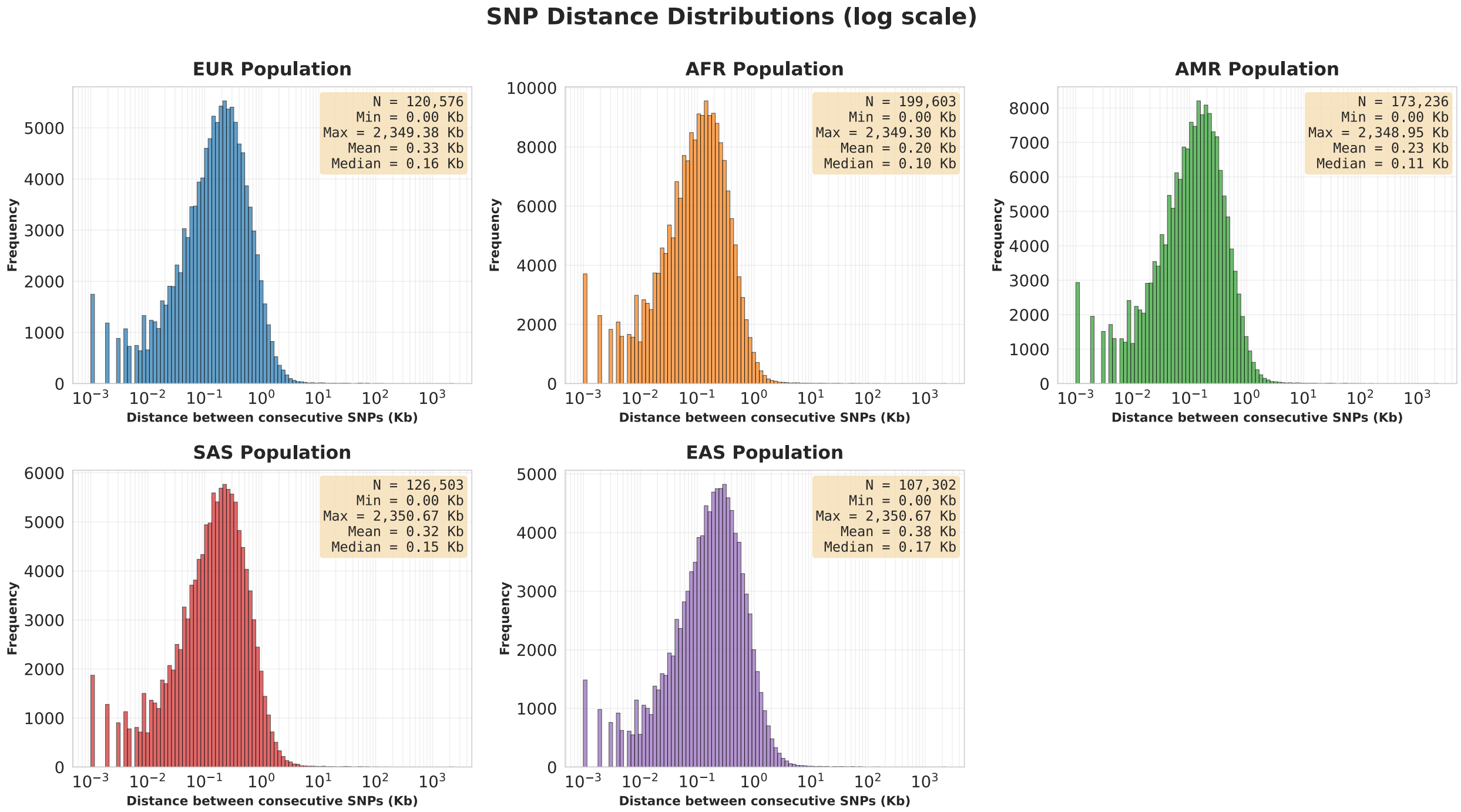

**Figure S1 | Distribution of distances between consecutive SNPs in 1KGP populations.** These log-scaled histograms illustrate adjacent-SNP distances on chromosome 22 for each of the five 1KGP populations. In all five groups, most consecutive SNPs fall within 1 Kb, defining the usual physical extent of short SNP sequences.

We further compared LD decay patterns between the 1KGP and LOS cohorts across matched ancestries (**Supplementary Fig. S2**). Consistent with established population-genetic patterns, European ancestry populations exhibit stronger and more extended LD than African ancestry populations. In both datasets and ancestries, LD decays rapidly with genomic distance, falling below $r^2 = 0.1$ within approximately 30 kb. These results indicate that a 128-SNP window provides sufficient genomic coverage to capture the majority of informative LD structure in both cohorts, despite differences in sample size and ancestry composition.

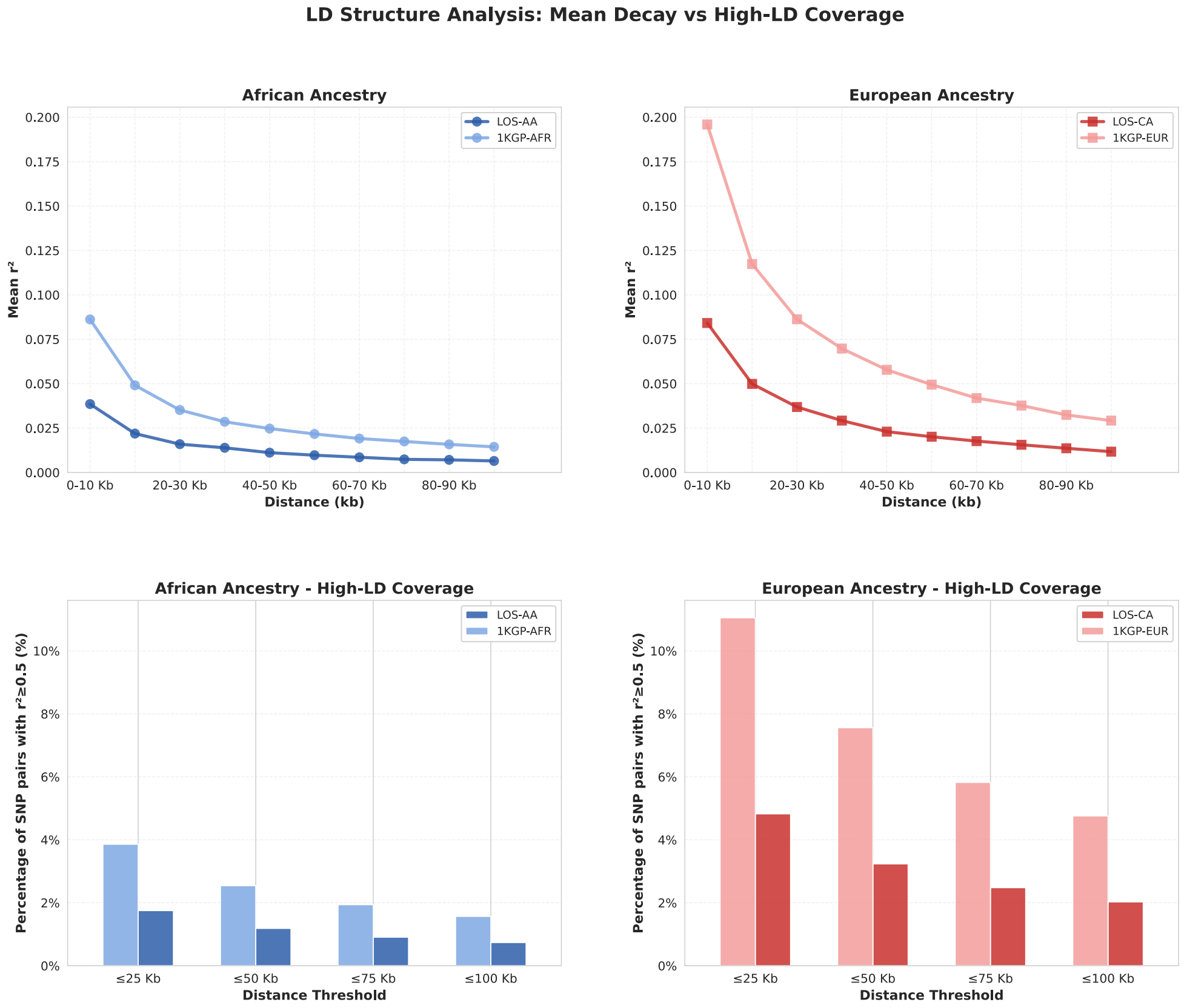

**Figure S2 | Cross-dataset comparison of LD decay and high-LD coverage between 1KGP and LOS cohorts.** Mean LD ($r^2$) decay curves (top) and the proportion of highly correlated SNP pairs ($r^2 \geq 0.5$) within distance thresholds (bottom) are shown for African and European ancestries. Within each ancestry, the 1KGP populations consistently display stronger and more extended LD than the corresponding LOS cohorts.

# Comprehensive results

**Tables S6–S35** summarize comprehensive performance results for all evaluated methods across LOS and 1KGP cohorts.

## Models' performance on LOS/mixed populations cohort

**Table S6. Beagle's performance on the mixed populations of LOS dataset**

| | MAF | #SNPs | Accuracy | $r^2$ | Precision | Recall | $F_1$ |
|---|---|---|---|---|---|---|---|
| 5% Missing | **0.1%~0.5%** | **80535** | 0.9969 | 0.8968 | 0.7493 | 0.7500 | 0.7497 |
| | **0.5%~1%** | **43321** | 0.9929 | 0.8978 | 0.7496 | 0.7490 | 0.7493 |
| | **1%~10%** | **87563** | 0.9707 | 0.9070 | 0.7500 | 0.7498 | 0.7499 |
| | **10%~20%** | **17259** | 0.8794 | 0.9053 | 0.7490 | 0.7489 | 0.7489 |
| | **20%~30%** | **11414** | 0.8183 | 0.8717 | 0.7489 | 0.7490 | 0.7489 |
| | **30%~40%** | **9073** | 0.7780 | 0.8412 | 0.7490 | 0.7491 | 0.7491 |
| | **40%~50%** | **8187** | 0.7591 | 0.8207 | 0.7493 | 0.7494 | 0.7494 |
| | **Overall** | **257352** | 0.9562 | 0.9191 | 0.7494 | 0.7494 | 0.7494 |
| 15% Missing | **MAF** | **#SNPs** | **Accuracy** | **$r^2$** | **Precision** | **Recall** | **$F_1$** |
| | **0.1%~0.5%** | **80535** | 0.9968 | 0.8834 | 0.7469 | 0.7442 | 0.7455 |
| | **0.5%~1%** | **43321** | 0.9928 | 0.8902 | 0.7471 | 0.7455 | 0.7463 |
| | **1%~10%** | **87563** | 0.9703 | 0.9029 | 0.7487 | 0.7479 | 0.7483 |
| | **10%~20%** | **17259** | 0.8789 | 0.9030 | 0.7488 | 0.7487 | 0.7487 |
| | **20%~30%** | **11414** | 0.8171 | 0.8688 | 0.7476 | 0.7478 | 0.7477 |
| | **30%~40%** | **9073** | 0.7766 | 0.8379 | 0.7475 | 0.7478 | 0.7476 |
| | **40%~50%** | **8187** | 0.7576 | 0.8172 | 0.7476 | 0.7479 | 0.7478 |
| | **Overall** | **257352** | 0.9558 | 0.9166 | 0.7485 | 0.7482 | 0.7483 |
| 25% Missing | **MAF** | **#SNPs** | **Accuracy** | **$r^2$** | **Precision** | **Recall** | **$F_1$** |
| | **0.1%~0.5%** | **80535** | 0.9967 | 0.8709 | 0.7452 | 0.7392 | 0.7422 |
| | **0.5%~1%** | **43321** | 0.9926 | 0.8813 | 0.7458 | 0.7423 | 0.7440 |
| | **1%~10%** | **87563** | 0.9697 | 0.8975 | 0.7463 | 0.7445 | 0.7454 |
| | **10%~20%** | **17259** | 0.8776 | 0.8997 | 0.7473 | 0.7469 | 0.7471 |
| | **20%~30%** | **11414** | 0.8158 | 0.8652 | 0.7462 | 0.7465 | 0.7463 |
| | **30%~40%** | **9073** | 0.7756 | 0.8342 | 0.7462 | 0.7469 | 0.7465 |
| | **40%~50%** | **8187** | 0.7567 | 0.8134 | 0.7463 | 0.7471 | 0.7467 |
| | **Overall** | **257352** | 0.9554 | 0.9135 | 0.7473 | 0.7466 | 0.7469 |
| 50% Missing | **MAF** | **#SNPs** | **Accuracy** | **$r^2$** | **Precision** | **Recall** | **$F_1$** |
| | **0.1%~0.5%** | **80535** | 0.9962 | 0.8130 | 0.7392 | 0.7085 | 0.7228 |
| | **0.5%~1%** | **43321** | 0.9919 | 0.8433 | 0.7404 | 0.7231 | 0.7314 |
| | **1%~10%** | **87563** | 0.9679 | 0.8759 | 0.7414 | 0.7327 | 0.7369 |
| | **10%~20%** | **17259** | 0.8727 | 0.8850 | 0.7415 | 0.7391 | 0.7402 |
| | **20%~30%** | **11414** | 0.8096 | 0.8485 | 0.7385 | 0.7394 | 0.7389 |
| | **30%~40%** | **9073** | 0.7685 | 0.8156 | 0.7367 | 0.7395 | 0.7380 |
| | **40%~50%** | **8187** | 0.7494 | 0.7946 | 0.7362 | 0.7397 | 0.7379 |
| | **Overall** | **257352** | 0.9534 | 0.8995 | 0.7421 | 0.7380 | 0.7400 |

**Table S7. SCDA's performance on the mixed populations of LOS dataset**

| | MAF | #SNPs | Accuracy | $r^2$ | Precision | Recall | $F_1$ |
|---|---|---|---|---|---|---|---|
| 5% Missing | 0.1%~0.5% | 80535 | 0.9993 | 0.9354 | 0.9916 | 0.9411 | 0.9651 |
| | 0.5%~1% | 43321 | 0.9988 | 0.9552 | 0.9914 | 0.9585 | 0.9744 |
| | 1%~10% | 87563 | 0.9977 | 0.9817 | 0.9902 | 0.9852 | 0.9877 |
| | 10%~20% | 17259 | 0.9943 | 0.9886 | 0.9904 | 0.9931 | 0.9917 |
| | 20%~30% | 11414 | 0.9923 | 0.9860 | 0.9904 | 0.9929 | 0.9917 |
| | 30%~40% | 9073 | 0.9915 | 0.9840 | 0.9909 | 0.9921 | 0.9915 |
| | 40%~50% | 8187 | 0.9913 | 0.9828 | 0.9912 | 0.9915 | 0.9913 |
| | Overall | 257352 | 0.9975 | 0.9881 | 0.9909 | 0.9906 | 0.9907 |
| 15% Missing | MAF | #SNPs | Accuracy | $r^2$ | Precision | Recall | $F_1$ |
| | 0.1%~0.5% | 80535 | 0.9986 | 0.8809 | 0.9865 | 0.8945 | 0.9359 |
| | 0.5%~1% | 43321 | 0.9977 | 0.9128 | 0.9866 | 0.9216 | 0.9519 |
| | 1%~10% | 87563 | 0.9950 | 0.9589 | 0.9858 | 0.9637 | 0.9745 |
| | 10%~20% | 17259 | 0.9898 | 0.9790 | 0.9863 | 0.9846 | 0.9855 |
| | 20%~30% | 11414 | 0.9873 | 0.9762 | 0.9860 | 0.9864 | 0.9862 |
| | 30%~40% | 9073 | 0.9861 | 0.9732 | 0.9858 | 0.9862 | 0.9860 |
| | 40%~50% | 8187 | 0.9860 | 0.9714 | 0.9859 | 0.9860 | 0.9859 |
| | Overall | 257352 | 0.9953 | 0.9772 | 0.9873 | 0.9790 | 0.9831 |
| 25% Missing | MAF | #SNPs | Accuracy | $r^2$ | Precision | Recall | $F_1$ |
| | 0.1%~0.5% | 80535 | 0.9982 | 0.8394 | 0.9769 | 0.8637 | 0.9130 |
| | 0.5%~1% | 43321 | 0.9964 | 0.8642 | 0.9792 | 0.8804 | 0.9244 |
| | 1%~10% | 87563 | 0.9914 | 0.9283 | 0.9802 | 0.9350 | 0.9566 |
| | 10%~20% | 17259 | 0.9824 | 0.9626 | 0.9805 | 0.9701 | 0.9752 |
| | 20%~30% | 11414 | 0.9787 | 0.9587 | 0.9791 | 0.9750 | 0.9770 |
| | 30%~40% | 9073 | 0.9769 | 0.9539 | 0.9776 | 0.9761 | 0.9768 |
| | 40%~50% | 8187 | 0.9767 | 0.9509 | 0.9769 | 0.9766 | 0.9767 |
| | Overall | 257352 | 0.9922 | 0.9614 | 0.9826 | 0.9627 | 0.9724 |
| 50% Missing | MAF | #SNPs | Accuracy | $r^2$ | Precision | Recall | $F_1$ |
| | 0.1%~0.5% | 80535 | 0.9954 | 0.5603 | 0.8282 | 0.7201 | 0.7607 |
| | 0.5%~1% | 43321 | 0.9911 | 0.6331 | 0.8847 | 0.7358 | 0.7931 |
| | 1%~10% | 87563 | 0.9735 | 0.7552 | 0.9403 | 0.7921 | 0.8532 |
| | 10%~20% | 17259 | 0.9232 | 0.8147 | 0.9420 | 0.8532 | 0.8916 |
| | 20%~30% | 11414 | 0.9004 | 0.7828 | 0.9283 | 0.8694 | 0.8939 |
| | 30%~40% | 9073 | 0.8885 | 0.7539 | 0.9132 | 0.8782 | 0.8904 |
| | 40%~50% | 8187 | 0.8846 | 0.7348 | 0.9038 | 0.8823 | 0.8870 |
| | Overall | 257352 | 0.9709 | 0.8292 | 0.9517 | 0.8483 | 0.8944 |

**Table S8. BiU-Net's performance on the mixed populations of LOS dataset**

| | MAF | #SNPs | Accuracy | $r^2$ | Precision | Recall | $F_1$ |
|---|---|---|---|---|---|---|---|
| 5% Missing | 0.1%~0.5% | 80535 | 0.9999 | 0.9937 | 0.9987 | 0.9946 | 0.9966 |
| | 0.5%~1% | 43321 | 0.9993 | 0.9721 | 0.9981 | 0.9741 | 0.9858 |
| | 1%~10% | 87563 | 0.9983 | 0.9860 | 0.9960 | 0.9874 | 0.9917 |
| | 10%~20% | 17259 | 0.9966 | 0.9931 | 0.9955 | 0.9948 | 0.9951 |
| | 20%~30% | 11414 | 0.9962 | 0.9927 | 0.9957 | 0.9958 | 0.9957 |
| | 30%~40% | 9073 | 0.9959 | 0.9921 | 0.9958 | 0.9959 | 0.9958 |
| | 40%~50% | 8187 | 0.9959 | 0.9918 | 0.9959 | 0.9959 | 0.9959 |
| | Overall | 257352 | 0.9986 | 0.9932 | 0.9962 | 0.9937 | 0.9949 |
| 15% Missing | MAF | #SNPs | Accuracy | $r^2$ | Precision | Recall | $F_1$ |
| | 0.1%~0.5% | 80535 | 0.9991 | 0.9256 | 0.9968 | 0.9311 | 0.9617 |
| | 0.5%~1% | 43321 | 0.9980 | 0.9259 | 0.9949 | 0.9312 | 0.9609 |
| | 1%~10% | 87563 | 0.9949 | 0.9580 | 0.9883 | 0.9617 | 0.9746 |
| | 10%~20% | 17259 | 0.9897 | 0.9787 | 0.9864 | 0.9839 | 0.9851 |
| | 20%~30% | 11414 | 0.9881 | 0.9776 | 0.9867 | 0.9868 | 0.9868 |
| | 30%~40% | 9073 | 0.9874 | 0.9757 | 0.9870 | 0.9873 | 0.9871 |
| | 40%~50% | 8187 | 0.9875 | 0.9747 | 0.9874 | 0.9875 | 0.9874 |
| | Overall | 257352 | 0.9956 | 0.9787 | 0.9886 | 0.9796 | 0.9841 |
| 25% Missing | MAF | #SNPs | Accuracy | $r^2$ | Precision | Recall | $F_1$ |
| | 0.1%~0.5% | 80535 | 0.9986 | 0.8805 | 0.9952 | 0.8888 | 0.9358 |
| | 0.5%~1% | 43321 | 0.9967 | 0.8788 | 0.9919 | 0.8866 | 0.9331 |
| | 1%~10% | 87563 | 0.9914 | 0.9287 | 0.9808 | 0.9342 | 0.9563 |
| | 10%~20% | 17259 | 0.9819 | 0.9628 | 0.9768 | 0.9712 | 0.9740 |
| | 20%~30% | 11414 | 0.9790 | 0.9605 | 0.9769 | 0.9765 | 0.9767 |
| | 30%~40% | 9073 | 0.9779 | 0.9573 | 0.9773 | 0.9776 | 0.9774 |
| | 40%~50% | 8187 | 0.9779 | 0.9553 | 0.9778 | 0.9779 | 0.9778 |
| | Overall | 257352 | 0.9925 | 0.9634 | 0.9806 | 0.9647 | 0.9725 |
| 50% Missing | MAF | #SNPs | Accuracy | $r^2$ | Precision | Recall | $F_1$ |
| | 0.1%~0.5% | 80535 | 0.9968 | 0.7273 | 0.9923 | 0.7426 | 0.8262 |
| | 0.5%~1% | 43321 | 0.9931 | 0.7436 | 0.9858 | 0.7559 | 0.8362 |
| | 1%~10% | 87563 | 0.9803 | 0.8359 | 0.9615 | 0.8428 | 0.8934 |
| | 10%~20% | 17259 | 0.9545 | 0.9046 | 0.9476 | 0.9222 | 0.9343 |
| | 20%~30% | 11414 | 0.9459 | 0.8958 | 0.9440 | 0.9366 | 0.9402 |
| | 30%~40% | 9073 | 0.9424 | 0.8861 | 0.9423 | 0.9407 | 0.9415 |
| | 40%~50% | 8187 | 0.9424 | 0.8807 | 0.9423 | 0.9422 | 0.9423 |
| | Overall | 257352 | 0.9818 | 0.9101 | 0.9563 | 0.9106 | 0.9322 |

**Table S9. STICI's performance on the mixed populations of LOS dataset**

| | MAF | #SNPs | Accuracy | $r^2$ | Precision | Recall | $F_1$ |
|---|---|---|---|---|---|---|---|
| 5% Missing | **0.1%~0.5%** | **80535** | 0.9969 | 0.7037 | 0.9180 | 0.7721 | 0.8298 |
| | **0.5%~1%** | **43321** | 0.9950 | 0.7938 | 0.9424 | 0.8309 | 0.8790 |
| | **1%~10%** | **87563** | 0.9896 | 0.9000 | 0.9651 | 0.9153 | 0.9389 |
| | **10%~20%** | **17259** | 0.9758 | 0.9374 | 0.9707 | 0.9562 | 0.9633 |
| | **20%~30%** | **11414** | 0.9701 | 0.9305 | 0.9699 | 0.9621 | 0.9659 |
| | **30%~40%** | **9073** | 0.9678 | 0.9246 | 0.9686 | 0.9654 | 0.9669 |
| | **40%~50%** | **8187** | 0.9674 | 0.9202 | 0.9680 | 0.9669 | 0.9673 |
| | **Overall** | **257352** | 0.9895 | 0.9394 | 0.9722 | 0.9460 | 0.9587 |
| 15% Missing | **MAF** | **#SNPs** | **Accuracy** | **$r^2$** | **Precision** | **Recall** | **$F_1$** |
| | **0.1%~0.5%** | **80535** | 0.9974 | 0.7602 | 0.9530 | 0.7995 | 0.8613 |
| | **0.5%~1%** | **43321** | 0.9959 | 0.8385 | 0.9644 | 0.8608 | 0.9064 |
| | **1%~10%** | **87563** | 0.9912 | 0.9196 | 0.9748 | 0.9298 | 0.9513 |
| | **10%~20%** | **17259** | 0.9802 | 0.9525 | 0.9770 | 0.9653 | 0.9711 |
| | **20%~30%** | **11414** | 0.9762 | 0.9482 | 0.9764 | 0.9709 | 0.9736 |
| | **30%~40%** | **9073** | 0.9742 | 0.9426 | 0.9751 | 0.9728 | 0.9739 |
| | **40%~50%** | **8187** | 0.9741 | 0.9391 | 0.9745 | 0.9738 | 0.9741 |
| | **Overall** | **257352** | 0.9914 | 0.9529 | 0.9792 | 0.9565 | 0.9676 |
| 25% Missing | **MAF** | **#SNPs** | **Accuracy** | **$r^2$** | **Precision** | **Recall** | **$F_1$** |
| | **0.1%~0.5%** | **80535** | 0.9975 | 0.7757 | 0.9658 | 0.8057 | 0.8702 |
| | **0.5%~1%** | **43321** | 0.9958 | 0.8382 | 0.9713 | 0.8569 | 0.9068 |
| | **1%~10%** | **87563** | 0.9905 | 0.9163 | 0.9763 | 0.9253 | 0.9495 |
| | **10%~20%** | **17259** | 0.9792 | 0.9526 | 0.9767 | 0.9643 | 0.9704 |
| | **20%~30%** | **11414** | 0.9754 | 0.9487 | 0.9758 | 0.9704 | 0.9731 |
| | **30%~40%** | **9073** | 0.9736 | 0.9433 | 0.9744 | 0.9724 | 0.9734 |
| | **40%~50%** | **8187** | 0.9735 | 0.9401 | 0.9738 | 0.9733 | 0.9735 |
| | **Overall** | **257352** | 0.9910 | 0.9527 | 0.9795 | 0.9554 | 0.9672 |
| 50% Missing | **MAF** | **#SNPs** | **Accuracy** | **$r^2$** | **Precision** | **Recall** | **$F_1$** |
| | **0.1%~0.5%** | **80535** | 0.9967 | 0.7139 | 0.9750 | 0.7377 | 0.8186 |
| | **0.5%~1%** | **43321** | 0.9940 | 0.7764 | 0.9718 | 0.7951 | 0.8640 |
| | **1%~10%** | **87563** | 0.9865 | 0.8857 | 0.9700 | 0.8955 | 0.9296 |
| | **10%~20%** | **17259** | 0.9723 | 0.9403 | 0.9684 | 0.9538 | 0.9609 |
| | **20%~30%** | **11414** | 0.9673 | 0.9354 | 0.9665 | 0.9620 | 0.9642 |
| | **30%~40%** | **9073** | 0.9646 | 0.9282 | 0.9647 | 0.9636 | 0.9641 |
| | **40%~50%** | **8187** | 0.9647 | 0.9251 | 0.9648 | 0.9646 | 0.9647 |
| | **Overall** | **257352** | 0.9877 | 0.9383 | 0.9718 | 0.9401 | 0.9554 |

**Table S10. GenoBERT's performance on the mixed populations of LOS dataset**

| | MAF | #SNPs | Accuracy | $r^2$ | Precision | Recall | $F_1$ |
|---|---|---|---|---|---|---|---|
| 5% Missing | 0.1%~0.5% | 80535 | 0.9999 | 0.9933 | 0.9987 | 0.9958 | 0.9972 |
| | 0.5%~1% | 43321 | 0.9996 | 0.9825 | 0.9982 | 0.9851 | 0.9916 |
| | 1%~10% | 87563 | 0.9992 | 0.9929 | 0.9982 | 0.9943 | 0.9962 |
| | 10%~20% | 17259 | 0.9986 | 0.9968 | 0.9983 | 0.9979 | 0.9981 |
| | 20%~30% | 11414 | 0.9985 | 0.9968 | 0.9984 | 0.9983 | 0.9983 |
| | 30%~40% | 9073 | 0.9984 | 0.9965 | 0.9983 | 0.9983 | 0.9983 |
| | 40%~50% | 8187 | 0.9984 | 0.9964 | 0.9984 | 0.9984 | 0.9984 |
| | Overall | 257352 | 0.9994 | 0.9966 | 0.9985 | 0.9972 | 0.9978 |
| 15% Missing | MAF | #SNPs | Accuracy | $r^2$ | Precision | Recall | $F_1$ |
| | 0.1%~0.5% | 80535 | 0.9993 | 0.9322 | 0.9949 | 0.9432 | 0.9677 |
| | 0.5%~1% | 43321 | 0.9987 | 0.9491 | 0.9943 | 0.9576 | 0.9753 |
| | 1%~10% | 87563 | 0.9975 | 0.9773 | 0.9943 | 0.9816 | 0.9879 |
| | 10%~20% | 17259 | 0.9954 | 0.9892 | 0.9946 | 0.9928 | 0.9937 |
| | 20%~30% | 11414 | 0.9949 | 0.9893 | 0.9947 | 0.9943 | 0.9945 |
| | 30%~40% | 9073 | 0.9946 | 0.9884 | 0.9946 | 0.9945 | 0.9946 |
| | 40%~50% | 8187 | 0.9946 | 0.9878 | 0.9946 | 0.9946 | 0.9946 |
| | Overall | 257352 | 0.9978 | 0.9883 | 0.9952 | 0.9898 | 0.9925 |
| 25% Missing | MAF | #SNPs | Accuracy | $r^2$ | Precision | Recall | $F_1$ |
| | 0.1%~0.5% | 80535 | 0.9988 | 0.8846 | 0.9899 | 0.9049 | 0.9437 |
| | 0.5%~1% | 43321 | 0.9978 | 0.9104 | 0.9897 | 0.9258 | 0.9558 |
| | 1%~10% | 87563 | 0.9954 | 0.9578 | 0.9902 | 0.9655 | 0.9776 |
| | 10%~20% | 17259 | 0.9910 | 0.9786 | 0.9903 | 0.9854 | 0.9878 |
| | 20%~30% | 11414 | 0.9900 | 0.9784 | 0.9900 | 0.9885 | 0.9893 |
| | 30%~40% | 9073 | 0.9894 | 0.9767 | 0.9896 | 0.9892 | 0.9894 |
| | 40%~50% | 8187 | 0.9894 | 0.9752 | 0.9895 | 0.9894 | 0.9894 |
| | Overall | 257352 | 0.9959 | 0.9778 | 0.9914 | 0.9807 | 0.9860 |
| 50% Missing | MAF | #SNPs | Accuracy | $r^2$ | Precision | Recall | $F_1$ |
| | 0.1%~0.5% | 80535 | 0.9968 | 0.6749 | 0.9596 | 0.7444 | 0.8252 |
| | 0.5%~1% | 43321 | 0.9939 | 0.7322 | 0.9670 | 0.7873 | 0.8610 |
| | 1%~10% | 87563 | 0.9842 | 0.8437 | 0.9747 | 0.8718 | 0.9182 |
| | 10%~20% | 17259 | 0.9601 | 0.8944 | 0.9693 | 0.9262 | 0.9465 |
| | 20%~30% | 11414 | 0.9512 | 0.8821 | 0.9617 | 0.9382 | 0.9491 |
| | 30%~40% | 9073 | 0.9468 | 0.8694 | 0.9543 | 0.9430 | 0.9478 |
| | 40%~50% | 8187 | 0.9456 | 0.8598 | 0.9497 | 0.9449 | 0.9464 |
| | Overall | 257352 | 0.9842 | 0.9029 | 0.9758 | 0.9177 | 0.9452 |

## Models' performance on LOS/AA cohort

**Table S11. Beagle's performance on the African American cohort of LOS dataset**

| | MAF | #SNPs | Accuracy | $r^2$ | Precision | Recall | $F_1$ |
|---|---|---|---|---|---|---|---|
| 5% Missing | 0.1%~0.5% | 72297 | 0.9966 | 0.9123 | 0.7516 | 0.7527 | 0.7522 |
| | 0.5%~1% | 41655 | 0.9928 | 0.9173 | 0.7509 | 0.7517 | 0.7513 |
| | 1%~10% | 120279 | 0.9653 | 0.9059 | 0.7504 | 0.7504 | 0.7504 |
| | 10%~20% | 30423 | 0.8802 | 0.8956 | 0.7503 | 0.7503 | 0.7503 |
| | 20%~30% | 17490 | 0.8194 | 0.8700 | 0.7497 | 0.7497 | 0.7497 |
| | 30%~40% | 13326 | 0.7797 | 0.8418 | 0.7501 | 0.7502 | 0.7501 |
| | 40%~50% | 12129 | 0.7591 | 0.8201 | 0.7498 | 0.7499 | 0.7498 |
| | Overall | 307599 | 0.9435 | 0.9138 | 0.7503 | 0.7503 | 0.7503 |
| 15% Missing | MAF | #SNPs | Accuracy | $r^2$ | Precision | Recall | $F_1$ |
| | 0.1%~0.5% | 72297 | 0.9965 | 0.9032 | 0.7459 | 0.7465 | 0.7462 |
| | 0.5%~1% | 41655 | 0.9925 | 0.9078 | 0.7462 | 0.7451 | 0.7457 |
| | 1%~10% | 120279 | 0.9646 | 0.9012 | 0.7474 | 0.7472 | 0.7473 |
| | 10%~20% | 30423 | 0.8788 | 0.8923 | 0.7481 | 0.7482 | 0.7481 |
| | 20%~30% | 17490 | 0.8175 | 0.8663 | 0.7475 | 0.7476 | 0.7476 |
| | 30%~40% | 13326 | 0.7779 | 0.8380 | 0.7481 | 0.7484 | 0.7482 |
| | 40%~50% | 12129 | 0.7569 | 0.8156 | 0.7472 | 0.7477 | 0.7475 |
| | Overall | 307599 | 0.9428 | 0.9107 | 0.7481 | 0.7480 | 0.7481 |
| 25% Missing | MAF | #SNPs | Accuracy | $r^2$ | Precision | Recall | $F_1$ |
| | 0.1%~0.5% | 72297 | 0.9963 | 0.8909 | 0.7418 | 0.7395 | 0.7406 |
| | 0.5%~1% | 41655 | 0.9923 | 0.8988 | 0.7422 | 0.7394 | 0.7408 |
| | 1%~10% | 120279 | 0.9638 | 0.8959 | 0.7436 | 0.7430 | 0.7433 |
| | 10%~20% | 30423 | 0.8765 | 0.8880 | 0.7443 | 0.7443 | 0.7443 |
| | 20%~30% | 17490 | 0.8147 | 0.8616 | 0.7439 | 0.7443 | 0.7441 |
| | 30%~40% | 13326 | 0.7744 | 0.8327 | 0.7441 | 0.7447 | 0.7444 |
| | 40%~50% | 12129 | 0.7532 | 0.8097 | 0.7430 | 0.7439 | 0.7434 |
| | Overall | 307599 | 0.9417 | 0.9069 | 0.7446 | 0.7444 | 0.7445 |
| 50% Missing | MAF | #SNPs | Accuracy | $r^2$ | Precision | Recall | $F_1$ |
| | 0.1%~0.5% | 72297 | 0.9959 | 0.8434 | 0.7403 | 0.7147 | 0.7267 |
| | 0.5%~1% | 41655 | 0.9916 | 0.8686 | 0.7399 | 0.7253 | 0.7323 |
| | 1%~10% | 120279 | 0.9624 | 0.8794 | 0.7416 | 0.7374 | 0.7394 |
| | 10%~20% | 30423 | 0.8739 | 0.8757 | 0.7422 | 0.7414 | 0.7418 |
| | 20%~30% | 17490 | 0.8105 | 0.8471 | 0.7391 | 0.7401 | 0.7396 |
| | 30%~40% | 13326 | 0.7688 | 0.8164 | 0.7370 | 0.7392 | 0.7381 |
| | 40%~50% | 12129 | 0.7485 | 0.7936 | 0.7362 | 0.7393 | 0.7377 |
| | Overall | 307599 | 0.9400 | 0.8951 | 0.7421 | 0.7401 | 0.7410 |

**Table S12. SCDA's performance on the African American cohort of LOS dataset**

| | MAF | #SNPs | Accuracy | $r^2$ | Precision | Recall | $F_1$ |
|---|---|---|---|---|---|---|---|
| 5% Missing | **0.1%~0.5%** | **72297** | 0.9944 | 0.5640 | 0.7086 | 0.9448 | 0.8016 |
| | **0.5%~1%** | **41655** | 0.9932 | 0.7058 | 0.8099 | 0.9366 | 0.8667 |
| | **1%~10%** | **120279** | 0.9850 | 0.8485 | 0.9233 | 0.9158 | 0.9195 |
| | **10%~20%** | **30423** | 0.9584 | 0.8691 | 0.9491 | 0.9172 | 0.9326 |
| | **20%~30%** | **17490** | 0.9405 | 0.8447 | 0.9457 | 0.9194 | 0.9316 |
| | **30%~40%** | **13326** | 0.9296 | 0.8169 | 0.9385 | 0.9221 | 0.9287 |
| | **40%~50%** | **12129** | 0.9241 | 0.7951 | 0.9311 | 0.9229 | 0.9247 |
| | **Overall** | **307599** | 0.9784 | 0.8730 | 0.9442 | 0.9198 | 0.9317 |
| 15% Missing | **MAF** | **#SNPs** | **Accuracy** | **$r^2$** | **Precision** | **Recall** | **$F_1$** |
| | **0.1%~0.5%** | **72297** | 0.9919 | 0.4260 | 0.6420 | 0.8997 | 0.7374 |
| | **0.5%~1%** | **41655** | 0.9891 | 0.5497 | 0.7522 | 0.8587 | 0.7998 |
| | **1%~10%** | **120279** | 0.9757 | 0.7383 | 0.9074 | 0.8570 | 0.8810 |
| | **10%~20%** | **30423** | 0.9306 | 0.7602 | 0.9398 | 0.8605 | 0.8963 |
| | **20%~30%** | **17490** | 0.8985 | 0.7183 | 0.9288 | 0.8619 | 0.8900 |
| | **30%~40%** | **13326** | 0.8782 | 0.6753 | 0.9126 | 0.8647 | 0.8809 |
| | **40%~50%** | **12129** | 0.8680 | 0.6451 | 0.8997 | 0.8656 | 0.8723 |
| | **Overall** | **307599** | 0.9640 | 0.7702 | 0.9347 | 0.8620 | 0.8960 |
| 25% Missing | **MAF** | **#SNPs** | **Accuracy** | **$r^2$** | **Precision** | **Recall** | **$F_1$** |
| | **0.1%~0.5%** | **72297** | 0.9902 | 0.3369 | 0.6033 | 0.8422 | 0.6906 |
| | **0.5%~1%** | **41655** | 0.9863 | 0.4464 | 0.7161 | 0.7946 | 0.7515 |
| | **1%~10%** | **120279** | 0.9670 | 0.6373 | 0.8953 | 0.7951 | 0.8405 |
| | **10%~20%** | **30423** | 0.9016 | 0.6551 | 0.9315 | 0.8000 | 0.8551 |
| | **20%~30%** | **17490** | 0.8540 | 0.6020 | 0.9143 | 0.8006 | 0.8433 |
| | **30%~40%** | **13326** | 0.8233 | 0.5518 | 0.8925 | 0.8034 | 0.8291 |
| | **40%~50%** | **12129** | 0.8080 | 0.5188 | 0.8771 | 0.8043 | 0.8172 |
| | **Overall** | **307599** | 0.9497 | 0.6713 | 0.9270 | 0.8008 | 0.8564 |
| 50% Missing | **MAF** | **#SNPs** | **Accuracy** | **$r^2$** | **Precision** | **Recall** | **$F_1$** |
| | **0.1%~0.5%** | **72297** | 0.9892 | 0.1980 | 0.5696 | 0.6551 | 0.6052 |
| | **0.5%~1%** | **41655** | 0.9830 | 0.2733 | 0.6853 | 0.6222 | 0.6502 |
| | **1%~10%** | **120279** | 0.9467 | 0.4118 | 0.8845 | 0.6244 | 0.7184 |
| | **10%~20%** | **30423** | 0.8225 | 0.4085 | 0.9156 | 0.6303 | 0.7188 |
| | **20%~30%** | **17490** | 0.7309 | 0.3508 | 0.8873 | 0.6295 | 0.6943 |
| | **30%~40%** | **13326** | 0.6699 | 0.3041 | 0.8592 | 0.6316 | 0.6705 |
| | **40%~50%** | **12129** | 0.6397 | 0.2773 | 0.8423 | 0.6326 | 0.6539 |
| | **Overall** | **307599** | 0.9130 | 0.4350 | 0.9179 | 0.6296 | 0.7289 |

**Table S13. BiU-Net's performance on the African American cohort of LOS dataset**

| | MAF | #SNPs | Accuracy | $r^2$ | Precision | Recall | $F_1$ |
|---|---|---|---|---|---|---|---|
| 5% Missing | **0.1%~0.5%** | **72297** | 1.0000 | 0.9976 | 0.9975 | 0.9989 | 0.9982 |
| | **0.5%~1%** | **41655** | 0.9999 | 0.9956 | 0.9982 | 0.9959 | 0.9971 |
| | **1%~10%** | **120279** | 0.9970 | 0.9788 | 0.9963 | 0.9792 | 0.9876 |
| | **10%~20%** | **30423** | 0.9939 | 0.9872 | 0.9933 | 0.9884 | 0.9908 |
| | **20%~30%** | **17490** | 0.9928 | 0.9882 | 0.9925 | 0.9909 | 0.9917 |
| | **30%~40%** | **13326** | 0.9927 | 0.9882 | 0.9926 | 0.9922 | 0.9924 |
| | **40%~50%** | **12129** | 0.9928 | 0.9880 | 0.9928 | 0.9927 | 0.9927 |
| | **Overall** | **307599** | 0.9972 | 0.9896 | 0.9945 | 0.9887 | 0.9916 |
| 15% Missing | **MAF** | **#SNPs** | **Accuracy** | **$r^2$** | **Precision** | **Recall** | **$F_1$** |
| | **0.1%~0.5%** | **72297** | 0.9994 | 0.9587 | 0.9897 | 0.9644 | 0.9767 |
| | **0.5%~1%** | **41655** | 0.9978 | 0.9316 | 0.9923 | 0.9260 | 0.9568 |
| | **1%~10%** | **120279** | 0.9906 | 0.9333 | 0.9875 | 0.9355 | 0.9600 |
| | **10%~20%** | **30423** | 0.9792 | 0.9549 | 0.9788 | 0.9602 | 0.9692 |
| | **20%~30%** | **17490** | 0.9752 | 0.9571 | 0.9754 | 0.9683 | 0.9718 |
| | **30%~40%** | **13326** | 0.9748 | 0.9571 | 0.9751 | 0.9729 | 0.9739 |
| | **40%~50%** | **12129** | 0.9752 | 0.9564 | 0.9752 | 0.9748 | 0.9750 |
| | **Overall** | **307599** | 0.9903 | 0.9632 | 0.9825 | 0.9611 | 0.9715 |
| 25% Missing | **MAF** | **#SNPs** | **Accuracy** | **$r^2$** | **Precision** | **Recall** | **$F_1$** |
| | **0.1%~0.5%** | **72297** | 0.9986 | 0.9009 | 0.9705 | 0.9199 | 0.9437 |
| | **0.5%~1%** | **41655** | 0.9960 | 0.8776 | 0.9789 | 0.8740 | 0.9201 |
| | **1%~10%** | **120279** | 0.9835 | 0.8813 | 0.9752 | 0.8885 | 0.9274 |
| | **10%~20%** | **30423** | 0.9600 | 0.9100 | 0.9614 | 0.9232 | 0.9411 |
| | **20%~30%** | **17490** | 0.9509 | 0.9106 | 0.9540 | 0.9367 | 0.9448 |
| | **30%~40%** | **13326** | 0.9495 | 0.9087 | 0.9517 | 0.9454 | 0.9482 |
| | **40%~50%** | **12129** | 0.9499 | 0.9058 | 0.9506 | 0.9490 | 0.9495 |
| | **Overall** | **307599** | 0.9818 | 0.9276 | 0.9678 | 0.9266 | 0.9462 |
| 50% Missing | **MAF** | **#SNPs** | **Accuracy** | **$r^2$** | **Precision** | **Recall** | **$F_1$** |
| | **0.1%~0.5%** | **72297** | 0.9946 | 0.5963 | 0.7870 | 0.7701 | 0.7747 |
| | **0.5%~1%** | **41655** | 0.9898 | 0.6583 | 0.8495 | 0.7366 | 0.7804 |
| | **1%~10%** | **120279** | 0.9605 | 0.6949 | 0.9035 | 0.7474 | 0.8079 |
| | **10%~20%** | **30423** | 0.8875 | 0.7211 | 0.8958 | 0.7825 | 0.8279 |
| | **20%~30%** | **17490** | 0.8476 | 0.6902 | 0.8747 | 0.8000 | 0.8286 |
| | **30%~40%** | **13326** | 0.8309 | 0.6577 | 0.8595 | 0.8151 | 0.8291 |
| | **40%~50%** | **12129** | 0.8255 | 0.6344 | 0.8490 | 0.8221 | 0.8268 |
| | **Overall** | **307599** | 0.9479 | 0.7597 | 0.9081 | 0.7925 | 0.8417 |

**Table S14. STICI's performance on the African American cohort of LOS dataset**

| | MAF | #SNPs | Accuracy | $r^2$ | Precision | Recall | $F_1$ |
|---|---|---|---|---|---|---|---|
| 5% Missing | **0.1%~0.5%** | **72297** | 0.9960 | 0.7324 | 0.9346 | 0.7184 | 0.7925 |
| | **0.5%~1%** | **41655** | 0.9943 | 0.8252 | 0.9474 | 0.8119 | 0.8680 |
| | **1%~10%** | **120279** | 0.9881 | 0.9047 | 0.9637 | 0.9173 | 0.9394 |
| | **10%~20%** | **30423** | 0.9763 | 0.9320 | 0.9683 | 0.9548 | 0.9614 |
| | **20%~30%** | **17490** | 0.9684 | 0.9230 | 0.9662 | 0.9593 | 0.9627 |
| | **30%~40%** | **13326** | 0.9651 | 0.9142 | 0.9653 | 0.9623 | 0.9638 |
| | **40%~50%** | **12129** | 0.9638 | 0.9074 | 0.9643 | 0.9633 | 0.9636 |
| | **Overall** | **307599** | 0.9866 | 0.9358 | 0.9693 | 0.9451 | 0.9569 |
| 15% Missing | **MAF** | **#SNPs** | **Accuracy** | **$r^2$** | **Precision** | **Recall** | **$F_1$** |
| | **0.1%~0.5%** | **72297** | 0.9966 | 0.7701 | 0.9503 | 0.7556 | 0.8269 |
| | **0.5%~1%** | **41655** | 0.9946 | 0.8344 | 0.9567 | 0.8213 | 0.8777 |
| | **1%~10%** | **120279** | 0.9886 | 0.9116 | 0.9687 | 0.9240 | 0.9453 |
| | **10%~20%** | **30423** | 0.9775 | 0.9392 | 0.9715 | 0.9591 | 0.9652 |
| | **20%~30%** | **17490** | 0.9706 | 0.9324 | 0.9693 | 0.9635 | 0.9664 |
| | **30%~40%** | **13326** | 0.9682 | 0.9262 | 0.9687 | 0.9664 | 0.9675 |
| | **40%~50%** | **12129** | 0.9673 | 0.9208 | 0.9677 | 0.9670 | 0.9673 |
| | **Overall** | **307599** | 0.9874 | 0.9427 | 0.9730 | 0.9508 | 0.9616 |
| 25% Missing | **MAF** | **#SNPs** | **Accuracy** | **$r^2$** | **Precision** | **Recall** | **$F_1$** |
| | **0.1%~0.5%** | **72297** | 0.9967 | 0.7812 | 0.9585 | 0.7653 | 0.8368 |
| | **0.5%~1%** | **41655** | 0.9944 | 0.8301 | 0.9606 | 0.8152 | 0.8747 |
| | **1%~10%** | **120279** | 0.9872 | 0.9037 | 0.9689 | 0.9165 | 0.9412 |
| | **10%~20%** | **30423** | 0.9749 | 0.9354 | 0.9698 | 0.9555 | 0.9625 |
| | **20%~30%** | **17490** | 0.9679 | 0.9299 | 0.9671 | 0.9609 | 0.9640 |
| | **30%~40%** | **13326** | 0.9657 | 0.9243 | 0.9662 | 0.9640 | 0.9651 |
| | **40%~50%** | **12129** | 0.9650 | 0.9193 | 0.9653 | 0.9647 | 0.9650 |
| | **Overall** | **307599** | 0.9863 | 0.9397 | 0.9719 | 0.9475 | 0.9594 |
| 50% Missing | **MAF** | **#SNPs** | **Accuracy** | **$r^2$** | **Precision** | **Recall** | **$F_1$** |
| | **0.1%~0.5%** | **72297** | 0.9961 | 0.7451 | 0.9701 | 0.7218 | 0.8026 |
| | **0.5%~1%** | **41655** | 0.9927 | 0.7825 | 0.9637 | 0.7571 | 0.8316 |
| | **1%~10%** | **120279** | 0.9814 | 0.8664 | 0.9611 | 0.8805 | 0.9171 |
| | **10%~20%** | **30423** | 0.9641 | 0.9139 | 0.9578 | 0.9381 | 0.9477 |
| | **20%~30%** | **17490** | 0.9548 | 0.9089 | 0.9532 | 0.9466 | 0.9498 |
| | **30%~40%** | **13326** | 0.9519 | 0.9024 | 0.9519 | 0.9504 | 0.9511 |
| | **40%~50%** | **12129** | 0.9520 | 0.8983 | 0.9520 | 0.9518 | 0.9519 |
| | **Overall** | **307599** | 0.9807 | 0.9202 | 0.9612 | 0.9283 | 0.9441 |

**Table S15. GenoBERT's performance on the African American cohort of LOS dataset**

| | MAF | #SNPs | Accuracy | $r^2$ | Precision | Recall | $F_1$ |
|---|---|---|---|---|---|---|---|
| 5% Missing | 0.1%~0.5% | 72297 | 1.0000 | 0.9979 | 0.9995 | 0.9988 | 0.9992 |
| | 0.5%~1% | 41655 | 0.9999 | 0.9963 | 0.9990 | 0.9974 | 0.9982 |
| | 1%~10% | 120279 | 0.9991 | 0.9933 | 0.9982 | 0.9944 | 0.9963 |
| | 10%~20% | 30423 | 0.9986 | 0.9965 | 0.9983 | 0.9977 | 0.9980 |
| | 20%~30% | 17490 | 0.9982 | 0.9963 | 0.9982 | 0.9979 | 0.9981 |
| | 30%~40% | 13326 | 0.9981 | 0.9961 | 0.9981 | 0.9981 | 0.9981 |
| | 40%~50% | 12129 | 0.9982 | 0.9960 | 0.9982 | 0.9982 | 0.9982 |
| | Overall | 307599 | 0.9992 | 0.9967 | 0.9985 | 0.9973 | 0.9979 |
| 15% Missing | MAF | #SNPs | Accuracy | $r^2$ | Precision | Recall | $F_1$ |
| | 0.1%~0.5% | 72297 | 0.9996 | 0.9676 | 0.9977 | 0.9687 | 0.9828 |
| | 0.5%~1% | 41655 | 0.9985 | 0.9518 | 0.9958 | 0.9494 | 0.9715 |
| | 1%~10% | 120279 | 0.9972 | 0.9790 | 0.9943 | 0.9828 | 0.9885 |
| | 10%~20% | 30423 | 0.9956 | 0.9893 | 0.9947 | 0.9930 | 0.9938 |
| | 20%~30% | 17490 | 0.9945 | 0.9885 | 0.9943 | 0.9937 | 0.9940 |
| | 30%~40% | 13326 | 0.9942 | 0.9879 | 0.9942 | 0.9941 | 0.9942 |
| | 40%~50% | 12129 | 0.9943 | 0.9874 | 0.9943 | 0.9943 | 0.9943 |
| | Overall | 307599 | 0.9974 | 0.9888 | 0.9950 | 0.9906 | 0.9928 |
| 25% Missing | MAF | #SNPs | Accuracy | $r^2$ | Precision | Recall | $F_1$ |
| | 0.1%~0.5% | 72297 | 0.9990 | 0.9263 | 0.9939 | 0.9285 | 0.9590 |
| | 0.5%~1% | 41655 | 0.9973 | 0.9154 | 0.9911 | 0.9121 | 0.9483 |
| | 1%~10% | 120279 | 0.9949 | 0.9620 | 0.9893 | 0.9693 | 0.9791 |
| | 10%~20% | 30423 | 0.9919 | 0.9797 | 0.9901 | 0.9869 | 0.9885 |
| | 20%~30% | 17490 | 0.9897 | 0.9785 | 0.9893 | 0.9882 | 0.9887 |
| | 30%~40% | 13326 | 0.9892 | 0.9772 | 0.9893 | 0.9890 | 0.9891 |
| | 40%~50% | 12129 | 0.9893 | 0.9765 | 0.9893 | 0.9893 | 0.9893 |
| | Overall | 307599 | 0.9951 | 0.9791 | 0.9907 | 0.9827 | 0.9867 |
| 50% Missing | MAF | #SNPs | Accuracy | $r^2$ | Precision | Recall | $F_1$ |
| | 0.1%~0.5% | 72297 | 0.9964 | 0.7075 | 0.9288 | 0.7819 | 0.8383 |
| | 0.5%~1% | 41655 | 0.9930 | 0.7565 | 0.9452 | 0.7876 | 0.8497 |
| | 1%~10% | 120279 | 0.9829 | 0.8621 | 0.9658 | 0.8922 | 0.9260 |
| | 10%~20% | 30423 | 0.9662 | 0.9076 | 0.9666 | 0.9397 | 0.9527 |
| | 20%~30% | 17490 | 0.9560 | 0.8987 | 0.9605 | 0.9460 | 0.9529 |
| | 30%~40% | 13326 | 0.9532 | 0.8903 | 0.9567 | 0.9507 | 0.9535 |
| | 40%~50% | 12129 | 0.9529 | 0.8854 | 0.9544 | 0.9525 | 0.9532 |
| | Overall | 307599 | 0.9818 | 0.9133 | 0.9689 | 0.9318 | 0.9496 |

## Models' performance on LOS/CA cohort

**Table S16. Beagle's performance on the Caucasian cohort of LOS dataset**

| | MAF | #SNPs | Accuracy | $r^2$ | Precision | Recall | $F_1$ |
|---|---|---|---|---|---|---|---|
| 5% Missing | 0.1%~0.5% | 52524 | 0.9973 | 0.9284 | 0.7477 | 0.7491 | 0.7484 |
| | 0.5%~1% | 20151 | 0.9933 | 0.9323 | 0.7456 | 0.7461 | 0.7458 |
| | 1%~10% | 54011 | 0.9623 | 0.9178 | 0.7498 | 0.7495 | 0.7497 |
| | 10%~20% | 19838 | 0.8751 | 0.8992 | 0.7496 | 0.7496 | 0.7496 |
| | 20%~30% | 15717 | 0.8144 | 0.8686 | 0.7494 | 0.7494 | 0.7494 |
| | 30%~40% | 13401 | 0.7738 | 0.8367 | 0.7488 | 0.7489 | 0.7488 |
| | 40%~50% | 12757 | 0.7537 | 0.8144 | 0.7486 | 0.7487 | 0.7486 |
| | Overall | 188399 | 0.9263 | 0.9138 | 0.7493 | 0.7492 | 0.7493 |
| 15% Missing | MAF | #SNPs | Accuracy | $r^2$ | Precision | Recall | $F_1$ |
| | 0.1%~0.5% | 52524 | 0.9972 | 0.9181 | 0.7447 | 0.7435 | 0.7441 |
| | 0.5%~1% | 20151 | 0.9931 | 0.9235 | 0.7477 | 0.7440 | 0.7458 |
| | 1%~10% | 54011 | 0.9617 | 0.9136 | 0.7486 | 0.7474 | 0.7480 |
| | 10%~20% | 19838 | 0.8749 | 0.8970 | 0.7501 | 0.7499 | 0.7500 |
| | 20%~30% | 15717 | 0.8141 | 0.8665 | 0.7492 | 0.7493 | 0.7493 |
| | 30%~40% | 13401 | 0.7737 | 0.8345 | 0.7487 | 0.7490 | 0.7489 |
| | 40%~50% | 12757 | 0.7542 | 0.8126 | 0.7489 | 0.7492 | 0.7490 |
| | Overall | 188399 | 0.9261 | 0.9118 | 0.7495 | 0.7493 | 0.7494 |
| 25% Missing | MAF | #SNPs | Accuracy | $r^2$ | Precision | Recall | $F_1$ |
| | 0.1%~0.5% | 52524 | 0.9971 | 0.9074 | 0.7452 | 0.7396 | 0.7424 |
| | 0.5%~1% | 20151 | 0.9929 | 0.9141 | 0.7443 | 0.7376 | 0.7409 |
| | 1%~10% | 54011 | 0.9614 | 0.9094 | 0.7491 | 0.7467 | 0.7479 |
| | 10%~20% | 19838 | 0.8736 | 0.8938 | 0.7485 | 0.7482 | 0.7484 |
| | 20%~30% | 15717 | 0.8126 | 0.8631 | 0.7476 | 0.7478 | 0.7477 |
| | 30%~40% | 13401 | 0.7721 | 0.8308 | 0.7469 | 0.7474 | 0.7471 |
| | 40%~50% | 12757 | 0.7524 | 0.8086 | 0.7467 | 0.7474 | 0.7470 |
| | Overall | 188399 | 0.9254 | 0.9091 | 0.7484 | 0.7479 | 0.7482 |
| 50% Missing | MAF | #SNPs | Accuracy | $r^2$ | Precision | Recall | $F_1$ |
| | 0.1%~0.5% | 52524 | 0.9966 | 0.8559 | 0.7368 | 0.6999 | 0.7167 |
| | 0.5%~1% | 20151 | 0.9918 | 0.8752 | 0.7359 | 0.7060 | 0.7199 |
| | 1%~10% | 54011 | 0.9583 | 0.8884 | 0.7396 | 0.7301 | 0.7347 |
| | 10%~20% | 19838 | 0.8673 | 0.8795 | 0.7391 | 0.7378 | 0.7384 |
| | 20%~30% | 15717 | 0.8047 | 0.8482 | 0.7376 | 0.7384 | 0.7380 |
| | 30%~40% | 13401 | 0.7633 | 0.8143 | 0.7359 | 0.7381 | 0.7370 |
| | 40%~50% | 12757 | 0.7424 | 0.7906 | 0.7344 | 0.7373 | 0.7358 |
| | Overall | 188399 | 0.9216 | 0.8964 | 0.7396 | 0.7374 | 0.7385 |

**Table S17. SCDA's performance on the Caucasian cohort of LOS dataset**

| | MAF | #SNPs | Accuracy | $r^2$ | Precision | Recall | $F_1$ |
|---|---|---|---|---|---|---|---|
| 5% Missing | **0.1%~0.5%** | **52524** | 0.9858 | 0.3219 | 0.5053 | 0.9166 | 0.6180 |
| | **0.5%~1%** | **20151** | 0.9844 | 0.5384 | 0.6652 | 0.9138 | 0.7601 |
| | **1%~10%** | **54011** | 0.9745 | 0.7977 | 0.8776 | 0.8994 | 0.8883 |
| | **10%~20%** | **19838** | 0.9453 | 0.8463 | 0.9277 | 0.9035 | 0.9152 |
| | **20%~30%** | **15717** | 0.9314 | 0.8334 | 0.9336 | 0.9124 | 0.9223 |
| | **30%~40%** | **13401** | 0.9229 | 0.8095 | 0.9293 | 0.9166 | 0.9216 |
| | **40%~50%** | **12757** | 0.9216 | 0.7970 | 0.9269 | 0.9205 | 0.9218 |
| | **Overall** | **188399** | 0.9648 | 0.8455 | 0.9213 | 0.9113 | 0.9162 |
| 15% Missing | **MAF** | **#SNPs** | **Accuracy** | **$r^2$** | **Precision** | **Recall** | **$F_1$** |
| | **0.1%~0.5%** | **52524** | 0.9813 | 0.2320 | 0.4558 | 0.8720 | 0.5577 |
| | **0.5%~1%** | **20151** | 0.9787 | 0.4109 | 0.6050 | 0.8467 | 0.6944 |
| | **1%~10%** | **54011** | 0.9640 | 0.6952 | 0.8552 | 0.8497 | 0.8524 |
| | **10%~20%** | **19838** | 0.9188 | 0.7460 | 0.9163 | 0.8547 | 0.8829 |
| | **20%~30%** | **15717** | 0.8929 | 0.7176 | 0.9156 | 0.8621 | 0.8847 |
| | **30%~40%** | **13401** | 0.8771 | 0.6816 | 0.9037 | 0.8667 | 0.8789 |
| | **40%~50%** | **12757** | 0.8721 | 0.6614 | 0.8967 | 0.8702 | 0.8753 |
| | **Overall** | **188399** | 0.9473 | 0.7486 | 0.9076 | 0.8612 | 0.8833 |
| 25% Missing | **MAF** | **#SNPs** | **Accuracy** | **$r^2$** | **Precision** | **Recall** | **$F_1$** |
| | **0.1%~0.5%** | **52524** | 0.9784 | 0.1780 | 0.4274 | 0.8244 | 0.5193 |
| | **0.5%~1%** | **20151** | 0.9750 | 0.3274 | 0.5686 | 0.7962 | 0.6516 |
| | **1%~10%** | **54011** | 0.9542 | 0.6009 | 0.8370 | 0.7961 | 0.8156 |
| | **10%~20%** | **19838** | 0.8904 | 0.6467 | 0.9057 | 0.8008 | 0.8456 |
| | **20%~30%** | **15717** | 0.8510 | 0.6074 | 0.8996 | 0.8068 | 0.8421 |
| | **30%~40%** | **13401** | 0.8270 | 0.5652 | 0.8828 | 0.8120 | 0.8320 |
| | **40%~50%** | **12757** | 0.8177 | 0.5407 | 0.8731 | 0.8148 | 0.8250 |
| | **Overall** | **188399** | 0.9296 | 0.6537 | 0.8955 | 0.8066 | 0.8469 |
| 50% Missing | **MAF** | **#SNPs** | **Accuracy** | **$r^2$** | **Precision** | **Recall** | **$F_1$** |
| | **0.1%~0.5%** | **52524** | 0.9783 | 0.0965 | 0.3948 | 0.6481 | 0.4610 |
| | **0.5%~1%** | **20151** | 0.9721 | 0.1842 | 0.5218 | 0.6240 | 0.5646 |
| | **1%~10%** | **54011** | 0.9320 | 0.3842 | 0.8127 | 0.6348 | 0.7052 |
| | **10%~20%** | **19838** | 0.8096 | 0.4033 | 0.8857 | 0.6400 | 0.7191 |
| | **20%~30%** | **15717** | 0.7286 | 0.3563 | 0.8688 | 0.6429 | 0.7012 |
| | **30%~40%** | **13401** | 0.6774 | 0.3160 | 0.8462 | 0.6478 | 0.6819 |
| | **40%~50%** | **12757** | 0.6550 | 0.2937 | 0.8345 | 0.6492 | 0.6695 |
| | **Overall** | **188399** | 0.8825 | 0.4200 | 0.8759 | 0.6430 | 0.7272 |

**Table S18. BiU-Net's performance on the Caucasian cohort of LOS dataset**

| | MAF | #SNPs | Accuracy | $r^2$ | Precision | Recall | $F_1$ |
|---|---|---|---|---|---|---|---|
| 5% Missing | **0.1%~0.5%** | **52524** | 1.0000 | 0.9970 | 0.9983 | 0.9972 | 0.9977 |
| | **0.5%~1%** | **20151** | 0.9998 | 0.9946 | 0.9978 | 0.9938 | 0.9958 |
| | **1%~10%** | **54011** | 0.9983 | 0.9907 | 0.9964 | 0.9910 | 0.9937 |
| | **10%~20%** | **19838** | 0.9978 | 0.9953 | 0.9971 | 0.9966 | 0.9969 |
| | **20%~30%** | **15717** | 0.9977 | 0.9957 | 0.9975 | 0.9975 | 0.9975 |
| | **30%~40%** | **13401** | 0.9977 | 0.9954 | 0.9976 | 0.9976 | 0.9976 |
| | **40%~50%** | **12757** | 0.9977 | 0.9953 | 0.9977 | 0.9977 | 0.9977 |
| | **Overall** | **188399** | 0.9987 | 0.9962 | 0.9976 | 0.9967 | 0.9972 |
| 15% Missing | **MAF** | **#SNPs** | **Accuracy** | **$r^2$** | **Precision** | **Recall** | **$F_1$** |
| | **0.1%~0.5%** | **52524** | 0.9995 | 0.9674 | 0.9962 | 0.9580 | 0.9764 |
| | **0.5%~1%** | **20151** | 0.9982 | 0.9535 | 0.9936 | 0.9354 | 0.9627 |
| | **1%~10%** | **54011** | 0.9950 | 0.9721 | 0.9896 | 0.9726 | 0.9809 |
| | **10%~20%** | **19838** | 0.9931 | 0.9855 | 0.9913 | 0.9895 | 0.9904 |
| | **20%~30%** | **15717** | 0.9929 | 0.9868 | 0.9924 | 0.9922 | 0.9923 |
| | **30%~40%** | **13401** | 0.9929 | 0.9861 | 0.9928 | 0.9928 | 0.9928 |
| | **40%~50%** | **12757** | 0.9929 | 0.9855 | 0.9929 | 0.9929 | 0.9929 |
| | **Overall** | **188399** | 0.9959 | 0.9876 | 0.9927 | 0.9892 | 0.9910 |
| 25% Missing | **MAF** | **#SNPs** | **Accuracy** | **$r^2$** | **Precision** | **Recall** | **$F_1$** |
| | **0.1%~0.5%** | **52524** | 0.9991 | 0.9392 | 0.9940 | 0.9193 | 0.9536 |
| | **0.5%~1%** | **20151** | 0.9972 | 0.9267 | 0.9897 | 0.8979 | 0.9390 |
| | **1%~10%** | **54011** | 0.9916 | 0.9529 | 0.9827 | 0.9531 | 0.9674 |
| | **10%~20%** | **19838** | 0.9879 | 0.9745 | 0.9850 | 0.9813 | 0.9832 |
| | **20%~30%** | **15717** | 0.9874 | 0.9766 | 0.9865 | 0.9860 | 0.9863 |
| | **30%~40%** | **13401** | 0.9873 | 0.9754 | 0.9872 | 0.9872 | 0.9872 |
| | **40%~50%** | **12757** | 0.9875 | 0.9746 | 0.9875 | 0.9875 | 0.9875 |
| | **Overall** | **188399** | 0.9930 | 0.9786 | 0.9875 | 0.9811 | 0.9843 |
| 50% Missing | **MAF** | **#SNPs** | **Accuracy** | **$r^2$** | **Precision** | **Recall** | **$F_1$** |
| | **0.1%~0.5%** | **52524** | 0.9976 | 0.8366 | 0.9872 | 0.7786 | 0.8548 |
| | **0.5%~1%** | **20151** | 0.9937 | 0.8384 | 0.9782 | 0.7700 | 0.8451 |
| | **1%~10%** | **54011** | 0.9802 | 0.8893 | 0.9629 | 0.8843 | 0.9196 |
| | **10%~20%** | **19838** | 0.9675 | 0.9315 | 0.9630 | 0.9467 | 0.9546 |
| | **20%~30%** | **15717** | 0.9650 | 0.9344 | 0.9644 | 0.9596 | 0.9619 |
| | **30%~40%** | **13401** | 0.9647 | 0.9313 | 0.9649 | 0.9638 | 0.9644 |
| | **40%~50%** | **12757** | 0.9650 | 0.9288 | 0.9650 | 0.9649 | 0.9650 |
| | **Overall** | **188399** | 0.9818 | 0.9441 | 0.9694 | 0.9487 | 0.9587 |

**Table S19. STICI's performance on the Caucasian cohort of LOS dataset**

| | MAF | #SNPs | Accuracy | $r^2$ | Precision | Recall | $F_1$ |
|---|---|---|---|---|---|---|---|
| 5% Missing | **0.1%~0.5%** | **52524** | 0.9968 | 0.7786 | 0.8809 | 0.7701 | 0.8129 |
| | **0.5%~1%** | **20151** | 0.9945 | 0.8566 | 0.9215 | 0.8313 | 0.8697 |
| | **1%~10%** | **54011** | 0.9891 | 0.9380 | 0.9649 | 0.9406 | 0.9522 |
| | **10%~20%** | **19838** | 0.9835 | 0.9642 | 0.9764 | 0.9737 | 0.9750 |
| | **20%~30%** | **15717** | 0.9793 | 0.9610 | 0.9776 | 0.9757 | 0.9766 |
| | **30%~40%** | **13401** | 0.9783 | 0.9591 | 0.9780 | 0.9775 | 0.9777 |
| | **40%~50%** | **12757** | 0.9774 | 0.9545 | 0.9775 | 0.9773 | 0.9774 |
| | **Overall** | **188399** | 0.9889 | 0.9657 | 0.9767 | 0.9695 | 0.9730 |
| 15% Missing | **MAF** | **#SNPs** | **Accuracy** | **$r^2$** | **Precision** | **Recall** | **$F_1$** |
| | **0.1%~0.5%** | **52524** | 0.9974 | 0.8188 | 0.9303 | 0.7869 | 0.8427 |
| | **0.5%~1%** | **20151** | 0.9950 | 0.8691 | 0.9482 | 0.8325 | 0.8810 |
| | **1%~10%** | **54011** | 0.9899 | 0.9422 | 0.9730 | 0.9443 | 0.9581 |
| | **10%~20%** | **19838** | 0.9852 | 0.9674 | 0.9807 | 0.9768 | 0.9787 |
| | **20%~30%** | **15717** | 0.9827 | 0.9663 | 0.9818 | 0.9804 | 0.9811 |
| | **30%~40%** | **13401** | 0.9821 | 0.9643 | 0.9820 | 0.9817 | 0.9819 |
| | **40%~50%** | **12757** | 0.9820 | 0.9615 | 0.9820 | 0.9820 | 0.9820 |
| | **Overall** | **188399** | 0.9903 | 0.9696 | 0.9819 | 0.9738 | 0.9778 |
| 25% Missing | **MAF** | **#SNPs** | **Accuracy** | **$r^2$** | **Precision** | **Recall** | **$F_1$** |
| | **0.1%~0.5%** | **52524** | 0.9975 | 0.8321 | 0.9515 | 0.7876 | 0.8506 |
| | **0.5%~1%** | **20151** | 0.9950 | 0.8721 | 0.9600 | 0.8291 | 0.8832 |
| | **1%~10%** | **54011** | 0.9890 | 0.9374 | 0.9747 | 0.9384 | 0.9557 |
| | **10%~20%** | **19838** | 0.9840 | 0.9647 | 0.9804 | 0.9750 | 0.9776 |
| | **20%~30%** | **15717** | 0.9822 | 0.9651 | 0.9815 | 0.9800 | 0.9807 |
| | **30%~40%** | **13401** | 0.9817 | 0.9631 | 0.9817 | 0.9815 | 0.9816 |
| | **40%~50%** | **12757** | 0.9818 | 0.9610 | 0.9818 | 0.9818 | 0.9818 |
| | **Overall** | **188399** | 0.9899 | 0.9683 | 0.9823 | 0.9727 | 0.9774 |
| 50% Missing | **MAF** | **#SNPs** | **Accuracy** | **$r^2$** | **Precision** | **Recall** | **$F_1$** |
| | **0.1%~0.5%** | **52524** | 0.9970 | 0.8000 | 0.9717 | 0.7285 | 0.8082 |
| | **0.5%~1%** | **20151** | 0.9936 | 0.8343 | 0.9695 | 0.7656 | 0.8393 |
| | **1%~10%** | **54011** | 0.9844 | 0.9114 | 0.9710 | 0.9095 | 0.9380 |
| | **10%~20%** | **19838** | 0.9775 | 0.9510 | 0.9736 | 0.9650 | 0.9692 |
| | **20%~30%** | **15717** | 0.9764 | 0.9545 | 0.9752 | 0.9740 | 0.9746 |
| | **30%~40%** | **13401** | 0.9758 | 0.9518 | 0.9756 | 0.9758 | 0.9757 |
| | **40%~50%** | **12757** | 0.9763 | 0.9507 | 0.9762 | 0.9764 | 0.9763 |
| | **Overall** | **188399** | 0.9863 | 0.9575 | 0.9771 | 0.9628 | 0.9698 |

**Table S20. GenoBERT's performance on the Caucasian cohort of LOS dataset**

| | MAF | #SNPs | Accuracy | $r^2$ | Precision | Recall | $F_1$ |
|---|---|---|---|---|---|---|---|
| 5% Missing | 0.1%~0.5% | 52524 | 0.9997 | 0.9627 | 0.9993 | 0.9718 | 0.9853 |
| | 0.5%~1% | 20151 | 0.9991 | 0.9584 | 0.9987 | 0.9692 | 0.9836 |
| | 1%~10% | 54011 | 0.9977 | 0.9810 | 0.9980 | 0.9854 | 0.9917 |
| | 10%~20% | 19838 | 0.9978 | 0.9929 | 0.9982 | 0.9961 | 0.9971 |
| | 20%~30% | 15717 | 0.9977 | 0.9933 | 0.9981 | 0.9972 | 0.9976 |
| | 30%~40% | 13401 | 0.9977 | 0.9935 | 0.9979 | 0.9976 | 0.9978 |
| | 40%~50% | 12757 | 0.9975 | 0.9926 | 0.9976 | 0.9975 | 0.9976 |
| | Overall | 188399 | 0.9984 | 0.9926 | 0.9986 | 0.9951 | 0.9968 |
| 15% Missing | MAF | #SNPs | Accuracy | $r^2$ | Precision | Recall | $F_1$ |
| | 0.1%~0.5% | 52524 | 0.9995 | 0.9541 | 0.9982 | 0.9531 | 0.9748 |
| | 0.5%~1% | 20151 | 0.9984 | 0.9469 | 0.9967 | 0.9400 | 0.9669 |
| | 1%~10% | 54011 | 0.9966 | 0.9772 | 0.9952 | 0.9800 | 0.9875 |
| | 10%~20% | 19838 | 0.9961 | 0.9902 | 0.9957 | 0.9940 | 0.9949 |
| | 20%~30% | 15717 | 0.9960 | 0.9909 | 0.9960 | 0.9956 | 0.9958 |
| | 30%~40% | 13401 | 0.9959 | 0.9906 | 0.9960 | 0.9959 | 0.9959 |
| | 40%~50% | 12757 | 0.9958 | 0.9897 | 0.9959 | 0.9958 | 0.9958 |
| | Overall | 188399 | 0.9974 | 0.9904 | 0.9964 | 0.9927 | 0.9946 |
| 25% Missing | MAF | #SNPs | Accuracy | $r^2$ | Precision | Recall | $F_1$ |
| | 0.1%~0.5% | 52524 | 0.9992 | 0.9386 | 0.9964 | 0.9272 | 0.9594 |
| | 0.5%~1% | 20151 | 0.9977 | 0.9342 | 0.9945 | 0.9166 | 0.9525 |
| | 1%~10% | 54011 | 0.9949 | 0.9688 | 0.9920 | 0.9708 | 0.9812 |
| | 10%~20% | 19838 | 0.9939 | 0.9856 | 0.9929 | 0.9908 | 0.9919 |
| | 20%~30% | 15717 | 0.9937 | 0.9867 | 0.9935 | 0.9932 | 0.9933 |
| | 30%~40% | 13401 | 0.9934 | 0.9857 | 0.9934 | 0.9935 | 0.9935 |
| | 40%~50% | 12757 | 0.9934 | 0.9848 | 0.9934 | 0.9934 | 0.9934 |
| | Overall | 188399 | 0.9960 | 0.9864 | 0.9939 | 0.9892 | 0.9915 |
| 50% Missing | MAF | #SNPs | Accuracy | $r^2$ | Precision | Recall | $F_1$ |
| | 0.1%~0.5% | 52524 | 0.9977 | 0.8191 | 0.9731 | 0.7971 | 0.8664 |
| | 0.5%~1% | 20151 | 0.9946 | 0.8442 | 0.9788 | 0.8086 | 0.8768 |
| | 1%~10% | 54011 | 0.9878 | 0.9246 | 0.9808 | 0.9288 | 0.9534 |
| | 10%~20% | 19838 | 0.9836 | 0.9606 | 0.9821 | 0.9742 | 0.9781 |
| | 20%~30% | 15717 | 0.9824 | 0.9621 | 0.9826 | 0.9804 | 0.9815 |
| | 30%~40% | 13401 | 0.9816 | 0.9590 | 0.9820 | 0.9815 | 0.9817 |
| | 40%~50% | 12757 | 0.9815 | 0.9564 | 0.9816 | 0.9815 | 0.9816 |
| | Overall | 188399 | 0.9895 | 0.9638 | 0.9845 | 0.9710 | 0.9777 |

## Models' performance on 1KGP/mixed populations cohort

**Table S21. Beagle's performance on the mixed populations of 1KGP dataset**

| | MAF | #SNPs | Accuracy | $r^2$ | Precision | Recall | $F_1$ |
|---|---|---|---|---|---|---|---|
| **5% Missing** | **0.1%~0.5%** | **62625** | 0.9961 | 0.8989 | 0.7477 | 0.7485 | 0.7480 |
| | **0.5%~1%** | **41501** | 0.9927 | 0.8981 | 0.7489 | 0.7494 | 0.7491 |
| | **1%~10%** | **90533** | 0.9703 | 0.9122 | 0.7499 | 0.7497 | 0.7498 |
| | **10%~20%** | **17410** | 0.8831 | 0.9095 | 0.7501 | 0.7500 | 0.7500 |
| | **20%~30%** | **11479** | 0.8238 | 0.8758 | 0.7484 | 0.7484 | 0.7484 |
| | **30%~40%** | **9555** | 0.7859 | 0.8506 | 0.7482 | 0.7482 | 0.7482 |
| | **40%~50%** | **8010** | 0.7666 | 0.8305 | 0.7484 | 0.7485 | 0.7485 |
| | **Overall** | **241113** | 0.9535 | 0.9212 | 0.7491 | 0.7491 | 0.7491 |
| **15% Missing** | **MAF** | **#SNPs** | **Accuracy** | **$r^2$** | **Precision** | **Recall** | **$F_1$** |
| | **0.1%~0.5%** | **62625** | 0.9960 | 0.8925 | 0.7454 | 0.7457 | 0.7455 |
| | **0.5%~1%** | **41501** | 0.9924 | 0.8890 | 0.7444 | 0.7430 | 0.7437 |
| | **1%~10%** | **90533** | 0.9699 | 0.9085 | 0.7484 | 0.7474 | 0.7479 |
| | **10%~20%** | **17410** | 0.8828 | 0.9079 | 0.7500 | 0.7498 | 0.7499 |
| | **20%~30%** | **11479** | 0.8244 | 0.8746 | 0.7493 | 0.7495 | 0.7494 |
| | **30%~40%** | **9555** | 0.7860 | 0.8491 | 0.7483 | 0.7485 | 0.7484 |
| | **40%~50%** | **8010** | 0.7671 | 0.8292 | 0.7488 | 0.7491 | 0.7489 |
| | **Overall** | **241113** | 0.9533 | 0.9194 | 0.7490 | 0.7487 | 0.7488 |
| **25% Missing** | **MAF** | **#SNPs** | **Accuracy** | **$r^2$** | **Precision** | **Recall** | **$F_1$** |
| | **0.1%~0.5%** | **62625** | 0.9958 | 0.8698 | 0.7411 | 0.7318 | 0.7363 |
| | **0.5%~1%** | **41501** | 0.9922 | 0.8778 | 0.7414 | 0.7367 | 0.7390 |
| | **1%~10%** | **90533** | 0.9691 | 0.9033 | 0.7447 | 0.7425 | 0.7436 |
| | **10%~20%** | **17410** | 0.8815 | 0.9052 | 0.7479 | 0.7474 | 0.7477 |
| | **20%~30%** | **11479** | 0.8221 | 0.8712 | 0.7462 | 0.7464 | 0.7463 |
| | **30%~40%** | **9555** | 0.7849 | 0.8463 | 0.7468 | 0.7473 | 0.7470 |
| | **40%~50%** | **8010** | 0.7663 | 0.8264 | 0.7476 | 0.7483 | 0.7480 |
| | **Overall** | **241113** | 0.9527 | 0.9163 | 0.7468 | 0.7458 | 0.7463 |
| **50% Missing** | **MAF** | **#SNPs** | **Accuracy** | **$r^2$** | **Precision** | **Recall** | **$F_1$** |
| | **0.1%~0.5%** | **62625** | 0.9951 | 0.8066 | 0.7336 | 0.6951 | 0.7124 |
| | **0.5%~1%** | **41501** | 0.9915 | 0.8428 | 0.7354 | 0.7172 | 0.7258 |
| | **1%~10%** | **90533** | 0.9674 | 0.8861 | 0.7387 | 0.7304 | 0.7344 |
| | **10%~20%** | **17410** | 0.8769 | 0.8936 | 0.7412 | 0.7390 | 0.7400 |
| | **20%~30%** | **11479** | 0.8166 | 0.8581 | 0.7388 | 0.7394 | 0.7390 |
| | **30%~40%** | **9555** | 0.7779 | 0.8322 | 0.7374 | 0.7396 | 0.7384 |
| | **40%~50%** | **8010** | 0.7593 | 0.8115 | 0.7381 | 0.7411 | 0.7396 |
| | **Overall** | **241113** | 0.9506 | 0.9048 | 0.7407 | 0.7365 | 0.7385 |

**Table S22. SCDA's performance on the mixed populations of 1KGP dataset**

| | MAF | #SNPs | Accuracy | $r^2$ | Precision | Recall | $F_1$ |
|---|---|---|---|---|---|---|---|
| 5% Missing | **0.1%~0.5%** | **62625** | 0.9945 | 0.6350 | 0.8414 | 0.7521 | 0.7891 |
| | **0.5%~1%** | **41501** | 0.9914 | 0.6813 | 0.8868 | 0.7726 | 0.8200 |
| | **1%~10%** | **90533** | 0.9839 | 0.8691 | 0.9401 | 0.9045 | 0.9217 |
| | **10%~20%** | **17410** | 0.9748 | 0.9465 | 0.9591 | 0.9709 | 0.9649 |
| | **20%~30%** | **11479** | 0.9710 | 0.9428 | 0.9642 | 0.9736 | 0.9687 |
| | **30%~40%** | **9555** | 0.9704 | 0.9411 | 0.9681 | 0.9730 | 0.9703 |
| | **40%~50%** | **8010** | 0.9691 | 0.9352 | 0.9684 | 0.9702 | 0.9690 |
| | **Overall** | **241113** | 0.9857 | 0.9332 | 0.9567 | 0.9523 | 0.9545 |
| 15% Missing | **MAF** | **#SNPs** | **Accuracy** | **$r^2$** | **Precision** | **Recall** | **$F_1$** |
| | **0.1%~0.5%** | **62625** | 0.9944 | 0.6246 | 0.8287 | 0.7590 | 0.7886 |
| | **0.5%~1%** | **41501** | 0.9907 | 0.6525 | 0.8723 | 0.7557 | 0.8034 |
| | **1%~10%** | **90533** | 0.9813 | 0.8458 | 0.9368 | 0.8815 | 0.9075 |
| | **10%~20%** | **17410** | 0.9692 | 0.9322 | 0.9566 | 0.9568 | 0.9567 |
| | **20%~30%** | **11479** | 0.9650 | 0.9285 | 0.9603 | 0.9635 | 0.9619 |
| | **30%~40%** | **9555** | 0.9644 | 0.9270 | 0.9630 | 0.9654 | 0.9641 |
| | **40%~50%** | **8010** | 0.9629 | 0.9196 | 0.9623 | 0.9635 | 0.9628 |
| | **Overall** | **241113** | 0.9834 | 0.9207 | 0.9545 | 0.9388 | 0.9465 |
| 25% Missing | **MAF** | **#SNPs** | **Accuracy** | **$r^2$** | **Precision** | **Recall** | **$F_1$** |
| | **0.1%~0.5%** | **62625** | 0.9934 | 0.5542 | 0.7875 | 0.7119 | 0.7421 |
| | **0.5%~1%** | **41501** | 0.9896 | 0.6051 | 0.8470 | 0.7262 | 0.7739 |
| | **1%~10%** | **90533** | 0.9773 | 0.8097 | 0.9302 | 0.8479 | 0.8852 |
| | **10%~20%** | **17410** | 0.9586 | 0.9043 | 0.9511 | 0.9327 | 0.9417 |
| | **20%~30%** | **11479** | 0.9529 | 0.8990 | 0.9524 | 0.9449 | 0.9486 |
| | **30%~40%** | **9555** | 0.9520 | 0.8958 | 0.9529 | 0.9505 | 0.9517 |
| | **40%~50%** | **8010** | 0.9499 | 0.8856 | 0.9503 | 0.9497 | 0.9500 |
| | **Overall** | **241113** | 0.9792 | 0.8967 | 0.9499 | 0.9154 | 0.9320 |
| 50% Missing | **MAF** | **#SNPs** | **Accuracy** | **$r^2$** | **Precision** | **Recall** | **$F_1$** |
| | **0.1%~0.5%** | **62625** | 0.9884 | 0.2546 | 0.5496 | 0.5863 | 0.5608 |
| | **0.5%~1%** | **41501** | 0.9835 | 0.3532 | 0.6662 | 0.6244 | 0.6369 |
| | **1%~10%** | **90533** | 0.9578 | 0.5985 | 0.8621 | 0.7027 | 0.7657 |
| | **10%~20%** | **17410** | 0.8819 | 0.6853 | 0.9080 | 0.7761 | 0.8293 |
| | **20%~30%** | **11479** | 0.8487 | 0.6417 | 0.8965 | 0.7969 | 0.8350 |
| | **30%~40%** | **9555** | 0.8298 | 0.6093 | 0.8800 | 0.8111 | 0.8324 |
| | **40%~50%** | **8010** | 0.8246 | 0.5888 | 0.8704 | 0.8188 | 0.8301 |
| | **Overall** | **241113** | 0.9500 | 0.6970 | 0.9080 | 0.7705 | 0.8292 |

**Table S23. BiU-Net's performance on the mixed populations of 1KGP dataset**

| | MAF | #SNPs | Accuracy | $r^2$ | Precision | Recall | $F_1$ |
|---|---|---|---|---|---|---|---|
| 5% Missing | **0.1%~0.5%** | **62625** | 1.0000 | 0.9966 | 0.9984 | 0.9984 | 0.9984 |
| | **0.5%~1%** | **41501** | 0.9999 | 0.9965 | 0.9975 | 0.9989 | 0.9982 |
| | **1%~10%** | **90533** | 0.9989 | 0.9915 | 0.9964 | 0.9928 | 0.9946 |
| | **10%~20%** | **17410** | 0.9977 | 0.9952 | 0.9967 | 0.9965 | 0.9966 |
| | **20%~30%** | **11479** | 0.9973 | 0.9947 | 0.9969 | 0.9970 | 0.9969 |
| | **30%~40%** | **9555** | 0.9974 | 0.9948 | 0.9973 | 0.9973 | 0.9973 |
| | **40%~50%** | **8010** | 0.9972 | 0.9943 | 0.9972 | 0.9972 | 0.9972 |
| | **Overall** | **241113** | 0.9991 | 0.9957 | 0.9972 | 0.9964 | 0.9968 |
| 15% Missing | **MAF** | **#SNPs** | **Accuracy** | **$r^2$** | **Precision** | **Recall** | **$F_1$** |
| | **0.1%~0.5%** | **62625** | 0.9995 | 0.9655 | 0.9952 | 0.9703 | 0.9824 |
| | **0.5%~1%** | **41501** | 0.9982 | 0.9344 | 0.9917 | 0.9437 | 0.9665 |
| | **1%~10%** | **90533** | 0.9961 | 0.9698 | 0.9891 | 0.9730 | 0.9810 |
| | **10%~20%** | **17410** | 0.9927 | 0.9849 | 0.9899 | 0.9889 | 0.9894 |
| | **20%~30%** | **11479** | 0.9915 | 0.9836 | 0.9904 | 0.9904 | 0.9904 |
| | **30%~40%** | **9555** | 0.9915 | 0.9836 | 0.9912 | 0.9913 | 0.9913 |
| | **40%~50%** | **8010** | 0.9912 | 0.9821 | 0.9911 | 0.9912 | 0.9911 |
| | **Overall** | **241113** | 0.9965 | 0.9844 | 0.9911 | 0.9856 | 0.9883 |
| 25% Missing | **MAF** | **#SNPs** | **Accuracy** | **$r^2$** | **Precision** | **Recall** | **$F_1$** |
| | **0.1%~0.5%** | **62625** | 0.9977 | 0.8495 | 0.9903 | 0.8582 | 0.9142 |
| | **0.5%~1%** | **41501** | 0.9962 | 0.8619 | 0.9845 | 0.8778 | 0.9248 |
| | **1%~10%** | **90533** | 0.9928 | 0.9443 | 0.9809 | 0.9492 | 0.9645 |
| | **10%~20%** | **17410** | 0.9866 | 0.9720 | 0.9819 | 0.9792 | 0.9806 |
| | **20%~30%** | **11479** | 0.9844 | 0.9699 | 0.9826 | 0.9822 | 0.9824 |
| | **30%~40%** | **9555** | 0.9844 | 0.9697 | 0.9839 | 0.9841 | 0.9840 |
| | **40%~50%** | **8010** | 0.9838 | 0.9669 | 0.9837 | 0.9838 | 0.9837 |
| | **Overall** | **241113** | 0.9932 | 0.9694 | 0.9840 | 0.9706 | 0.9772 |
| 50% Missing | **MAF** | **#SNPs** | **Accuracy** | **$r^2$** | **Precision** | **Recall** | **$F_1$** |
| | **0.1%~0.5%** | **62625** | 0.9949 | 0.6621 | 0.9731 | 0.6770 | 0.7594 |
| | **0.5%~1%** | **41501** | 0.9921 | 0.7132 | 0.9653 | 0.7406 | 0.8188 |
| | **1%~10%** | **90533** | 0.9827 | 0.8651 | 0.9576 | 0.8725 | 0.9108 |
| | **10%~20%** | **17410** | 0.9639 | 0.9236 | 0.9550 | 0.9403 | 0.9475 |
| | **20%~30%** | **11479** | 0.9568 | 0.9147 | 0.9539 | 0.9490 | 0.9514 |
| | **30%~40%** | **9555** | 0.9561 | 0.9127 | 0.9556 | 0.9546 | 0.9551 |
| | **40%~50%** | **8010** | 0.9546 | 0.9053 | 0.9545 | 0.9543 | 0.9544 |
| | **Overall** | **241113** | 0.9829 | 0.9218 | 0.9606 | 0.9238 | 0.9414 |

**Table S24. STICI's performance on the mixed populations of 1KGP dataset**

| | MAF | #SNPs | Accuracy | $r^2$ | Precision | Recall | $F_1$ |
|---|---|---|---|---|---|---|---|
| 5% Missing | **0.1%~0.5%** | **62625** | 0.9943 | 0.5955 | 0.8671 | 0.6642 | 0.7239 |
| | **0.5%~1%** | **41501** | 0.9922 | 0.6983 | 0.9037 | 0.7512 | 0.8072 |
| | **1%~10%** | **90533** | 0.9843 | 0.8594 | 0.9481 | 0.8785 | 0.9100 |
| | **10%~20%** | **17410** | 0.9678 | 0.9101 | 0.9603 | 0.9416 | 0.9506 |
| | **20%~30%** | **11479** | 0.9610 | 0.8988 | 0.9586 | 0.9499 | 0.9541 |
| | **30%~40%** | **9555** | 0.9560 | 0.8846 | 0.9574 | 0.9521 | 0.9545 |
| | **40%~50%** | **8010** | 0.9528 | 0.8657 | 0.9542 | 0.9519 | 0.9528 |
| | **Overall** | **241113** | 0.9838 | 0.9089 | 0.9596 | 0.9227 | 0.9402 |
| 15% Missing | **MAF** | **#SNPs** | **Accuracy** | **$r^2$** | **Precision** | **Recall** | **$F_1$** |
| | **0.1%~0.5%** | **62625** | 0.9954 | 0.6740 | 0.9120 | 0.7200 | 0.7846 |
| | **0.5%~1%** | **41501** | 0.9933 | 0.7427 | 0.9286 | 0.7850 | 0.8409 |
| | **1%~10%** | **90533** | 0.9869 | 0.8840 | 0.9597 | 0.8998 | 0.9276 |
| | **10%~20%** | **17410** | 0.9732 | 0.9283 | 0.9671 | 0.9530 | 0.9599 |
| | **20%~30%** | **11479** | 0.9674 | 0.9198 | 0.9657 | 0.9593 | 0.9624 |
| | **30%~40%** | **9555** | 0.9648 | 0.9119 | 0.9655 | 0.9623 | 0.9639 |
| | **40%~50%** | **8010** | 0.9623 | 0.8991 | 0.9630 | 0.9618 | 0.9623 |
| | **Overall** | **241113** | 0.9866 | 0.9273 | 0.9676 | 0.9374 | 0.9519 |
| 25% Missing | **MAF** | **#SNPs** | **Accuracy** | **$r^2$** | **Precision** | **Recall** | **$F_1$** |
| | **0.1%~0.5%** | **62625** | 0.9952 | 0.6693 | 0.9309 | 0.7042 | 0.7766 |
| | **0.5%~1%** | **41501** | 0.9932 | 0.7401 | 0.9406 | 0.7780 | 0.8406 |
| | **1%~10%** | **90533** | 0.9865 | 0.8837 | 0.9633 | 0.8971 | 0.9277 |
| | **10%~20%** | **17410** | 0.9719 | 0.9287 | 0.9668 | 0.9512 | 0.9588 |
| | **20%~30%** | **11479** | 0.9661 | 0.9202 | 0.9650 | 0.9581 | 0.9615 |
| | **30%~40%** | **9555** | 0.9641 | 0.9142 | 0.9649 | 0.9619 | 0.9633 |
| | **40%~50%** | **8010** | 0.9620 | 0.9044 | 0.9626 | 0.9615 | 0.9620 |
| | **Overall** | **241113** | 0.9862 | 0.9279 | 0.9687 | 0.9359 | 0.9517 |
| 50% Missing | **MAF** | **#SNPs** | **Accuracy** | **$r^2$** | **Precision** | **Recall** | **$F_1$** |
| | **0.1%~0.5%** | **62625** | 0.9944 | 0.6203 | 0.9553 | 0.6398 | 0.7165 |
| | **0.5%~1%** | **41501** | 0.9917 | 0.6911 | 0.9551 | 0.7225 | 0.8007 |
| | **1%~10%** | **90533** | 0.9820 | 0.8515 | 0.9604 | 0.8608 | 0.9048 |
| | **10%~20%** | **17410** | 0.9615 | 0.9111 | 0.9562 | 0.9328 | 0.9440 |
| | **20%~30%** | **11479** | 0.9533 | 0.8996 | 0.9521 | 0.9430 | 0.9474 |
| | **30%~40%** | **9555** | 0.9519 | 0.8960 | 0.9523 | 0.9496 | 0.9509 |
| | **40%~50%** | **8010** | 0.9496 | 0.8866 | 0.9498 | 0.9491 | 0.9494 |
| | **Overall** | **241113** | 0.9818 | 0.9108 | 0.9607 | 0.9152 | 0.9368 |

**Table S25. GenoBERT's performance on the mixed populations of 1KGP dataset**

| | MAF | #SNPs | Accuracy | $r^2$ | Precision | Recall | $F_1$ |
|---|---|---|---|---|---|---|---|
| **5% Missing** | **0.1%~0.5%** | **62625** | 1.0000 | 0.9947 | 0.9992 | 0.9975 | 0.9984 |
| | **0.5%~1%** | **41501** | 0.9999 | 0.9946 | 0.9990 | 0.9975 | 0.9983 |
| | **1%~10%** | **90533** | 0.9993 | 0.9935 | 0.9988 | 0.9945 | 0.9967 |
| | **10%~20%** | **17410** | 0.9985 | 0.9961 | 0.9986 | 0.9973 | 0.9980 |
| | **20%~30%** | **11479** | 0.9983 | 0.9958 | 0.9985 | 0.9978 | 0.9981 |
| | **30%~40%** | **9555** | 0.9983 | 0.9961 | 0.9985 | 0.9982 | 0.9983 |
| | **40%~50%** | **8010** | 0.9983 | 0.9958 | 0.9983 | 0.9983 | 0.9983 |
| | **Overall** | **241113** | 0.9994 | 0.9966 | 0.9989 | 0.9972 | 0.9981 |
| **15% Missing** | **MAF** | **#SNPs** | **Accuracy** | **$r^2$** | **Precision** | **Recall** | **$F_1$** |
| | **0.1%~0.5%** | **62625** | 0.9996 | 0.9622 | 0.9968 | 0.9715 | 0.9840 |
| | **0.5%~1%** | **41501** | 0.9987 | 0.9456 | 0.9960 | 0.9564 | 0.9756 |
| | **1%~10%** | **90533** | 0.9976 | 0.9779 | 0.9961 | 0.9811 | 0.9885 |
| | **10%~20%** | **17410** | 0.9955 | 0.9884 | 0.9956 | 0.9921 | 0.9939 |
| | **20%~30%** | **11479** | 0.9948 | 0.9875 | 0.9953 | 0.9935 | 0.9944 |
| | **30%~40%** | **9555** | 0.9949 | 0.9882 | 0.9952 | 0.9946 | 0.9949 |
| | **40%~50%** | **8010** | 0.9948 | 0.9873 | 0.9949 | 0.9947 | 0.9948 |
| | **Overall** | **241113** | 0.9978 | 0.9881 | 0.9964 | 0.9897 | 0.9930 |
| **25% Missing** | **MAF** | **#SNPs** | **Accuracy** | **$r^2$** | **Precision** | **Recall** | **$F_1$** |
| | **0.1%~0.5%** | **62625** | 0.9980 | 0.8508 | 0.9913 | 0.8705 | 0.9237 |
| | **0.5%~1%** | **41501** | 0.9973 | 0.8860 | 0.9911 | 0.9067 | 0.9457 |
| | **1%~10%** | **90533** | 0.9955 | 0.9593 | 0.9922 | 0.9654 | 0.9785 |
| | **10%~20%** | **17410** | 0.9917 | 0.9789 | 0.9916 | 0.9859 | 0.9887 |
| | **20%~30%** | **11479** | 0.9904 | 0.9770 | 0.9911 | 0.9882 | 0.9897 |
| | **30%~40%** | **9555** | 0.9906 | 0.9781 | 0.9911 | 0.9901 | 0.9906 |
| | **40%~50%** | **8010** | 0.9903 | 0.9763 | 0.9906 | 0.9902 | 0.9904 |
| | **Overall** | **241113** | 0.9956 | 0.9765 | 0.9929 | 0.9793 | 0.9860 |
| **50% Missing** | **MAF** | **#SNPs** | **Accuracy** | **$r^2$** | **Precision** | **Recall** | **$F_1$** |
| | **0.1%~0.5%** | **62625** | 0.9947 | 0.5882 | 0.9172 | 0.6825 | 0.7599 |
| | **0.5%~1%** | **41501** | 0.9931 | 0.7018 | 0.9489 | 0.7786 | 0.8479 |
| | **1%~10%** | **90533** | 0.9870 | 0.8771 | 0.9730 | 0.9033 | 0.9359 |
| | **10%~20%** | **17410** | 0.9724 | 0.9244 | 0.9739 | 0.9523 | 0.9628 |
| | **20%~30%** | **11479** | 0.9673 | 0.9170 | 0.9713 | 0.9593 | 0.9651 |
| | **30%~40%** | **9555** | 0.9663 | 0.9148 | 0.9693 | 0.9642 | 0.9666 |
| | **40%~50%** | **8010** | 0.9653 | 0.9082 | 0.9669 | 0.9647 | 0.9656 |
| | **Overall** | **241113** | 0.9865 | 0.9233 | 0.9768 | 0.9384 | 0.9569 |

## Models' performance on 1KGP/EUR cohort

**Table S26. STICI's performance on the European cohort of 1KGP dataset**

| | MAF | #SNPs | Accuracy | $r^2$ | Precision | Recall | $F_1$ |
|---|---|---|---|---|---|---|---|
| 5% Missing | 0.1%~0.5% | 3664 | 0.9819 | 0.5288 | 0.8590 | 0.6348 | 0.6905 |
| | 0.5%~1% | 4519 | 0.9810 | 0.7666 | 0.8789 | 0.7080 | 0.7649 |
| | 1%~10% | 45930 | 0.9798 | 0.9050 | 0.9439 | 0.9028 | 0.9219 |
| | 10%~20% | 21922 | 0.9763 | 0.9500 | 0.9663 | 0.9579 | 0.9620 |
| | 20%~30% | 16560 | 0.9739 | 0.9528 | 0.9711 | 0.9684 | 0.9697 |
| | 30%~40% | 14516 | 0.9718 | 0.9470 | 0.9712 | 0.9703 | 0.9707 |
| | 40%~50% | 13466 | 0.9709 | 0.9442 | 0.9708 | 0.9707 | 0.9708 |
| | Overall | 120577 | 0.9765 | 0.9501 | 0.9672 | 0.9573 | 0.9621 |
| 15% Missing | MAF | #SNPs | Accuracy | $r^2$ | Precision | Recall | $F_1$ |
| | 0.1%~0.5% | 3664 | 0.9823 | 0.5354 | 0.8629 | 0.6435 | 0.7005 |
| | 0.5%~1% | 4519 | 0.9819 | 0.7772 | 0.8872 | 0.7224 | 0.7792 |
| | 1%~10% | 45930 | 0.9792 | 0.9021 | 0.9456 | 0.9007 | 0.9216 |
| | 10%~20% | 21922 | 0.9761 | 0.9487 | 0.9671 | 0.9583 | 0.9626 |
| | 20%~30% | 16560 | 0.9743 | 0.9531 | 0.9719 | 0.9694 | 0.9706 |
| | 30%~40% | 14516 | 0.9729 | 0.9492 | 0.9724 | 0.9717 | 0.9720 |
| | 40%~50% | 13466 | 0.9724 | 0.9468 | 0.9723 | 0.9723 | 0.9723 |
| | Overall | 120577 | 0.9766 | 0.9501 | 0.9684 | 0.9582 | 0.9632 |
| 25% Missing | MAF | #SNPs | Accuracy | $r^2$ | Precision | Recall | $F_1$ |
| | 0.1%~0.5% | 3664 | 0.9819 | 0.5205 | 0.8752 | 0.6314 | 0.6899 |
| | 0.5%~1% | 4519 | 0.9805 | 0.7596 | 0.8913 | 0.6938 | 0.7565 |
| | 1%~10% | 45930 | 0.9771 | 0.8912 | 0.9463 | 0.8898 | 0.9158 |
| | 10%~20% | 21922 | 0.9743 | 0.9439 | 0.9663 | 0.9560 | 0.9611 |
| | 20%~30% | 16560 | 0.9732 | 0.9507 | 0.9710 | 0.9687 | 0.9698 |
| | 30%~40% | 14516 | 0.9725 | 0.9479 | 0.9720 | 0.9715 | 0.9718 |
| | 40%~50% | 13466 | 0.9720 | 0.9449 | 0.9719 | 0.9719 | 0.9719 |
| | Overall | 120577 | 0.9752 | 0.9464 | 0.9679 | 0.9562 | 0.9619 |
| 50% Missing | MAF | #SNPs | Accuracy | $r^2$ | Precision | Recall | $F_1$ |
| | 0.1%~0.5% | 3664 | 0.9839 | 0.5700 | 0.9238 | 0.6638 | 0.7334 |
| | 0.5%~1% | 4519 | 0.9826 | 0.7847 | 0.9269 | 0.7187 | 0.7870 |
| | 1%~10% | 45930 | 0.9738 | 0.8739 | 0.9475 | 0.8720 | 0.9059 |
| | 10%~20% | 21922 | 0.9665 | 0.9251 | 0.9594 | 0.9437 | 0.9513 |
| | 20%~30% | 16560 | 0.9659 | 0.9367 | 0.9637 | 0.9613 | 0.9625 |
| | 30%~40% | 14516 | 0.9665 | 0.9354 | 0.9660 | 0.9659 | 0.9659 |
| | 40%~50% | 13466 | 0.9666 | 0.9324 | 0.9666 | 0.9666 | 0.9666 |
| | Overall | 120577 | 0.9703 | 0.9348 | 0.9631 | 0.9487 | 0.9557 |

**Table S27. GenoBERT's performance on the European cohort of 1KGP dataset**

| | MAF | #SNPs | Accuracy | $r^2$ | Precision | Recall | $F_1$ |
|---|---|---|---|---|---|---|---|
| 5% Missing | 0.1%~0.5% | 3664 | 0.9992 | 0.9780 | 0.9950 | 0.9854 | 0.9901 |
| | 0.5%~1% | 4519 | 0.9993 | 0.9865 | 0.9948 | 0.9929 | 0.9939 |
| | 1%~10% | 45930 | 0.9989 | 0.9938 | 0.9964 | 0.9964 | 0.9964 |
| | 10%~20% | 21922 | 0.9975 | 0.9938 | 0.9973 | 0.9957 | 0.9965 |
| | 20%~30% | 16560 | 0.9975 | 0.9946 | 0.9976 | 0.9970 | 0.9973 |
| | 30%~40% | 14516 | 0.9971 | 0.9936 | 0.9973 | 0.9970 | 0.9971 |
| | 40%~50% | 13466 | 0.9970 | 0.9929 | 0.9970 | 0.9970 | 0.9970 |
| | Overall | 120577 | 0.9980 | 0.9951 | 0.9976 | 0.9967 | 0.9972 |
| 15% Missing | MAF | #SNPs | Accuracy | $r^2$ | Precision | Recall | $F_1$ |
| | 0.1%~0.5% | 3664 | 0.9990 | 0.9729 | 0.9930 | 0.9820 | 0.9874 |
| | 0.5%~1% | 4519 | 0.9987 | 0.9765 | 0.9925 | 0.9840 | 0.9882 |
| | 1%~10% | 45930 | 0.9957 | 0.9775 | 0.9915 | 0.9813 | 0.9863 |
| | 10%~20% | 21922 | 0.9948 | 0.9875 | 0.9941 | 0.9916 | 0.9928 |
| | 20%~30% | 16560 | 0.9948 | 0.9890 | 0.9948 | 0.9941 | 0.9945 |
| | 30%~40% | 14516 | 0.9948 | 0.9885 | 0.9949 | 0.9946 | 0.9948 |
| | 40%~50% | 13466 | 0.9945 | 0.9872 | 0.9946 | 0.9945 | 0.9945 |
| | Overall | 120577 | 0.9954 | 0.9887 | 0.9947 | 0.9923 | 0.9935 |
| 25% Missing | MAF | #SNPs | Accuracy | $r^2$ | Precision | Recall | $F_1$ |
| | 0.1%~0.5% | 3664 | 0.9953 | 0.8704 | 0.9898 | 0.9020 | 0.9417 |
| | 0.5%~1% | 4519 | 0.9930 | 0.9034 | 0.9881 | 0.8867 | 0.9318 |
| | 1%~10% | 45930 | 0.9913 | 0.9555 | 0.9856 | 0.9591 | 0.9720 |
| | 10%~20% | 21922 | 0.9916 | 0.9799 | 0.9902 | 0.9866 | 0.9884 |
| | 20%~30% | 16560 | 0.9919 | 0.9835 | 0.9916 | 0.9910 | 0.9913 |
| | 30%~40% | 14516 | 0.9922 | 0.9832 | 0.9923 | 0.9921 | 0.9922 |
| | 40%~50% | 13466 | 0.9919 | 0.9818 | 0.9919 | 0.9919 | 0.9919 |
| | Overall | 120577 | 0.9918 | 0.9806 | 0.9910 | 0.9861 | 0.9885 |
| 50% Missing | MAF | #SNPs | Accuracy | $r^2$ | Precision | Recall | $F_1$ |
| | 0.1%~0.5% | 3664 | 0.9942 | 0.8381 | 0.9723 | 0.8894 | 0.9261 |
| | 0.5%~1% | 4519 | 0.9908 | 0.8695 | 0.9719 | 0.8585 | 0.9077 |
| | 1%~10% | 45930 | 0.9837 | 0.9174 | 0.9700 | 0.9246 | 0.9461 |
| | 10%~20% | 21922 | 0.9812 | 0.9559 | 0.9780 | 0.9699 | 0.9739 |
| | 20%~30% | 16560 | 0.9816 | 0.9636 | 0.9810 | 0.9793 | 0.9802 |
| | 30%~40% | 14516 | 0.9819 | 0.9622 | 0.9820 | 0.9816 | 0.9818 |
| | 40%~50% | 13466 | 0.9817 | 0.9598 | 0.9817 | 0.9817 | 0.9817 |
| | Overall | 120577 | 0.9831 | 0.9605 | 0.9803 | 0.9716 | 0.9759 |

## Models' performance on 1KGP/AFR cohort

**Table S28. STICI's performance on the African cohort of 1KGP dataset**

| | MAF | #SNPs | Accuracy | $r^2$ | Precision | Recall | $F_1$ |
|---|---|---|---|---|---|---|---|
| 5% Missing | **0.1%~0.5%** | **7140** | 0.9839 | 0.5652 | 0.8876 | 0.6572 | 0.7209 |
| | **0.5%~1%** | **10402** | 0.9850 | 0.8197 | 0.9142 | 0.7520 | 0.8118 |
| | **1%~10%** | **100338** | 0.9797 | 0.8896 | 0.9344 | 0.8869 | 0.9087 |
| | **10%~20%** | **34755** | 0.9640 | 0.9160 | 0.9441 | 0.9305 | 0.9369 |
| | **20%~30%** | **19414** | 0.9568 | 0.9174 | 0.9494 | 0.9450 | 0.9470 |
| | **30%~40%** | **14704** | 0.9522 | 0.9040 | 0.9506 | 0.9486 | 0.9494 |
| | **40%~50%** | **12851** | 0.9514 | 0.9007 | 0.9513 | 0.9509 | 0.9509 |
| | **Overall** | **199604** | 0.9713 | 0.9227 | 0.9475 | 0.9292 | 0.9378 |
| 15% Missing | **MAF** | **#SNPs** | **Accuracy** | **$r^2$** | **Precision** | **Recall** | **$F_1$** |
| | **0.1%~0.5%** | **7140** | 0.9849 | 0.5906 | 0.9011 | 0.6788 | 0.7454 |
| | **0.5%~1%** | **10402** | 0.9861 | 0.8313 | 0.9222 | 0.7710 | 0.8289 |
| | **1%~10%** | **100338** | 0.9778 | 0.8770 | 0.9367 | 0.8779 | 0.9048 |
| | **10%~20%** | **34755** | 0.9618 | 0.9090 | 0.9447 | 0.9279 | 0.9359 |
| | **20%~30%** | **19414** | 0.9541 | 0.9107 | 0.9481 | 0.9427 | 0.9453 |
| | **30%~40%** | **14704** | 0.9506 | 0.9003 | 0.9493 | 0.9474 | 0.9482 |
| | **40%~50%** | **12851** | 0.9498 | 0.8970 | 0.9497 | 0.9493 | 0.9494 |
| | **Overall** | **199604** | 0.9696 | 0.9168 | 0.9479 | 0.9264 | 0.9367 |
| 25% Missing | **MAF** | **#SNPs** | **Accuracy** | **$r^2$** | **Precision** | **Recall** | **$F_1$** |
| | **0.1%~0.5%** | **7140** | 0.9859 | 0.6161 | 0.9138 | 0.6982 | 0.7669 |
| | **0.5%~1%** | **10402** | 0.9868 | 0.8395 | 0.9291 | 0.7828 | 0.8401 |
| | **1%~10%** | **100338** | 0.9764 | 0.8679 | 0.9385 | 0.8716 | 0.9020 |
| | **10%~20%** | **34755** | 0.9580 | 0.8988 | 0.9432 | 0.9222 | 0.9323 |
| | **20%~30%** | **19414** | 0.9498 | 0.9017 | 0.9450 | 0.9381 | 0.9415 |
| | **30%~40%** | **14704** | 0.9466 | 0.8925 | 0.9456 | 0.9435 | 0.9445 |
| | **40%~50%** | **12851** | 0.9457 | 0.8885 | 0.9457 | 0.9453 | 0.9455 |
| | **Overall** | **199604** | 0.9673 | 0.9099 | 0.9468 | 0.9221 | 0.9339 |
| 50% Missing | **MAF** | **#SNPs** | **Accuracy** | **$r^2$** | **Precision** | **Recall** | **$F_1$** |
| | **0.1%~0.5%** | **7140** | 0.9877 | 0.6612 | 0.9487 | 0.7348 | 0.8090 |
| | **0.5%~1%** | **10402** | 0.9862 | 0.8301 | 0.9483 | 0.7694 | 0.8359 |
| | **1%~10%** | **100338** | 0.9691 | 0.8255 | 0.9385 | 0.8319 | 0.8777 |
| | **10%~20%** | **34755** | 0.9398 | 0.8540 | 0.9290 | 0.8901 | 0.9083 |
| | **20%~30%** | **19414** | 0.9292 | 0.8602 | 0.9259 | 0.9147 | 0.9201 |
| | **30%~40%** | **14704** | 0.9266 | 0.8544 | 0.9259 | 0.9236 | 0.9247 |
| | **40%~50%** | **12851** | 0.9251 | 0.8466 | 0.9250 | 0.9248 | 0.9249 |
| | **Overall** | **199604** | 0.9557 | 0.8770 | 0.9346 | 0.8965 | 0.9145 |

**Table S29. GenoBERT's performance on the African cohort of 1KGP dataset**

| | MAF | #SNPs | Accuracy | $r^2$ | Precision | Recall | $F_1$ |
|---|---|---|---|---|---|---|---|
| 5% Missing | 0.1%~0.5% | 7140 | 0.9988 | 0.9589 | 0.9845 | 0.9832 | 0.9838 |
| | 0.5%~1% | 10402 | 0.9989 | 0.9838 | 0.9860 | 0.9948 | 0.9903 |
| | 1%~10% | 100338 | 0.9972 | 0.9832 | 0.9884 | 0.9926 | 0.9905 |
| | 10%~20% | 34755 | 0.9935 | 0.9839 | 0.9915 | 0.9902 | 0.9908 |
| | 20%~30% | 19414 | 0.9937 | 0.9874 | 0.9930 | 0.9933 | 0.9932 |
| | 30%~40% | 14704 | 0.9935 | 0.9867 | 0.9935 | 0.9934 | 0.9935 |
| | 40%~50% | 12851 | 0.9936 | 0.9861 | 0.9936 | 0.9936 | 0.9936 |
| | Overall | 199604 | 0.9959 | 0.9880 | 0.9922 | 0.9930 | 0.9926 |
| 15% Missing | MAF | #SNPs | Accuracy | $r^2$ | Precision | Recall | $F_1$ |
| | 0.1%~0.5% | 7140 | 0.9984 | 0.9475 | 0.9763 | 0.9808 | 0.9786 |
| | 0.5%~1% | 10402 | 0.9978 | 0.9692 | 0.9756 | 0.9853 | 0.9804 |
| | 1%~10% | 100338 | 0.9907 | 0.9467 | 0.9743 | 0.9634 | 0.9688 |
| | 10%~20% | 34755 | 0.9852 | 0.9637 | 0.9802 | 0.9781 | 0.9792 |
| | 20%~30% | 19414 | 0.9845 | 0.9689 | 0.9830 | 0.9832 | 0.9831 |
| | 30%~40% | 14704 | 0.9844 | 0.9683 | 0.9841 | 0.9843 | 0.9842 |
| | 40%~50% | 12851 | 0.9846 | 0.9675 | 0.9845 | 0.9846 | 0.9846 |
| | Overall | 199604 | 0.9889 | 0.9688 | 0.9817 | 0.9793 | 0.9805 |
| 25% Missing | MAF | #SNPs | Accuracy | $r^2$ | Precision | Recall | $F_1$ |
| | 0.1%~0.5% | 7140 | 0.9980 | 0.9369 | 0.9785 | 0.9693 | 0.9738 |
| | 0.5%~1% | 10402 | 0.9964 | 0.9512 | 0.9718 | 0.9627 | 0.9672 |
| | 1%~10% | 100338 | 0.9862 | 0.9212 | 0.9643 | 0.9424 | 0.9531 |
| | 10%~20% | 34755 | 0.9777 | 0.9459 | 0.9705 | 0.9666 | 0.9686 |
| | 20%~30% | 19414 | 0.9761 | 0.9529 | 0.9740 | 0.9739 | 0.9740 |
| | 30%~40% | 14704 | 0.9759 | 0.9516 | 0.9756 | 0.9757 | 0.9756 |
| | 40%~50% | 12851 | 0.9761 | 0.9500 | 0.9761 | 0.9761 | 0.9761 |
| | Overall | 199604 | 0.9833 | 0.9533 | 0.9731 | 0.9677 | 0.9704 |
| 50% Missing | MAF | #SNPs | Accuracy | $r^2$ | Precision | Recall | $F_1$ |
| | 0.1%~0.5% | 7140 | 0.9946 | 0.8375 | 0.9666 | 0.8989 | 0.9301 |
| | 0.5%~1% | 10402 | 0.9905 | 0.8746 | 0.9619 | 0.8619 | 0.9057 |
| | 1%~10% | 100338 | 0.9739 | 0.8502 | 0.9431 | 0.8770 | 0.9074 |
| | 10%~20% | 34755 | 0.9561 | 0.8926 | 0.9461 | 0.9287 | 0.9372 |
| | 20%~30% | 19414 | 0.9512 | 0.9031 | 0.9489 | 0.9442 | 0.9465 |
| | 30%~40% | 14704 | 0.9495 | 0.8976 | 0.9493 | 0.9483 | 0.9488 |
| | 40%~50% | 12851 | 0.9491 | 0.8936 | 0.9492 | 0.9490 | 0.9491 |
| | Overall | 199604 | 0.9668 | 0.9066 | 0.9510 | 0.9298 | 0.9401 |

## Models' performance on 1KGP/AMR cohort

**Table S30. STICI's performance on the admixed American cohort of 1KGP dataset**

| | MAF | #SNPs | Accuracy | $r^2$ | Precision | Recall | $F_1$ |
|---|---|---|---|---|---|---|---|
| 5% Missing | **0.1%~0.5%** | **9403** | 0.9811 | 0.6372 | 0.7590 | 0.6811 | 0.6991 |
| | **0.5%~1%** | **11925** | 0.9800 | 0.6837 | 0.7648 | 0.7314 | 0.7367 |
| | **1%~10%** | **82174** | 0.9710 | 0.8659 | 0.8591 | 0.8508 | 0.8535 |
| | **10%~20%** | **24826** | 0.9415 | 0.8979 | 0.8989 | 0.8965 | 0.8976 |
| | **20%~30%** | **16815** | 0.9270 | 0.8940 | 0.9093 | 0.9112 | 0.9102 |
| | **30%~40%** | **15134** | 0.9191 | 0.8699 | 0.9129 | 0.9139 | 0.9134 |
| | **40%~50%** | **12960** | 0.9191 | 0.8651 | 0.9172 | 0.9179 | 0.9175 |
| | **Overall** | **173237** | 0.9552 | 0.9059 | 0.9015 | 0.8988 | 0.9000 |
| 15% Missing | **MAF** | **#SNPs** | **Accuracy** | **$r^2$** | **Precision** | **Recall** | **$F_1$** |
| | **0.1%~0.5%** | **9403** | 0.9823 | 0.6634 | 0.7831 | 0.6899 | 0.7136 |
| | **0.5%~1%** | **11925** | 0.9813 | 0.7031 | 0.7846 | 0.7370 | 0.7488 |
| | **1%~10%** | **82174** | 0.9687 | 0.8428 | 0.8626 | 0.8329 | 0.8457 |
| | **10%~20%** | **24826** | 0.9376 | 0.8838 | 0.8979 | 0.8889 | 0.8933 |
| | **20%~30%** | **16815** | 0.9232 | 0.8812 | 0.9074 | 0.9060 | 0.9067 |
| | **30%~40%** | **15134** | 0.9162 | 0.8607 | 0.9108 | 0.9107 | 0.9107 |
| | **40%~50%** | **12960** | 0.9162 | 0.8549 | 0.9144 | 0.9149 | 0.9146 |
| | **Overall** | **173237** | 0.9529 | 0.8951 | 0.9024 | 0.8917 | 0.8969 |
| 25% Missing | **MAF** | **#SNPs** | **Accuracy** | **$r^2$** | **Precision** | **Recall** | **$F_1$** |
| | **0.1%~0.5%** | **9403** | 0.9833 | 0.6852 | 0.8090 | 0.6945 | 0.7268 |
| | **0.5%~1%** | **11925** | 0.9821 | 0.7046 | 0.8074 | 0.7329 | 0.7562 |
| | **1%~10%** | **82174** | 0.9669 | 0.8225 | 0.8686 | 0.8166 | 0.8395 |
| | **10%~20%** | **24826** | 0.9329 | 0.8692 | 0.8960 | 0.8795 | 0.8875 |
| | **20%~30%** | **16815** | 0.9192 | 0.8695 | 0.9052 | 0.9004 | 0.9028 |
| | **30%~40%** | **15134** | 0.9129 | 0.8524 | 0.9082 | 0.9070 | 0.9076 |
| | **40%~50%** | **12960** | 0.9125 | 0.8454 | 0.9109 | 0.9110 | 0.9110 |
| | **Overall** | **173237** | 0.9505 | 0.8850 | 0.9031 | 0.8843 | 0.8934 |
| 50% Missing | **MAF** | **#SNPs** | **Accuracy** | **$r^2$** | **Precision** | **Recall** | **$F_1$** |
| | **0.1%~0.5%** | **9403** | 0.9838 | 0.6738 | 0.8597 | 0.6823 | 0.7344 |
| | **0.5%~1%** | **11925** | 0.9811 | 0.6382 | 0.8643 | 0.6856 | 0.7426 |
| | **1%~10%** | **82174** | 0.9598 | 0.7597 | 0.8850 | 0.7602 | 0.8109 |
| | **10%~20%** | **24826** | 0.9160 | 0.8238 | 0.8867 | 0.8458 | 0.8649 |
| | **20%~30%** | **16815** | 0.9012 | 0.8298 | 0.8904 | 0.8780 | 0.8839 |
| | **30%~40%** | **15134** | 0.8967 | 0.8193 | 0.8931 | 0.8897 | 0.8913 |
| | **40%~50%** | **12960** | 0.8954 | 0.8088 | 0.8941 | 0.8936 | 0.8938 |
| | **Overall** | **173237** | 0.9403 | 0.8503 | 0.8982 | 0.8559 | 0.8758 |

**Table S31. GenoBERT's performance on the mixed American cohort of 1KGP dataset**

| | MAF | #SNPs | Accuracy | $r^2$ | Precision | Recall | $F_1$ |
|---|---|---|---|---|---|---|---|
| 5% Missing | 0.1%~0.5% | 9403 | 0.9984 | 0.9509 | 0.9750 | 0.9769 | 0.9756 |
| | 0.5%~1% | 11925 | 0.9985 | 0.9603 | 0.9778 | 0.9838 | 0.9805 |
| | 1%~10% | 82174 | 0.9970 | 0.9772 | 0.9876 | 0.9890 | 0.9883 |
| | 10%~20% | 24826 | 0.9908 | 0.9759 | 0.9911 | 0.9834 | 0.9872 |
| | 20%~30% | 16815 | 0.9897 | 0.9771 | 0.9915 | 0.9868 | 0.9891 |
| | 30%~40% | 15134 | 0.9879 | 0.9731 | 0.9897 | 0.9864 | 0.9880 |
| | 40%~50% | 12960 | 0.9878 | 0.9713 | 0.9886 | 0.9873 | 0.9879 |
| | Overall | 173237 | 0.9941 | 0.9812 | 0.9922 | 0.9871 | 0.9896 |
| 15% Missing | MAF | #SNPs | Accuracy | $r^2$ | Precision | Recall | $F_1$ |
| | 0.1%~0.5% | 9403 | 0.9966 | 0.9114 | 0.9432 | 0.9779 | 0.9596 |
| | 0.5%~1% | 11925 | 0.9961 | 0.9125 | 0.9453 | 0.9769 | 0.9603 |
| | 1%~10% | 82174 | 0.9903 | 0.9366 | 0.9675 | 0.9620 | 0.9647 |
| | 10%~20% | 24826 | 0.9825 | 0.9580 | 0.9803 | 0.9713 | 0.9758 |
| | 20%~30% | 16815 | 0.9804 | 0.9599 | 0.9822 | 0.9760 | 0.9791 |
| | 30%~40% | 15134 | 0.9784 | 0.9553 | 0.9806 | 0.9766 | 0.9785 |
| | 40%~50% | 12960 | 0.9773 | 0.9497 | 0.9785 | 0.9766 | 0.9774 |
| | Overall | 173237 | 0.9870 | 0.9624 | 0.9811 | 0.9736 | 0.9773 |
| 25% Missing | MAF | #SNPs | Accuracy | $r^2$ | Precision | Recall | $F_1$ |
| | 0.1%~0.5% | 9403 | 0.9948 | 0.8619 | 0.9153 | 0.9662 | 0.9395 |
| | 0.5%~1% | 11925 | 0.9928 | 0.8398 | 0.9144 | 0.9478 | 0.9304 |
| | 1%~10% | 82174 | 0.9850 | 0.9031 | 0.9487 | 0.9424 | 0.9455 |
| | 10%~20% | 24826 | 0.9771 | 0.9468 | 0.9704 | 0.9656 | 0.9680 |
| | 20%~30% | 16815 | 0.9764 | 0.9533 | 0.9758 | 0.9733 | 0.9745 |
| | 30%~40% | 15134 | 0.9760 | 0.9519 | 0.9765 | 0.9751 | 0.9758 |
| | 40%~50% | 12960 | 0.9747 | 0.9455 | 0.9750 | 0.9744 | 0.9747 |
| | Overall | 173237 | 0.9826 | 0.9512 | 0.9712 | 0.9675 | 0.9693 |
| 50% Missing | MAF | #SNPs | Accuracy | $r^2$ | Precision | Recall | $F_1$ |
| | 0.1%~0.5% | 9403 | 0.9930 | 0.8051 | 0.9686 | 0.8704 | 0.9139 |
| | 0.5%~1% | 11925 | 0.9884 | 0.7210 | 0.9546 | 0.8195 | 0.8761 |
| | 1%~10% | 82174 | 0.9766 | 0.8439 | 0.9465 | 0.8753 | 0.9080 |
| | 10%~20% | 24826 | 0.9628 | 0.9123 | 0.9561 | 0.9359 | 0.9457 |
| | 20%~30% | 16815 | 0.9602 | 0.9202 | 0.9598 | 0.9517 | 0.9557 |
| | 30%~40% | 15134 | 0.9600 | 0.9182 | 0.9608 | 0.9573 | 0.9590 |
| | 40%~50% | 12960 | 0.9571 | 0.9066 | 0.9576 | 0.9562 | 0.9569 |
| | Overall | 173237 | 0.9718 | 0.9195 | 0.9606 | 0.9356 | 0.9477 |

## Models' performance on 1KGP/SAS cohort

**Table S32. STICI's performance on the South Asian cohort of 1KGP dataset**

| | MAF | #SNPs | Accuracy | $r^2$ | Precision | Recall | $F_1$ |
|---|---|---|---|---|---|---|---|
| 5% Missing | 0.1%~0.5% | 3695 | 0.9797 | 0.7655 | 0.8847 | 0.6335 | 0.6952 |
| | 0.5%~1% | 5229 | 0.9810 | 0.6896 | 0.8774 | 0.6990 | 0.7559 |
| | 1%~10% | 49481 | 0.9820 | 0.9084 | 0.9504 | 0.9077 | 0.9275 |
| | 10%~20% | 22056 | 0.9757 | 0.9446 | 0.9658 | 0.9565 | 0.9610 |
| | 20%~30% | 17336 | 0.9732 | 0.9418 | 0.9702 | 0.9666 | 0.9684 |
| | 30%~40% | 15352 | 0.9711 | 0.9335 | 0.9703 | 0.9690 | 0.9696 |
| | 40%~50% | 13355 | 0.9704 | 0.9293 | 0.9703 | 0.9699 | 0.9700 |
| | Overall | 126504 | 0.9770 | 0.9433 | 0.9675 | 0.9561 | 0.9617 |
| 15% Missing | MAF | #SNPs | Accuracy | $r^2$ | Precision | Recall | $F_1$ |
| | 0.1%~0.5% | 3695 | 0.9804 | 0.7725 | 0.8926 | 0.6472 | 0.7119 |
| | 0.5%~1% | 5229 | 0.9820 | 0.7076 | 0.8903 | 0.7162 | 0.7744 |
| | 1%~10% | 49481 | 0.9810 | 0.9033 | 0.9503 | 0.9050 | 0.9262 |
| | 10%~20% | 22056 | 0.9747 | 0.9425 | 0.9660 | 0.9561 | 0.9610 |
| | 20%~30% | 17336 | 0.9730 | 0.9421 | 0.9703 | 0.9675 | 0.9689 |
| | 30%~40% | 15352 | 0.9714 | 0.9358 | 0.9707 | 0.9698 | 0.9703 |
| | 40%~50% | 13355 | 0.9712 | 0.9329 | 0.9711 | 0.9708 | 0.9710 |
| | Overall | 126504 | 0.9766 | 0.9430 | 0.9679 | 0.9567 | 0.9622 |
| 25% Missing | MAF | #SNPs | Accuracy | $r^2$ | Precision | Recall | $F_1$ |
| | 0.1%~0.5% | 3695 | 0.9807 | 0.7767 | 0.9017 | 0.6501 | 0.7168 |
| | 0.5%~1% | 5229 | 0.9824 | 0.7170 | 0.8962 | 0.7215 | 0.7807 |
| | 1%~10% | 49481 | 0.9803 | 0.9002 | 0.9505 | 0.9023 | 0.9247 |
| | 10%~20% | 22056 | 0.9730 | 0.9393 | 0.9648 | 0.9540 | 0.9593 |
| | 20%~30% | 17336 | 0.9718 | 0.9407 | 0.9691 | 0.9664 | 0.9678 |
| | 30%~40% | 15352 | 0.9706 | 0.9353 | 0.9700 | 0.9693 | 0.9696 |
| | 40%~50% | 13355 | 0.9706 | 0.9327 | 0.9705 | 0.9703 | 0.9704 |
| | Overall | 126504 | 0.9757 | 0.9416 | 0.9672 | 0.9558 | 0.9614 |
| 50% Missing | MAF | #SNPs | Accuracy | $r^2$ | Precision | Recall | $F_1$ |
| | 0.1%~0.5% | 3695 | 0.9817 | 0.7890 | 0.9275 | 0.6614 | 0.7327 |
| | 0.5%~1% | 5229 | 0.9828 | 0.7271 | 0.9257 | 0.7192 | 0.7879 |
| | 1%~10% | 49481 | 0.9751 | 0.8758 | 0.9486 | 0.8755 | 0.9085 |
| | 10%~20% | 22056 | 0.9636 | 0.9190 | 0.9566 | 0.9396 | 0.9478 |
| | 20%~30% | 17336 | 0.9634 | 0.9255 | 0.9606 | 0.9582 | 0.9594 |
| | 30%~40% | 15352 | 0.9624 | 0.9211 | 0.9617 | 0.9616 | 0.9616 |
| | 40%~50% | 13355 | 0.9633 | 0.9204 | 0.9631 | 0.9632 | 0.9632 |
| | Overall | 126504 | 0.9692 | 0.9278 | 0.9605 | 0.9456 | 0.9529 |

**Table S33. GenoBERT's performance on the South Asian cohort of 1KGP dataset**

| | MAF | #SNPs | Accuracy | $r^2$ | Precision | Recall | $F_1$ |
|---|---|---|---|---|---|---|---|
| 5% Missing | 0.1%~0.5% | 3695 | 0.9996 | 0.9941 | 0.9964 | 0.9967 | 0.9966 |
| | 0.5%~1% | 5229 | 0.9996 | 0.9923 | 0.9953 | 0.9976 | 0.9965 |
| | 1%~10% | 49481 | 0.9992 | 0.9950 | 0.9973 | 0.9974 | 0.9973 |
| | 10%~20% | 22056 | 0.9976 | 0.9940 | 0.9978 | 0.9957 | 0.9968 |
| | 20%~30% | 17336 | 0.9966 | 0.9925 | 0.9973 | 0.9956 | 0.9964 |
| | 30%~40% | 15352 | 0.9951 | 0.9894 | 0.9955 | 0.9946 | 0.9951 |
| | 40%~50% | 13355 | 0.9957 | 0.9897 | 0.9959 | 0.9955 | 0.9957 |
| | Overall | 126504 | 0.9977 | 0.9940 | 0.9976 | 0.9957 | 0.9967 |
| 15% Missing | MAF | #SNPs | Accuracy | $r^2$ | Precision | Recall | $F_1$ |
| | 0.1%~0.5% | 3695 | 0.9993 | 0.9881 | 0.9924 | 0.9935 | 0.9929 |
| | 0.5%~1% | 5229 | 0.9991 | 0.9821 | 0.9903 | 0.9923 | 0.9913 |
| | 1%~10% | 49481 | 0.9966 | 0.9815 | 0.9925 | 0.9858 | 0.9891 |
| | 10%~20% | 22056 | 0.9947 | 0.9875 | 0.9945 | 0.9912 | 0.9928 |
| | 20%~30% | 17336 | 0.9945 | 0.9878 | 0.9948 | 0.9934 | 0.9941 |
| | 30%~40% | 15352 | 0.9937 | 0.9860 | 0.9941 | 0.9934 | 0.9938 |
| | 40%~50% | 13355 | 0.9939 | 0.9855 | 0.9941 | 0.9938 | 0.9940 |
| | Overall | 126504 | 0.9955 | 0.9886 | 0.9950 | 0.9922 | 0.9936 |
| 25% Missing | MAF | #SNPs | Accuracy | $r^2$ | Precision | Recall | $F_1$ |
| | 0.1%~0.5% | 3695 | 0.9985 | 0.9785 | 0.9911 | 0.9800 | 0.9855 |
| | 0.5%~1% | 5229 | 0.9980 | 0.9662 | 0.9879 | 0.9755 | 0.9816 |
| | 1%~10% | 49481 | 0.9952 | 0.9745 | 0.9897 | 0.9791 | 0.9843 |
| | 10%~20% | 22056 | 0.9932 | 0.9846 | 0.9925 | 0.9892 | 0.9908 |
| | 20%~30% | 17336 | 0.9935 | 0.9862 | 0.9936 | 0.9925 | 0.9930 |
| | 30%~40% | 15352 | 0.9932 | 0.9851 | 0.9934 | 0.9929 | 0.9932 |
| | 40%~50% | 13355 | 0.9933 | 0.9846 | 0.9934 | 0.9932 | 0.9933 |
| | Overall | 126504 | 0.9944 | 0.9862 | 0.9934 | 0.9904 | 0.9919 |
| 50% Missing | MAF | #SNPs | Accuracy | $r^2$ | Precision | Recall | $F_1$ |
| | 0.1%~0.5% | 3695 | 0.9946 | 0.9291 | 0.9833 | 0.9091 | 0.9430 |
| | 0.5%~1% | 5229 | 0.9927 | 0.8797 | 0.9739 | 0.8919 | 0.9288 |
| | 1%~10% | 49481 | 0.9879 | 0.9373 | 0.9772 | 0.9446 | 0.9603 |
| | 10%~20% | 22056 | 0.9842 | 0.9642 | 0.9823 | 0.9751 | 0.9787 |
| | 20%~30% | 17336 | 0.9849 | 0.9679 | 0.9847 | 0.9828 | 0.9837 |
| | 30%~40% | 15352 | 0.9844 | 0.9664 | 0.9847 | 0.9839 | 0.9843 |
| | 40%~50% | 13355 | 0.9844 | 0.9643 | 0.9846 | 0.9843 | 0.9845 |
| | Overall | 126504 | 0.9865 | 0.9671 | 0.9842 | 0.9768 | 0.9804 |

## Models' performance on 1KGP/EAS cohort

**Table S34. STICI's performance on the East Asian cohort of 1KGP dataset**

| | MAF | #SNPs | Accuracy | $r^2$ | Precision | Recall | $F_1$ |
|---|---|---|---|---|---|---|---|
| 5% Missing | **0.1%~0.5%** | **3486** | 0.9792 | 0.8365 | 0.7288 | 0.6255 | 0.6617 |
| | **0.5%~1%** | **4236** | 0.9796 | 0.8892 | 0.7952 | 0.7184 | 0.7478 |
| | **1%~10%** | **39158** | 0.9643 | 0.9074 | 0.8577 | 0.8438 | 0.8504 |
| | **10%~20%** | **19749** | 0.9353 | 0.9024 | 0.8961 | 0.8884 | 0.8922 |
| | **20%~30%** | **14357** | 0.9195 | 0.8870 | 0.9036 | 0.9015 | 0.9025 |
| | **30%~40%** | **13453** | 0.9083 | 0.8674 | 0.9043 | 0.9020 | 0.9029 |
| | **40%~50%** | **12864** | 0.9059 | 0.8557 | 0.9053 | 0.9051 | 0.9050 |
| | **Overall** | **107303** | 0.9400 | 0.9074 | 0.9002 | 0.8939 | 0.8970 |
| 15% Missing | **MAF** | **#SNPs** | **Accuracy** | **$r^2$** | **Precision** | **Recall** | **$F_1$** |
| | **0.1%~0.5%** | **3486** | 0.9796 | 0.8430 | 0.7387 | 0.6381 | 0.6731 |
| | **0.5%~1%** | **4236** | 0.9792 | 0.8826 | 0.7950 | 0.7217 | 0.7502 |
| | **1%~10%** | **39158** | 0.9621 | 0.8953 | 0.8582 | 0.8369 | 0.8468 |
| | **10%~20%** | **19749** | 0.9353 | 0.9009 | 0.8985 | 0.8904 | 0.8944 |
| | **20%~30%** | **14357** | 0.9212 | 0.8854 | 0.9071 | 0.9046 | 0.9058 |
| | **30%~40%** | **13453** | 0.9118 | 0.8675 | 0.9084 | 0.9064 | 0.9073 |
| | **40%~50%** | **12864** | 0.9103 | 0.8575 | 0.9098 | 0.9096 | 0.9095 |
| | **Overall** | **107303** | 0.9404 | 0.9046 | 0.9038 | 0.8965 | 0.9001 |
| 25% Missing | **MAF** | **#SNPs** | **Accuracy** | **$r^2$** | **Precision** | **Recall** | **$F_1$** |
| | **0.1%~0.5%** | **3486** | 0.9797 | 0.8411 | 0.7540 | 0.6274 | 0.6673 |
| | **0.5%~1%** | **4236** | 0.9772 | 0.8592 | 0.7959 | 0.6813 | 0.7216 |
| | **1%~10%** | **39158** | 0.9596 | 0.8810 | 0.8619 | 0.8233 | 0.8410 |
| | **10%~20%** | **19749** | 0.9347 | 0.8970 | 0.9012 | 0.8912 | 0.8961 |
| | **20%~30%** | **14357** | 0.9231 | 0.8833 | 0.9107 | 0.9079 | 0.9093 |
| | **30%~40%** | **13453** | 0.9159 | 0.8683 | 0.9130 | 0.9112 | 0.9120 |
| | **40%~50%** | **12864** | 0.9158 | 0.8617 | 0.9153 | 0.9152 | 0.9151 |
| | **Overall** | **107303** | 0.9407 | 0.9010 | 0.9082 | 0.8981 | 0.9031 |
| 50% Missing | **MAF** | **#SNPs** | **Accuracy** | **$r^2$** | **Precision** | **Recall** | **$F_1$** |
| | **0.1%~0.5%** | **3486** | 0.9820 | 0.8628 | 0.8077 | 0.6412 | 0.6901 |
| | **0.5%~1%** | **4236** | 0.9783 | 0.8626 | 0.8301 | 0.6756 | 0.7265 |
| | **1%~10%** | **39158** | 0.9572 | 0.8654 | 0.8786 | 0.8022 | 0.8354 |
| | **10%~20%** | **19749** | 0.9297 | 0.8827 | 0.9037 | 0.8838 | 0.8933 |
| | **20%~30%** | **14357** | 0.9206 | 0.8718 | 0.9108 | 0.9065 | 0.9086 |
| | **30%~40%** | **13453** | 0.9174 | 0.8631 | 0.9149 | 0.9143 | 0.9146 |
| | **40%~50%** | **12864** | 0.9195 | 0.8622 | 0.9192 | 0.9191 | 0.9191 |
| | **Overall** | **107303** | 0.9394 | 0.8930 | 0.9129 | 0.8969 | 0.9046 |

**Table S35. GenoBERT's performance on the East Asian cohort of 1KGP dataset**

| | MAF | #SNPs | Accuracy | $r^2$ | Precision | Recall | $F_1$ |
|---|---|---|---|---|---|---|---|
| 5% Missing | 0.1%~0.5% | 3486 | 0.9997 | 0.9970 | 0.9970 | 0.9970 | 0.9970 |
| | 0.5%~1% | 4236 | 0.9998 | 0.9976 | 0.9974 | 0.9986 | 0.9980 |
| | 1%~10% | 39158 | 0.9991 | 0.9961 | 0.9969 | 0.9970 | 0.9970 |
| | 10%~20% | 19749 | 0.9976 | 0.9954 | 0.9975 | 0.9961 | 0.9968 |
| | 20%~30% | 14357 | 0.9978 | 0.9953 | 0.9978 | 0.9974 | 0.9976 |
| | 30%~40% | 13453 | 0.9975 | 0.9946 | 0.9975 | 0.9974 | 0.9974 |
| | 40%~50% | 12864 | 0.9978 | 0.9947 | 0.9978 | 0.9978 | 0.9978 |
| | Overall | 107303 | 0.9983 | 0.9962 | 0.9979 | 0.9973 | 0.9976 |
| 15% Missing | MAF | #SNPs | Accuracy | $r^2$ | Precision | Recall | $F_1$ |
| | 0.1%~0.5% | 3486 | 0.9996 | 0.9958 | 0.9968 | 0.9947 | 0.9958 |
| | 0.5%~1% | 4236 | 0.9993 | 0.9941 | 0.9968 | 0.9914 | 0.9941 |
| | 1%~10% | 39158 | 0.9963 | 0.9856 | 0.9941 | 0.9835 | 0.9887 |
| | 10%~20% | 19749 | 0.9950 | 0.9898 | 0.9949 | 0.9921 | 0.9935 |
| | 20%~30% | 14357 | 0.9950 | 0.9898 | 0.9951 | 0.9943 | 0.9947 |
| | 30%~40% | 13453 | 0.9951 | 0.9890 | 0.9952 | 0.9950 | 0.9951 |
| | 40%~50% | 12864 | 0.9953 | 0.9889 | 0.9953 | 0.9953 | 0.9953 |
| | Overall | 107303 | 0.9958 | 0.9907 | 0.9954 | 0.9932 | 0.9943 |
| 25% Missing | MAF | #SNPs | Accuracy | $r^2$ | Precision | Recall | $F_1$ |
| | 0.1%~0.5% | 3486 | 0.9965 | 0.9669 | 0.9953 | 0.9271 | 0.9586 |
| | 0.5%~1% | 4236 | 0.9943 | 0.9556 | 0.9934 | 0.9090 | 0.9471 |
| | 1%~10% | 39158 | 0.9922 | 0.9708 | 0.9900 | 0.9632 | 0.9762 |
| | 10%~20% | 19749 | 0.9917 | 0.9832 | 0.9911 | 0.9870 | 0.9891 |
| | 20%~30% | 14357 | 0.9921 | 0.9841 | 0.9919 | 0.9911 | 0.9915 |
| | 30%~40% | 13453 | 0.9921 | 0.9831 | 0.9922 | 0.9920 | 0.9921 |
| | 40%~50% | 12864 | 0.9922 | 0.9823 | 0.9923 | 0.9922 | 0.9922 |
| | Overall | 107303 | 0.9923 | 0.9833 | 0.9919 | 0.9872 | 0.9895 |
| 50% Missing | MAF | #SNPs | Accuracy | $r^2$ | Precision | Recall | $F_1$ |
| | 0.1%~0.5% | 3486 | 0.9956 | 0.9561 | 0.9877 | 0.9137 | 0.9475 |
| | 0.5%~1% | 4236 | 0.9917 | 0.9350 | 0.9836 | 0.8734 | 0.9212 |
| | 1%~10% | 39158 | 0.9849 | 0.9436 | 0.9785 | 0.9301 | 0.9529 |
| | 10%~20% | 19749 | 0.9811 | 0.9617 | 0.9798 | 0.9705 | 0.9750 |
| | 20%~30% | 14357 | 0.9817 | 0.9629 | 0.9815 | 0.9793 | 0.9804 |
| | 30%~40% | 13453 | 0.9820 | 0.9616 | 0.9821 | 0.9817 | 0.9819 |
| | 40%~50% | 12864 | 0.9820 | 0.9596 | 0.9820 | 0.9820 | 0.9820 |
| | Overall | 107303 | 0.9837 | 0.9645 | 0.9821 | 0.9732 | 0.9776 |

# Datasets specifications

**Table S36.** provides the summary and specifications of the LOS dataset.

**Table S36. Specification for the LOS dataset**

| | Overall | Female | Male | p-value | African American | Caucasian | p-value |
|---|---|---|---|---|---|---|---|
| **#Samples (%)** | 7675 | 3611(47.05) | 4062 (52.93) | | 3103 (40.43) | 4405 (57.39) | |
| **Age, yr (mean (SD))** | 43.75 (14.40) | 44.97 (15.54) | 42.66 (13.23) | <.001 | 44.41 (13.14) | 43.45 (15.19) | <.001 |
| **Weight, kg (mean (SD))** | 79.37 (19.46) | 74.67 (20.26) | 83.54 (17.72) | <.001 | 83.82 (20.48) | 76.67 (18.19) | <.001 |

**Table S37.** provides the summary and specifications for both datasets after preprocessing steps:

**Table S37. Specification of cohort composition after preprocessing for the LOS and 1KGP datasets.**

| | #Samples | | | | | | | | |
|---|---|---|---|---|---|---|---|---|---|
| **Training** | 6001 | 3522 | 2479 | 2035 | 417 | 410 | 404 | 393 | 411 |
| **Validation** | 749 | 440 | 309 | 250 | 51 | 50 | 50 | 49 | 50 |
| **Test** | 754 | 442 | 312 | 263 | 54 | 53 | 52 | 50 | 54 |
| **#SNPs** | 257352 | 188399 | 307599 | 241113 | 120577 | 199604 | 173237 | 126504 | 107303 |
| **Population** | **ALL** | **CA** | **AA** | **ALL** | **EUR** | **AFR** | **AMR** | **SAS** | **EAS** |
| **Dataset** | | **LOS** | | | | **1KGP** | | | |

Shown are the numbers of samples in the training, validation, and test splits, along with the number of retained SNPs, stratified by population group.